\documentclass{article}
\usepackage[margin=1in]{geometry}
\usepackage{graphicx} 
\usepackage{caption}
\usepackage{subcaption}
\usepackage{tabularx}
\usepackage{xcolor}
\usepackage{bm}
\usepackage{amsmath, amssymb}
\usepackage{physics}
\usepackage{multirow}
\usepackage{authblk}
\usepackage[T1]{fontenc}
\usepackage[hidelinks]{hyperref}
\usepackage{cite}
\hypersetup{
  colorlinks=true,
  linktoc=all,
  citecolor=purple,
  linkcolor=blue,
}

\DeclareMathOperator*{\argmin}{arg\,min}

\makeatletter
\def\@fnsymbol#1{*}
\makeatother

\begin{document}

\begin{titlepage}
\centering
\vspace*{\fill}

{\LARGE \bfseries Toward Quantum-Enabled Biomarker Discovery: An Outlook from Q4Bio\par}
\vspace{1.25em}

{\large
Dhirpal Shah$^{1}$, Mariesa Teo$^{2}$, Ryan A.~Robinett$^{1}$, Sophia Madejski$^{2}$, Zachary Morrell$^{1}$,\\
Siddhi Ramesh$^{3}$, Colin Campbell$^{4}$, Bharath Thotakura$^{4}$, Victory Omole$^{4}$, Ben Hall$^{4}$,\\
Aram W.~Harrow$^{5}$, Teague Tomesh$^{4\ast}$, Alexander T.~Pearson$^{6,7\ast}$,\\
Frederic T.~Chong$^{1,4\ast}$, Samantha J.~Riesenfeld$^{2,6,7,8,9\ast}$\par}
\vspace{1em}

{\small
$^{1}$ Department of Computer Science, University of Chicago, Chicago, IL, USA\\
$^{2}$ Pritzker School of Molecular Engineering, University of Chicago, Chicago, IL, USA\\
$^{3}$ Pritzker School of Medicine, University of Chicago, Chicago, IL, USA\\
$^{4}$ Infleqtion, Chicago, IL, USA\\
$^{5}$ Center for Theoretical Physics, Massachusetts Institute of Technology, Cambridge, MA, USA\\
$^{6}$ Department of Medicine, University of Chicago, Chicago, IL, USA\\
$^{7}$ Chan Zuckerberg Biohub Chicago LLC, Chicago, IL, USA\\
$^{8}$ NSF-Simons National Institute for Theory and Mathematics in Biology, Chicago IL 60611\\
$^{9}$ Institute for Biophysical Dynamics, University of Chicago, Chicago, IL 60637\\
\par
}
\vspace{1em}

{\small
$^{\ast}$\;Corresponding authors:\;
\href{mailto:teague.tomesh@infleqtion.com}{teague.tomesh@infleqtion.com}\;|\;
\href{mailto:frederic.chong@infleqtion.com}{frederic.chong@infleqtion.com}\;|\;
\href{mailto:apearson5@uchicago.edu}{apearson5@uchicago.edu}\;|\;
\href{mailto:sriesenfeld@uchicago.edu}{sriesenfeld@uchicago.edu}\par}
\vspace{1em}

{\small \today\par}

\vspace*{\fill}
\end{titlepage}

\clearpage
\addcontentsline{toc}{section}{Executive Summary}
\section*{Executive Summary}\label{sec:execsum}

We present a case study in \textit{empirical quantum advantage (EQA)} for a clinically relevant task: biomarker discovery in precision oncology. We target a practical EQA crossover where a quantum-in-the-loop subroutine measurably outperforms strong classical baselines on the same instances under realistic time or shot budgets. Because classical methods keep improving, this crossover is a moving target. Our co-design approach aims to keep the advantage window open by shaping the problem, algorithm, and use of hardware together.

\textbf{Biomarker discovery:} We focus on biomarker discovery where models must pick clinically informative feature sets from high-dimensional, multimodal biological data. Biomarkers can guide clinical diagnosis, treatment, and research, and enable patients to avoid unneeded, potentially harmful, therapies. Small feature sets avoid overparameterization risks due to small sample numbers, support biological interpretation, and can be easier to translate clinically, but for many patients, current biomarkers fail or offer no guidance.  

\textbf{Higher-order PCBO + Hyper-RQAOA:} Formulating biomarker discovery as feature selection, we cast it as a higher-order polynomial constrained binary optimization (PCBO) problem, \texttt{entropy-cubo}, that captures first\nobreakdashes-, second\nobreakdashes-, and third-order interactions among features. For tractability on current devices, we adapt the recursive quantum approximate optimization algorithm (RQAOA) into hyper-RQAOA (HRQAOA) to transfer parameters learned on small, classically simulable subproblems to initialize larger circuits and then recursively fix variables. By eliminating expensive variational loops on large instances, HRQAOA reduces the number of quantum evaluations by orders of magnitude while preserving solution quality. On benchmarks, the results are competitive compared to strong classical baselines, even when the PCBO is sparsified.

\textbf{Hardware hybrid co-design:} Effectively bridging the algorithm and hardware requires further co-design. We discretize gene expression and pathomics features into compact alphabets that stabilize information theoretic scoring and reduce quantum encoding cost. We sparsify the Hamiltonian to reduce circuit depth and two-qubit gate usage, and we map high-weight third-order terms onto heavy-hexagon topology so as to minimize SWAPs on IBM Heron-class systems. In initial hardware runs, we compare sampler-based and estimator-based routes, with results that underscore the impact of targeting expectation values on these devices. The estimator with twirled readout error extinction (TREX) and zero-noise extrapolation (ZNE) mitigation materially improves agreement with noiseless baselines and yields more reliable edge-fixing.

\textbf{Crossover and resource analysis:} Our scaling study shows that exact solvers and strong heuristics face growing runtimes on dense, third-order \texttt{entropy-cubo} problems beyond $N\approx100$. By contrast, HRQAOA can shrink such instances via a few edge-fixing rounds (polynomial overhead) before a classical finish. The crossover point for EQA depends less on qubit count than on how fast a fault-tolerant device can build the correlation dictionary, i.e., estimate and rank many low-weight Pauli expectations. Hence, shot throughput, constant-depth Pauli-phasor compilation, and connectivity are the binding dials. In the near term, we target that runtime by removing variational loops with parameter transfer, trading depth for width, and choosing codes that curb $T$-gate and routing overhead. While advances in classical heuristics may shift the crossover, as platforms approach $\mathcal{O}(10^2)$ logical qubits (per public roadmaps), these cross-layer optimizations delineate a plausible regime where the hybrid method could beat exact solvers.

\textbf{Scientific dividends and clinical focus:} Scientifically, this pipeline  reframes biomarker discovery to efficiently exploit complex biological structure and interactions, rather than focusing on individual markers or large models. It is already delivering practical dividends: unexpectedly compact, interpretable feature panels, robust cross-dataset performance, and a clear route from hardware-aware design to clinical translation. After using tissue-of-origin cancer classification as a testbed, we focus on treatment-response prediction in head-and-neck cancer, using curated, multimodal data from an institutional cohort, scaling circuit depth and rounds where hardware allows. Regardless of where the exact crossover is, the co-design methods here -- problem shaping, parameter transfer, topology-aware sparsification, and mitigation-first estimation -- offer a template for turning real devices into accelerators within clinically meaningful workflows.

\textbf{Q4Bio program impact:} This work was launched through the Wellcome Leap Quantum for Bio (Q4Bio) program, as an interdepartmental, multi-institutional collaboration that also involves close coordination with IBM researchers on algorithm–hardware co-design. Q4Bio program milestones and periodic reviews helped keep a steady cadence from concept to functioning pipeline and early hardware trials. Funding and compute resources, including a UChicago–IBM grant and priority access to IBM cloud QPUs, supported large-scale simulations, cohort curation, and method validation. Q4Bio provided a practical framework and infrastructure that, compared with the usual research cycle, led to more coordination and faster progress.

\clearpage
\tableofcontents

\clearpage
\section{Introduction}\label{sec:introduction}
Demonstrating the utility of quantum computation in solving real-world problems is currently one of the central goals of the field. It calls for empirical quantum advantage (EQA), which entails measurable gains on actual hardware over state-of-the-art classical methods. Because classical algorithms keep improving, EQA is a moving target, and the crossover point can shift even if an advantage persists. With that lens, we treat quantum processors as domain-specific accelerators within a hybrid quantum-classical algorithmic framework. Constructing and implementing such a hybrid algorithm requires co-design across the stack: first, co-design of the application and algorithm in the choice of a useful task and an appropriate quantum subroutine for which we expect to see improvement over classical methods, and second, co-design of the algorithm and hardware in working within device constraints to construct an approach that is implementable on hardware, while both retaining the quantum-classical separation of the algorithm and the usefulness of the application. In this work, we present a perspective on developing and realizing a hybrid quantum-classical pipeline for a target biomedical application.

Biomarker discovery -- the identification of biological features, processes, or patterns that signal the risk, existence, subtype, progression, or treatment response of a disease -- is a critical area of cancer research~\cite{dakal2024emerging}. Cancer biomarkers could be clinically deployed for cancer screening, diagnosis, risk stratification, treatment selection, and surveillance. Biomarkers generally seek to discriminate distinct biological phenomena into classes that are relevant to clinical care, often anchoring to molecular states. The promise of \emph{precision oncology} -- identifying the right treatment for the right patient at the right time -- is that selecting a treatment based on a biomarker in order to target a molecular state will improve cancer outcomes. Unfortunately, the majority of clinical trials are not successful, including trials selecting for current biomarkers. Moreover, current biomarkers are not relevant for many patients (e.g., their tumors do not exhibit mutations that guide treatment). Nonetheless, the current global market size for cancer biomarkers is estimated to be in the range of US\$15–20 billion dollars annually~\cite{dacosta_byfield_2025_biomarker}. Advances in computing technology are also changing how biomarkers are conceptualized. Although pathology images are routinely used for diagnosis, precision biomarkers have, until recently, consisted of proteins, gene mutations, or other molecules. Now the biomarker concept can include images processed with computational tools or multiple data modalities jointly analyzed. In 2025, the first AI-enabled digital pathology companion diagnostic -- a biomarker specifically developed to identify tumors suitable for a certain medical treatment -- was granted Food and Drug Administration (FDA) Breakthrough designation, paving the way for more computational biomarkers to come~\cite{roche2025_trop2_bdd}. However, current classical computing approaches are limited in their ability to process and interpret complex multimodal data, restricting much of the present classical biomarker discovery effort to single-mode analysis. Even single-modality biomarkers have not successfully leveraged complex information encoded in genomic and transcriptomic data, likely due, in part, to the relative scarcity of samples compared to other settings. 

Quantum computing provides promising (though currently limited) application-specific tools to help address these computational gaps. Employing the principle of co-design, we develop a hybrid quantum-classical algorithm, integrated within a clinical data analysis workflow for biomarker discovery. Given samples, a feature set, and a target function (e.g., classification) of the samples, the algorithm selects a small set of features most relevant for predicting the target. Co-design, characterized by the free-flow of information up and down the software-hardware stack to inform design decisions, has proven to be effective in areas of quantum computing, such as tailoring the design of error correcting codes to the hardware's qubit connectivity or modifying quantum algorithm design~\cite{tomesh2021quantum}. In this article, we describe specific examples of how we  employed and leveraged co-design throughout the project. 

This research project began in 2023 as part of the Wellcome Leap Quantum for Bio (Q4Bio) Challenge Program, which specifically aimed at bringing together cross-disciplinary teams of researchers to answer the question: can scientists develop new algorithms that deliver near-term quantum advantage for human health applications?
Our team, composed of quantum computing experts, computational biologists, and clinical oncologists, addressed this challenge by developing hybrid algorithms for feature selection on multimodal cancer datasets.
The central aspect of our approach is framing the feature selection problem as a polynomial constrained binary optimization problem (PCBO), enabling us to apply the recursive quantum approximate optimization algorithm (RQAOA)~\cite{bravyi2020obstacles} as a core subroutine within our hybrid algorithm. We tailor this algorithm to current hardware capabilities, developing methods for addressing a finite shot budget, imperfect gates and readout, and limited circuit depths. 

\begin{figure}
      \centering
      \includegraphics[width=0.8\linewidth]{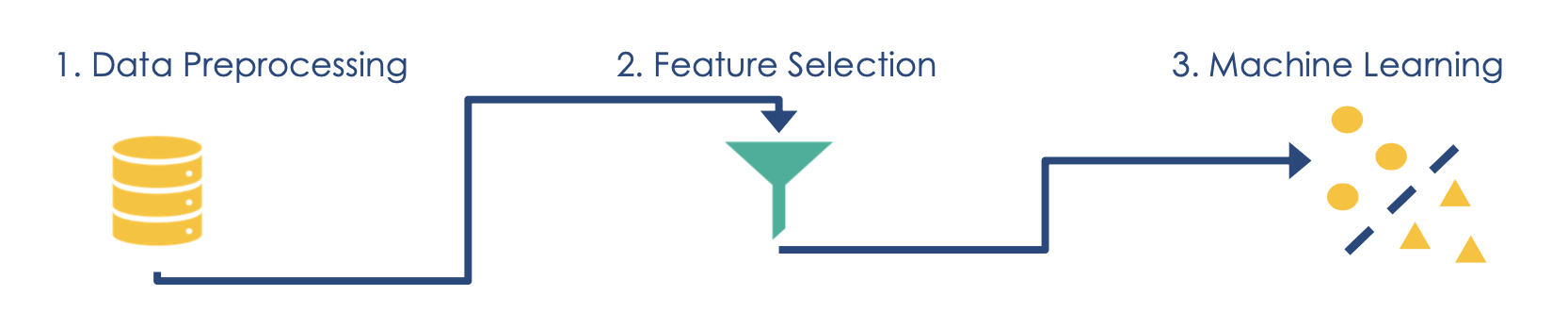}
      \caption{Overview of a generic data analysis pipeline. Our team explored potential applications of quantum computers within each step, but ultimately focused efforts on developing hybrid feature-selection algorithms.}
      \label{fig:pipeline}
\end{figure}

Our process of designing an effective hybrid quantum-classical algorithm focused first on identifying the subroutine with high possibility for quantum improvement, within the context of a problem with potentially impactful applications. 
Starting by considering the entire computational pipeline    (Fig.~\ref{fig:pipeline}) that encompasses multimodal cancer data analysis, we identified possible applications of quantum computers at each point. For example, within the data preprocessing step we considered the application of quantum algorithms for constructing coreset summaries of the pathomics data modality~\cite{harrow2020small, tomesh2021coreset}. Within the machine learning step, we evaluated the efficacy of quantum machine learning (QML) models on various classification tasks compared to classical models.
Ultimately, we found the feature selection step to be a promising application area, given the strengths and weaknesses of quantum computers, with potential for demonstrating EQA with the near-term and intermediate-scale fault-tolerant quantum systems expected to be brought online in the future. 

From the perspective of combinatorial optimization, the task of feature selection is to search through a vast number of possible feature combinations and select the subset that most accurately predicts the desired learning outcome. This task is well suited to quantum computers in that it takes relatively few bits to define the problem, for instance, the membership of a set of $N$ features in the selected subset may be represented by the $\{0,1\}$ state of $N$ qubits, but the number of possible combinations to search through quickly explodes.
In our work, we co-design the quantum and classical portions of the feature selection algorithm such that the quantum processor is relegated to a very specific task -- producing low energy samples from a particular distribution -- that it can efficiently execute while the classical processor carries out the rest of the algorithm. The question is whether this hybrid algorithm can outperform a state-of-the-art classical approach for generic instances of this feature selection problem. 
Since we do not yet have access to intermediate-scale fault-tolerant quantum computers of sufficient capabilities to empirically test this hypothesis against strong classical optimizers, such as Gurobi~\cite{gurobi2024}, we focus on studying the algorithm's performance in simulation, benchmarking the scalability of classical solvers as a function of problem size, and extrapolating the resources required for a quantum computer to compete at the necessary scale. We find that dense third-order PCBOs yield compact, interpretable panels, RQAOA with parameter transfer reduces instance size at polynomial cost, and classical runtimes rise sharply beyond $N\approx100$, suggesting a plausible crossover once correlation-dictionary estimation can be executed efficiently.

In this outlook paper, we discuss the role of co-design in
the development of our hybrid algorithm (Section~\ref{sec:algorithm}). Moving further down the stack, we discuss methods for physically implementing our algorithm on current quantum devices, balancing device constraints while maintaining a problem with quantum-classical separation and practical relevance (Section~\ref{sec:hardware}). Looking towards implementations on future devices, we discuss the quantum hardware capabilities necessary to implement a hybrid algorithm that is competitive with the classical state of the art, and how these requirements align with current device roadmaps (Section~\ref{sec:resources}). Finally, we identify the potential implications of our pipeline for the study of cancer treatment response, as well as further applications beyond oncology (Section~\ref{sec:applications}). While individual components of this pipeline might require adaptation as advances continue to be made in classical and quantum computing, our work highlights the critical role of co-design in bridging hardware capabilities to useful applications.

\section{Designing a Hybrid Pipeline for Cancer Biomarker Discovery} \label{sec:algorithm}

\begin{figure*}[t]
    \centering
    \includegraphics[width=0.85\linewidth]{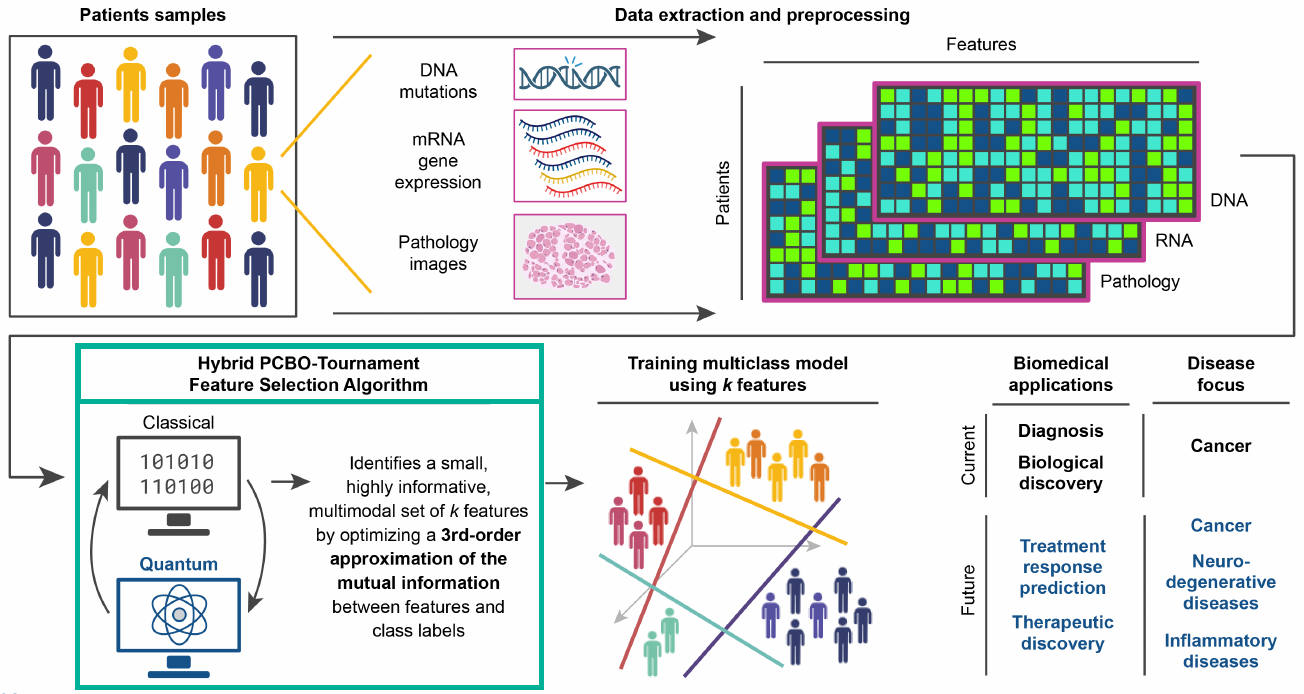}
    \caption{End-to-end overview of the hybrid data processing pipeline developed by our team for biomarker discovery and scaling towards EQA for clinical applications.}
    \label{fig:overview}
\end{figure*}

Quantum processors can be viewed as domain-specific accelerators embedded in larger classical workflows. The canonical example is Shor’s algorithm, where a quantum period-finding subroutine leverages coherent superposition, entanglement, and interference, while the surrounding steps are classical~\cite{Shor_1997}. The general lesson is to target subroutines that can exploit large superpositions and carefully engineered interference, while acknowledging that outcomes must be exposed through a small number of measurements -- an inherent limitation formalized by Holevo’s bound~\cite{holevo1973bounds}.

In this section, we apply that principle to biomarker discovery. As shown in Fig.~\ref{fig:overview}, classical preprocessing harmonizes multimodal data; a quantum-in-the-loop subroutine tackles the combinatorial core of feature selection; and standard classifiers assess predictive utility downstream. Guided by co-design, we surveyed several placements for the quantum component and ultimately focus on feature selection, formulating it as a PCBO to capture higher-order interactions while remaining compatible with hardware-aware quantum heuristics. The remainder of the section introduces the downstream learning tasks used for evaluation, details the preprocessing needed to stabilize information-theoretic scoring, and develops the PCBO formulations and quantum solvers we study.

\subsection{Applications and Downstream Machine Learning}

\begin{table}[t]
\centering
\begin{tabular}{|p{0.2\textwidth}|p{0.4\textwidth}|p{0.3\textwidth}|}
\hline
\rule{0pt}{4ex} 
{\centering \textbf{Dataset name} \par} &  {\centering \textbf{Data Summary} \par}  & {\centering  \textbf{Target learning problem} \par} \\
\hline
\texttt{Pansquamous-1 (TCGA)}  & 1,427 squamous cell carcinoma patients; 18,499 binary DNA features; 13,525 integer mRNA features  & 3-tissue classification \\
\hline
\texttt{Pancan (TCGA)} & 8,013 cancer patients; 19,533 binary DNA features; 20,530 integer mRNA features  & 30-tissue classification \\
\hline
\texttt{Pansquamous-2 (TCGA)} & 660 squamous cell carcinoma patients; 2,720 binary DNA features; 18,879 integer mRNA features; 2,046 integer pathomics features  & 3-tissue classification  \\
\hline
\texttt{Pancan-Val (TCGA)}~\cite{tomczak2015review} & 7,171 cancer patients; 3,576 binary DNA features; 18,881 integer mRNA features; 768 integer pathomics features  & 30-cancer tissue of origin classification  \\
\hline
\texttt{Pancan-Val (CPTAC)}~\cite{li2023proteogenomic} & 1024 cancer patients; 46,811 binary DNA features; 32,531 integer mRNA features; 768 integer pathomics features  & 8-cancer tissue of origin classification \\
\hline
\texttt{UC-Cohort} & 84 cancer patients (patient data acquisition currently ongoing); 16,802 integer mRNA features, 5 clinical variables  & 2-class RECIST 1.1 status prediction \\
\hline
\end{tabular}
\caption{Overview of the multimodal datasets and learning tasks derived from the public TCGA~\cite{tomczak2015review} and CPTAC~\cite{li2023proteogenomic} online data repositories, as well as the prospective dataset curated at the University of Chicago during the course of the Q4Bio program.}
\label{tab:final-p2-datasets}
\end{table}

Our initial efforts focus on methodologically important, but clinically relatively well-established, learning problems, such as the prediction of positive or negative status of human papillomavirus (HPV) infection, a common cause of certain cancers, from RNA-sequencing (RNA-seq) data from patient tumors~\cite{hieromnimon_building_2025}.
We keep the machine learning (ML) models simple, focusing on logistic regression and support vector classifiers (SVCs), partly to highlight the impact of changes within the feature selection and data preprocessing steps of the pipeline. Simple ML models also align with our goals for interpretability, generalizability, and clinical deployability. Using one cohort of samples, classical baseline algorithms or our hybrid algorithm select feature sets from the same pool of features, and then a logistic regression or SVC model is trained on the cohort using the selected features. The trained models are evaluated on a held-out or cross-dataset cohort, with train-test data strictly separated to avoid leakage. We compare performance across models using balanced accuracy and weighted F1 scores, which account for imbalanced classes, yielding a robust metric for assessing whether a given feature selection algorithm improves predictive performance.

As our research progressed, we deliberately increased the complexity of the learning problems to challenge and refine our algorithms, while maintaining sufficiently trustworthy benchmark data. Specifically, we formulated multiple different supervised learning tasks centered on predicting the cancer tissue of origin. Using the data available in The Cancer Genome Atlas (TCGA) and the Clinical Proteomic Tumor Analysis Consortium (CPTAC), we generated multi-class classification problems containing 3, 8 and 30 different tissues of origin. These multi-class tissue of origin learning problems represent an excellent development environment for feature selection algorithms since they are well-supported by the TCGA and CPTAC datasets and, while still being relatively well-studied in prior work, they are substantially more difficult than the binary HPV classification task. Finally, in Phase 3 of the Q4Bio project, we are now taking the hybrid feature selection algorithms, developed on the simpler tissue of origin datasets, and applying them to the more complex and clinically interesting task of predicting treatment outcome. Not only are the biological mechanisms underpinning this learning problem expected to be more complex, much of the challenge stems from the lack of well-annotated and harmonized datasets -- a challenge we addressed head-on with our uniquely interdisciplinary team (Sec.~\ref{sec:applications}).

We analyze multimodal cohorts from TCGA, CPTAC, and a prospective UChicago set (Table~\ref{tab:final-p2-datasets}). DNA features are gene-level mutation indicators (0/1). Raw, bulk RNA-seq data are processed via read alignment and summarization to estimate the transcript counts of each gene in each sample. The counts data are adjusted using size-factor estimation, batch-aware normalization, and log-counts-per-million (log-CPM) transformation using \texttt{PyDESeq2}~\cite{muzellec2023pydeseq2}, with gene filtering to remove consistently lowly expressed transcripts; when using TCGA’s Genome Data Analysis Center (GDAC) outputs, we optionally discretize the $\log_{2}$ \texttt{scaled\_estimate} values from Illumina HiSeq.\footnote{\url{https://gdac.broadinstitute.org/runs/stddata__2016_01_28/samples_report/index.html}} For the UChicago cohort only, we applied an additional sample-level quality control screen based on coverage (number of genes with raw counts $>10$): samples were binned as \emph{low} ($<5{,}000$), \emph{mid} ($5{,}000$--$12{,}000$), or \emph{high} ($\ge 12{,}000$). \emph{Low} samples were excluded; all \emph{high} samples and a limited subset of \emph{mid} samples were retained, after which normalization and embeddings were recomputed on the filtered set.

To compress information and stabilize information-theoretic quantities, mRNA features are quantile-binned to a five-level alphabet (Sec.~\ref{sec:mrna-preprocessing}). Pathomics uses SlideFlow to tile whole-slide images and integrate Prov-GigaPath embeddings, aggregated to slide-level vectors and discretized into four levels (Sec.~\ref{sec:path-preprocessing}). The resulting small-alphabet multimodal matrix (binary DNA, 5-level mRNA, 4-level pathomics) is the input to our information-theoretic scoring, PCBO formulations, and hybrid selector, with train/test handling and site stratification described in Sec.~\ref{sec:algorithm}.

\subsubsection{An Outlook on the Potential for QML Advantage on Multimodal Cancer Datasets}

Although the main course of our Q4Bio project followed the development of hybrid feature selection algorithms, we also pursued parallel efforts exploring potential applications of quantum machine learning (QML) to multimodal cancer datasets. Our goal was to assess under what conditions QML could outperform classical models in the third stage of our pipeline (Fig.~\ref{fig:pipeline}). Our approach~\cite{teo2025k}
centers on \emph{strong $k$‑contextuality}, a dataset property that leads to a provable lower bound on the memory resources a classical model needs to represent such data.

Contextuality, a long-range correlational structure originally studied in quantum foundations theory~\cite{shimony1984contextual} and explored as a resource for quantum computation~\cite{howard2014contextuality}, has been shown to be the source of memory‑based separations between classical and quantum learning models~\cite{gao2022enhancing, anschuetz2022interpretable}. Building on this, we introduce strong $k$-contextuality as a measure of contextuality, and prove that a classical generative model requires at least $k$ latent states to represent a dataset with strong $k$-contextuality to finite relative entropy. Crucially, strong $k$-contextuality does not result in a similar memory bound for a quantum generative model. Defining the contextuality number $k^\ast$ as the largest $k$ such that a dataset is still strongly $k$-contextual, we develop and benchmark algorithms for finding the (approximate) contextuality number of a given sequential learning task. Applying these algorithms to learning tasks on real and test datasets, we compare the performance of resource-constrained classical and quantum-enhanced Bayesian models, showing empirically that the estimated contextuality number is a good heuristic for a quantum-classical separation in performance~\cite{teo2025k}.

These methods can also be applied to learning tasks on multimodal cancer datasets. As an example learning task from \texttt{Pancan}, we took the $n$ most correlated DNA–mRNA feature pairs, constructing a scenario where given the existence of mutations in the DNA features (0 or 1), a learning model would have to predict mRNA expression levels (discretized as integers 0–4). For $n=5$, we found that the estimated contextuality number of this task was $k^\ast = 92$, implying any classical learner needs at least 7 bits of memory (since $2^n \ge 92$) for accurate modeling. While this toy example remains classically tractable, we expect $k^\ast$ to grow rapidly as the model includes more multimodal features. An interesting direction for future work would be to compare the performance of classical and quantum generative models on this task, further investigating the relationship between contextuality number and quantum-classical separations.

\subsection{Custom Data Preprocessing}\label{sec:data-preprocessing}

\begin{figure}[t]
    \centering
    \includegraphics[width=0.5\linewidth,trim={0 0 0 10mm},clip]{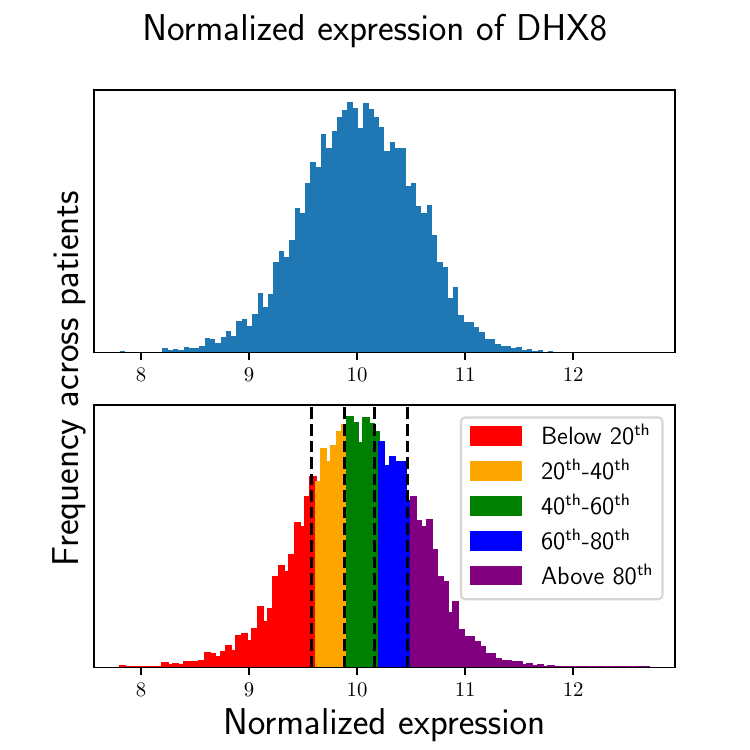}
    \caption{Example of the mRNA discretization for the DHX8 gene feature in the \texttt{Pancan} data. The continuous expression values (top) were each assigned one of five categorical values (bottom) according to percentile bins defined by multiples of 20 percent.}
    \label{fig:mrna-example}
\end{figure}

Machine‑learning analyses of multimodal cancer data must contend with the challenge of ``data scarcity'', characterized by high‑dimensional, low‑sample‑size datasets. In this setting, many estimators exhibit high variance and are prone to overfitting, degrading out‑of‑sample performance~\cite{sakamoto_1987, wainwright2019high}. This motivates feature selection and feature extraction: beyond improving the generalizability of trained models, these steps can isolate compact sets of variables most relevant to a given biological endpoint, thereby enabling biomarker discovery. Our emphasis on data preprocessing and feature selection is also motivated from a computational perspective, for both classical and quantum workflows. By mapping each modality to a small‑alphabet, integer‑valued feature space, we stabilize and accelerate the computation of information‑theoretic quantities (e.g., mutual and interaction information) and reduce the cost of encoding classical features into quantum states.

Our custom preprocessing techniques harmonize genomic (DNA), transcriptomic (mRNA), and pathomic (tumor images) data modalities into discretized representations tailored for information‑theoretic scoring and resource‑efficient learning. 
Specifically, the mRNA expression values are quantile-binned into five-level integers (details in Sec.~\ref{sec:mrna-preprocessing}), and pathomics uses slide-level embeddings from a deep-learning pipeline, discretized into four-level integers (details in Sec.~\ref{sec:path-preprocessing}). The DNA mutations are already encoded as binary indicators ($0/1$) corresponding to the presence or absence of single-nucleotide polymorphisms (SNPs). This shared discrete format stabilizes mutual-information and higher-order estimates and maps naturally to both classical pipelines and near-term quantum workflows.

\subsubsection{mRNA Quantile Discretization}\label{sec:mrna-preprocessing}

The expression levels of different mRNA species are measured using high-throughput transcriptomic profiling, primarily RNA-seq. In this approach, RNA is extracted from tumor samples, converted into cDNA, and sequenced using high-throughput short-read sequencing instruments (e.g., Illumina). Sequencing reads are then mapped to reference transcripts and normalized to quantify gene expression. This technology has become the standard for cancer genomics studies, including TCGA~\cite{wang2009rna}.

For each gene, we compute empirical quintiles at the 20th, 40th, 60th, and 80th percentiles across all available patients and map continuous expression values to an integer in $\{0,1,2,3,4\}$. 
This procedure is easy to compute, invariant to monotone transformations, and, under unimodal gene‑wise distributions, is robust to train/test splits. An example for the gene DHX8 in the \texttt{Pancan} dataset is shown in Figure \ref{fig:mrna-example}.

While this preprocessing technique was originally designed for the purpose of more easily computing mutual information between features, 
these discretizations have empirically supported the selection of compact, informative multimodal feature sets that accurately predict tissue of origin, rivaling or surpassing state‑of‑the‑art feature selection methods on the non-quantile-binned data.  The size and high-dimensionality of high-throughput gene expression data collection techniques, along with its inherent biological and technical noise, coupled with the complexity of genetic and transcriptomic interactions and their effect on tissues, make the selection of small, informative feature sets for tasks like tissue of origin and cancer prognosis notoriously difficult~\cite{sun2014feature}. Therefore, the contribution of this preprocessing technique is not merely a computational convenience, but also a sort of regularization to ensure better selection of informative feature sets using feature selection algorithms on both classical and quantum hardware.

\subsubsection{Pathomics Embeddings and Discretization}\label{sec:path-preprocessing}

\begin{figure}[t]
    \begin{subfigure}[t]{0.48\textwidth}
        \centering
        \includegraphics[width=\textwidth]{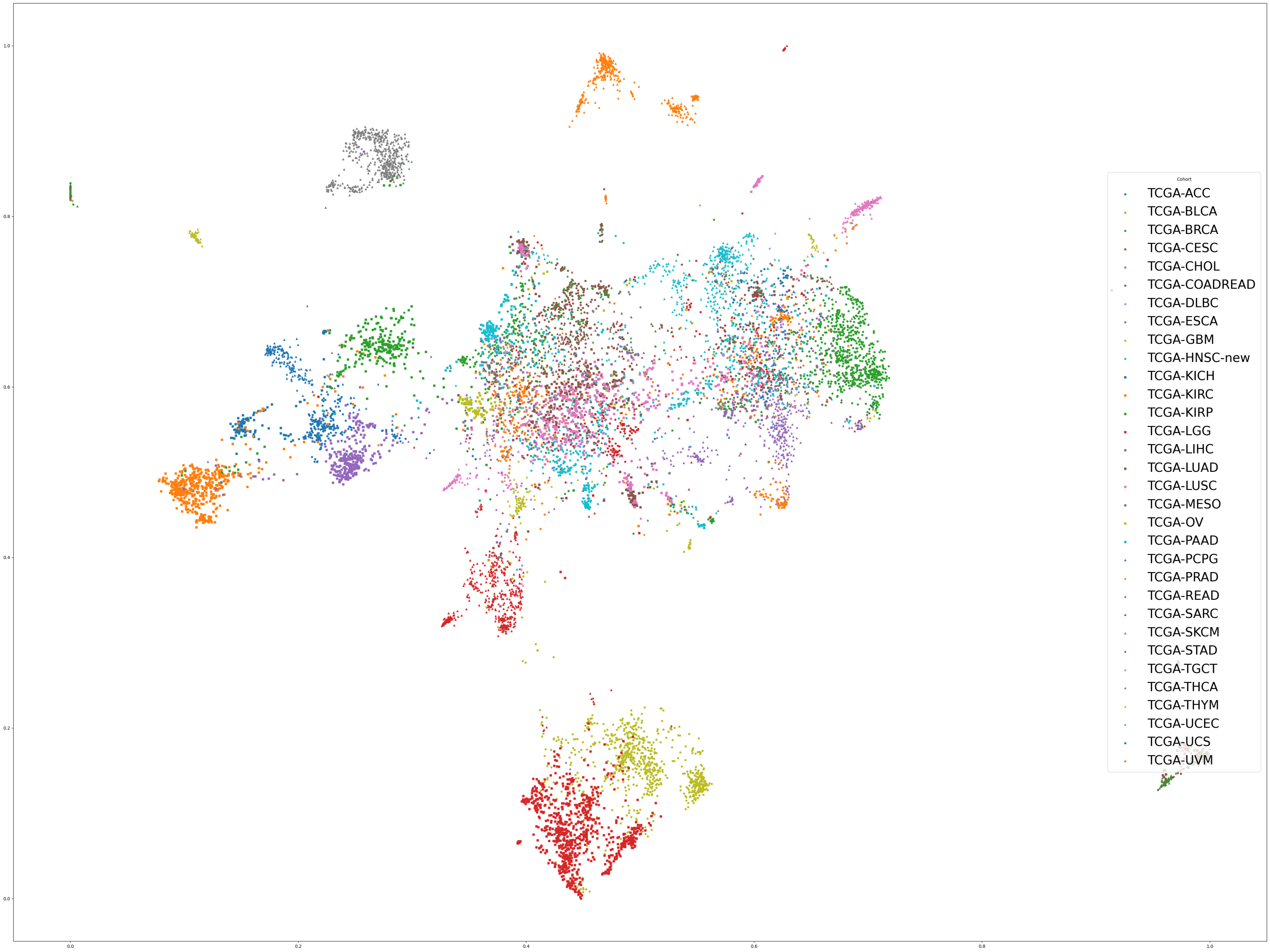}
        \caption{}
    \label{fig:slide-continuous}
    \end{subfigure}
    \hfill
    \begin{subfigure}[t]{0.48\textwidth}
        \centering
        \includegraphics[width=\textwidth]{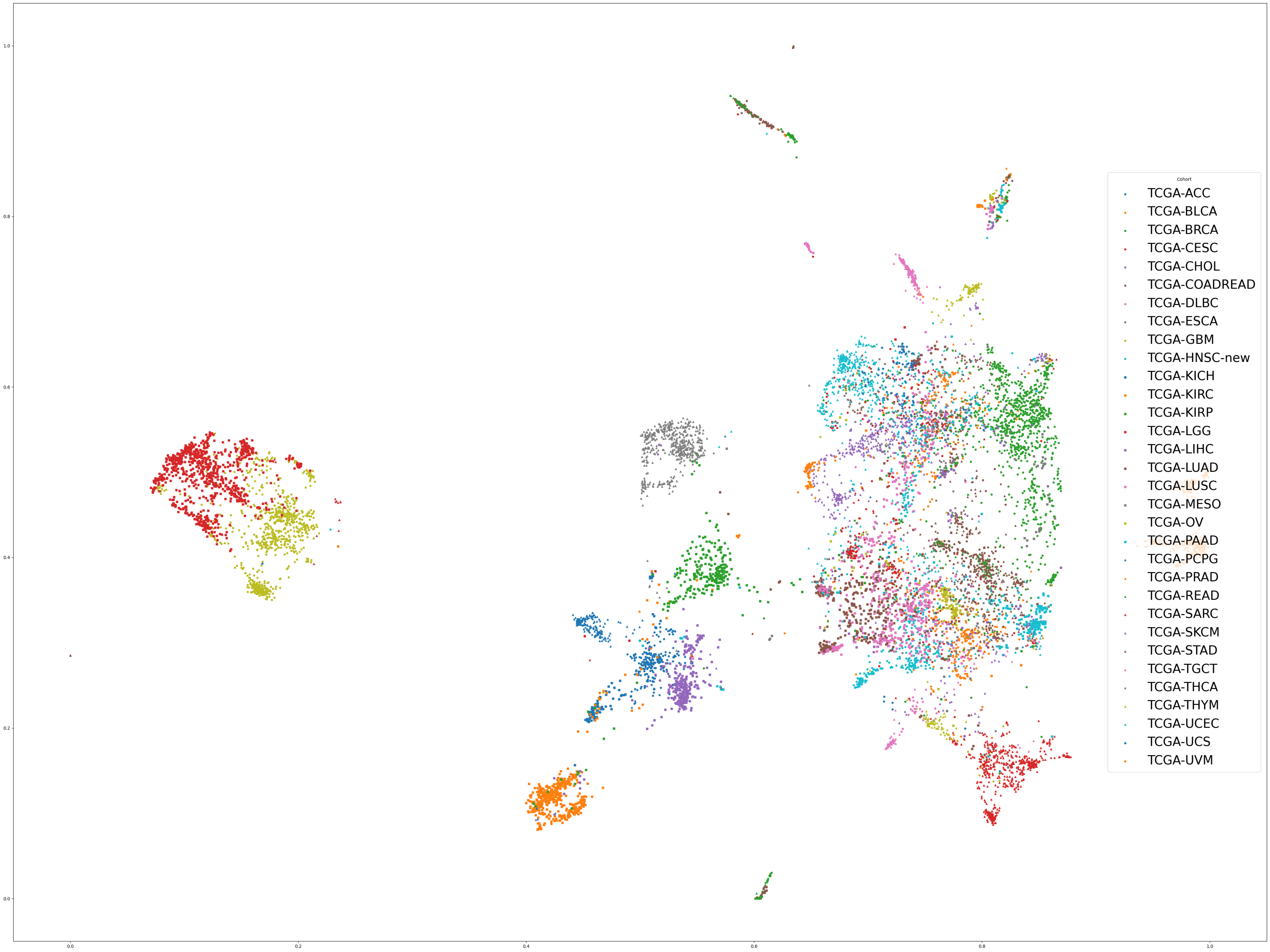}
        \caption{}
    \label{fig:slide-discrete}
    \end{subfigure}
    \caption{UMAP of continuous slide-level features (left) and discretized slide-level features (right) in the pan-cancer TCGA dataset, using Prov-GigaPath feature extractor.}
    \label{fig:tcga-path}
\end{figure}

To address the variable number of pathomics features per patient, we leveraged and customized several deep learning models, including a deep learning pipeline developed in prior work by our team~\cite{dolezal2024slideflow}. The approach proved effective for standardizing pathomics features for feature selection algorithms.

First, to generate feature vectors for each patient, we leveraged the efficient whole-slide image processing pipeline in SlideFlow~\cite{dolezal2024slideflow}, a state-of-the-art computer vision platform developed by our team for optimizing histopathology-based biomarker development. 

We integrated our SlideFlow output with Microsoft Research's Prov-Gigapath. Released in May 2024, Prov-GigaPath is a state-of-the-art whole-slide foundation model pretrained on a vast, diverse dataset of real-world pathology images~\cite{xu2024whole}. To align with the preprocessing requirements of Prov-GigaPath while maintaining our ability to efficiently compute and customize their pipeline, we implemented support for their model pipeline within Slideflow. With the pre-processed tiles, we used Prov-GigaPath to generate tile-level embeddings that capture rich histological features learned by the model during its pretraining. To generate slide-level embeddings from these tile-level features, we implemented the aggregation method of Prov-GigaPath within Slideflow. Using a second transformer network, called LongNetViT, this method contextualizes embeddings across all tiles to create an aggregate slide-level embedding. 

To reduce the bit requirements for computing mutual information between features and potential quantum embedding representations, we studied the distributional patterns of the features across slides within each cancer subtype and evaluated how well a variety of discretization techniques retained underlying information for downstream classification tasks. We found quartile discretization to be an effective trade-off between information retention and bit reduction.

The final output of our pipeline for each patient is one representative pathomic feature vector of size 1 x 768, containing integer values in the range 0–3, which indicate low, medium, high, and very high values associated with each feature axis of the tissue image, ultimately enabling normalized, condensed, and informative custom representations of pathomics data for downstream tasks.

\subsection{Feature Selection}

Dimensionality reduction of raw input data is a crucial aspect of nearly all machine learning pipelines. It is particularly important in data-scarce regimes where the number of samples is significantly smaller than the number of features. In such scenarios, large models with many parameters are prone to overfitting due to insufficient data to properly constrain the model parameters~\cite{sakamoto_1987, wainwright2019high}. Feature selection and extraction techniques can mitigate this issue by reducing noise and amplifying useful signal within the data, thereby improving the performance and generalization of downstream learning algorithms~\cite{guyon_2006}. When the downstream learning algorithm is a QML model, dimensionality reduction serves an additional purpose of addressing the data-loading problem. By reducing the effective size of the input data, the computational burden of loading classical data onto a quantum computer can be alleviated~\cite{harrow2020small, tomesh2021coreset}.

Feature selection is a type of dimensionality reduction technique that aims to identify a minimal set of highly informative features that maximize the performance of a machine learning model on a given learning task. Within a machine learning pipeline, feature selection is essential for improving model performance by removing irrelevant or noisy data and including the most informative features. It also enhances the generalization capacity of multimodal models by reducing the risk of overfitting.
Compared to other dimensionality reduction methods, such as principal component analysis (PCA) and autoencoders, which produce complex linear or nonlinear combinations of features, feature selection can result in more interpretable solutions, as it only selects which features will participate in the learning problem without manipulating the features themselves~\cite{ros_2024, guyon_2006}.

There are several general strategies for solving the feature selection problem that vary in how they evaluate the desirability of features and whether or not the downstream learning model is included in that evaluation~\cite{guyon_2006}.
\begin{itemize}
\item \textit{Filter methods} use a statistical measure or intrinsic characteristic (e.g., mutual information or Pearson correlation) to evaluate the quality of a feature subset without incorporating the downstream model or learning algorithm itself.
\item \textit{Wrapper methods} use a search algorithm to traverse the space of possible feature sets, evaluate the subsets using the downstream learning algorithm, and then select the best-performing subset.
\item \textit{Embedded methods} incorporate the feature selection process into the learning algorithm itself.
\end{itemize}

We focus on developing filter methods that leverage hybrid quantum-classical computational resources to select high-quality subsets of features from the multiple modalities present within our dataset. Filter methods are chosen for their efficiency, as they can be much faster than wrapper methods when the learning model is computationally demanding. Additionally, filter methods aim to select generally informative features, rather than features specific to a particular learning model, enhancing their generalizability.

\subsubsection{Information Theory and Feature Selection}
We consider hybrid filter methods based on the correlations -- quantified by information theoretic quantities such as the entropy, mutual information, and interaction information -- amongst features, as well as between features and the class labels, in a supervised learning setting.
The feature selection task, over $F$ total features, can be expressed as a PCBO problem of the form:
\begin{equation} \label{eqn:pcbo}
\begin{split}
    \boldsymbol{x}^* = &\argmin_{\boldsymbol{x}} Q(\bm{x}) \\
    =&\argmin_{\boldsymbol{x}} \left(\sum_i^F C(i) x_i +  \sum_{i<j}^F C(i,j) x_i x_j + \sum_{i < j < k}^F C(i,j,k) x_i x_j x_k + \dots \right) \\
    & \text{subject to the constraints:}\; \abs{\boldsymbol{x}}_1 = n \in \mathbb{Z}, \quad \boldsymbol{x} \in \mathbb{Z}_2^F
\end{split}
\end{equation}
Here, the aim is to ``turn on'' $n$ bits of the binary vector $\boldsymbol{x}$, where the indices of the $x_i=1$ bits correspond to the features $f_i$ in the selected feature set. The functions $C(\cdot)$ yield the coefficients defining the optimization problem. Below, we discuss several specific examples that use the entropy, mutual information, and interaction information between sets of features and class labels to define the coefficient values.
One favorable aspect of this approach to feature selection is that it allows us to explicitly encode first\nobreakdashes-, second\nobreakdashes-, third\nobreakdashes-, and higher-order relationships between the multimodal features into the formulation of the feature selection problem itself -- corresponding to the coefficients $C(\cdot)$. 
The key question our work seeks to answer is: can higher-order PCBO problems enable superior identification of synergistic features that might be overlooked by other approaches, and can quantum computers help mitigate the increased computational complexity of solving these higher-order feature selection problems?

Using information theoretic quantities to define the feature selection problem is well-motivated from a biological perspective, where prior work has demonstrated the effectiveness of mutual information (MI) for measuring feature relevance and reducing redundancy among features in biomedical datasets~\cite{watkinson2008identification, watkinson2009inference}.
MI as a filter technique can identify a broad set of potentially informative features, which can then be refined using more selective methods. Combining MI with other feature selection approaches has been shown to maximize the advantages of individual techniques while mitigating their limitations, leading to more robust and biologically relevant feature sets for tasks, such as cancer biomarker identification~\cite{sun2014feature, abdelwahab2022feature}.


For example, consider a supervised learning problem on a given dataset with $R$ rows, $N$ feature (i.e., column) vectors $f_i \in \mathbb{Z}^R$, and a target vector containing the class labels $y \in \mathbb{Z}^R$. For the biological datasets we consider, we pre-process the features and class label such that they always contain discrete integer values from some restricted alphabet $\mathcal{F}, \mathcal{Y}$. For example, in the ``Pancan'' learning problem from TCGA~\cite{tomczak2015review} that we have considered previously, for any given patient $m$ in the dataset, the features may take on any value $f_i^m \in \{0, 1, 2,3,4\} = \mathcal{F}$ and the label may have a value $y^m \in \{0,1,..., 29\} = \mathcal{Y}$, corresponding to one of 30 different tissues of origin. 

Treating a feature $f_i$ as a random variable, its Shannon entropy $H(f_i)$ can be computed as~\cite{ash2012information}:
\begin{equation}
    H(f_i) = \sum_{f_i \in \mathcal{F}} -p_\mathcal{F}(f_i) \log_2 p_\mathcal{F}(f_i).
\end{equation}
When considering multiple features, we may also compute their joint entropy
\begin{equation}
    H(f_1, \dots, f_N) = -\sum_{f_1^n \in \mathcal{F}} \dots \sum_{f_N^n \in \mathcal{F}} P(f_1^n, \dots, f_N^n) \log_2[P(f_1^n, \dots, f_N^n)],
\end{equation}
and conditional entropy
\begin{equation}
    H(f_1, \dots, f_N | y) = H(f_1, \dots, f_N, y) - H(y).
\end{equation}

The mutual information between two random variables $A$ and $B$ is a symmetric measure of the information obtained about $A$ after observing $B$, or vice versa~\cite{ash2012information}. Mutual information is useful for capturing both the redundancy between pairs of features as well as the importance of a feature with respect to a specific class label.
It is defined as
\begin{equation}
    I(A;B) \equiv H(A) - H(A|B) = H(B) - H(B|A),
\end{equation}
in terms of entropy, and may also be expressed via probabilities as:
\begin{equation}
 I(A;B) = \sum_{a \in A, b \in B} P_{(A,B)}(a,b) \log_2 \frac{P_{(A,B)}(a,b)}{P_A(a) P_B(b)}, 
\end{equation}
where the summation is taken over all the allowed values, $a$ and $b$, of the random variables, respectively.
If $I(A;B)=0$, then the two variables are independent.
High mutual information between two features indicates redundancy -- one can often be dropped with little loss. Conversely, a feature with strong mutual information with the class label is likely to boost model performance when included.

It is also possible that a combination of features can be highly informative without any of them individually having high mutual information with the class label, e.g., in the case of synergistic interactions among these features. Synergistic interactions between $k$ variables $X_1,\ldots,X_k$ can be measured by means of the interaction information~\cite{kuo_1962}, which is given by the recursion
\begin{equation}
    I\left(X_1;\ldots;X_k\right)=
    I\left(X_1;\ldots;X_{k-1}\right)-I\left(X_1,\ldots,X_{k-1}\mid X_k\right),
\end{equation}
where the conditional mutual information can be written in terms of joint and conditional entropies as
\begin{equation}
    I(X;Y \mid Z) = H(X \mid Z) + H(Y \mid Z) - H(X, Y \mid Z).
\end{equation}
While the mutual information between two random variables is strictly nonnegative, the interaction information can take on positive, zero, \textit{and} negative values. 
Negative interaction information represents a synergistic relation between the random variables. 
Previous research has demonstrated the effectiveness of interaction information in uncovering complex relationships and dependencies among genes, making it a particularly valuable tool for developing hybrid algorithms for feature selection on multimodal cancer data~\cite{watkinson2008identification, watkinson2009inference, tapia2018neurotransmitter, baudot2019topological}.

As indicated by the PCBO definition in Equation \ref{eqn:pcbo}, the problem may generally include higher-order terms corresponding to interactions between arbitrary numbers of features -- this leads to a delicate trade-off between the computational cost of solving the problem and the biological accuracy of the problem definition. On one hand, higher-order feature correlations in feature selection are biologically relevant, for example, due to complex genetic regulatory networks~\cite{watkinson2008identification, watkinson2009inference}. On the other hand, solving PCBOs containing these higher-order interactions significantly increases computational complexity~\cite{chen2019suboptimality, gamarnik2021overlap}. Therefore, our approach seeks to leverage non-local quantum optimization algorithms to obtain higher-quality solutions to the higher-order PCBO problems that better match the underlying biology, resulting in the selection of more robust and informative feature sets.


\subsubsection{Quantum Algorithms for Solving PCBOs}

Casting feature selection in terms of a PCBO problem exposes it to several quantum algorithmic approaches, including variants of Grover's search~\cite{grover1996fast}, the quantum approximate optimization algorithm (QAOA)~\cite{farhi2014quantum}, and quantum annealing~\cite{morstyn2022annealing, ferrari2022towards}. The potential for a quantum advantage, beyond the generic square root speedup achieved by Grover’s search algorithm~\cite{babbush2021focus}, is complicated by the fact that algorithms such as QAOA and quantum annealing are heuristic algorithms without rigorous bounds on their expected performance in many specific cases of interest. However, for our purposes, the question is rather empirical: on fixed, real instances, does a quantum subroutine reduce the problem to a size classical solvers can handle well? And, what magnitude of quantum resources are required to reach problem sizes where classical solvers begin to struggle?

The information theoretic quantities introduced above also appear within previous works studying quantum algorithms for feature selection~\cite{grossi2022mixed, ferrari2022towards, mucke2023feature}.
For example, Equation \ref{eqn:mrmr_def} shows a specific PCBO definition, referred to as minimum-redundancy maximum-relevancy (\texttt{mRmR})~\cite{mucke2023feature}, with real-valued hyperparameter $\lambda$.
\begin{equation}\label{eqn:mrmr_def}
 Q_{\texttt{mRmR}}(\bm{x}, \lambda) = -\lambda \sum_i I(f_i; y) x_i + \sum_{i,j} I(f_i; f_j) x_i x_j
\end{equation}
Because the objective is to minimize $Q(\bm{x}, \lambda)$, Equation~\ref{eqn:mrmr_def} can be understood intuitively to reward the selection of features $f_i$ that are highly correlated with the labels $y$ of the learning problem, and to punish the selection of pairs of highly correlated features.
Similarly, in Ref.~\cite{ferrari2022towards}, the \texttt{miqubo} formulation was proposed, weighting feature sets according to their correlation with the label and their conditional mutual information:
\begin{equation}\label{eqn:miqubo_def}
Q_{\texttt{miqubo}}(\bm{x}, \lambda) = -\lambda \sum_i I(f_i; y) x_i - \sum_{i,j} I(f_i; y | f_j) x_i x_j.
\end{equation}
Here, it is important that the feature selection optimization problem is constrained to select a specific number of features $\abs{\boldsymbol{x}}_1 = N$ to appear in the final feature set. Without this constraint, the \texttt{miqubo} definition has a trivial global optimum where all features are included in the feature set.

Over the course of our Q4Bio project, much of our effort was spent formulating and evaluating different PCBO variants. 
For example, we tested an additional second-order PCBO, referred to as  \texttt{full-qubo}, defined as:
\begin{equation}\label{eqn:fullqubo_def}
Q_{\texttt{full-qubo}}(\bm{x}, \lambda) = -\lambda \sum_i I(f_i; y) x_i + \sum_{i,j} (I(f_i; f_j) - I(f_i; y | f_j)) x_i x_j.
\end{equation}

We also developed higher-order PCBOs containing first\nobreakdashes-, second\nobreakdashes-, and third-order terms.
In particular, the \texttt{entropy-cubo} formulation, defined in Eq.~\ref{eq:entropy-cubo}, has demonstrated competitive performance compared to classical algorithms such as feature-importance~\cite{sklearn2024featimp}. The \texttt{entropy-cubo} PCBO is based on an approximation of the mutual information between $k$ selected features $f_i$ and the target class label $y$ (for example, the cancer tissue of origin). Intuitively, we want to select $k$ features that possess the maximum amount of mutual information with the target class label, that is why we multiply each term of Eq.~\ref{eq:entropy-cubo} by $-1$ so that we can talk about finding the \textit{minimum} energy state of the optimization problem. 

\begin{align}
    Q_{\texttt{entropy-cubo}}(\vec{x}, \vec{\alpha}, k) &= \frac{-\alpha_1}{k}\sum_{j=1}^k\bigg(H\left(f_{i_j}\right)-H\left(f_{i_j} | y\right)\bigg) x_{i_j} \notag\\
    &\hspace{0.3cm}-\frac{\alpha_2}{\binom{k}{2}}\sum_{\left\{j_1,j_2\right\}\subset[1,k]}\bigg(H(f_{i_{j_1}},f_{i_{j_2}})-H(f_{i_{j_1}},f_{i_{j_2}} | y)\bigg) x_{i_{j_1}} x_{i_{j_2}} \label{eq:entropy-cubo} \\ 
    &\hspace{0.3cm}-\frac{\alpha_3}{\binom{k}{3}}\sum_{\left\{j_1,j_2,j_3\right\}\subset[1,k]}\bigg(H\left(f_{i_{j_1}},f_{i_{j_2}},f_{i_{j_3}}\right)-H\left(f_{i_{j_1}},f_{i_{j_2}},f_{i_{j_3}} | y\right)\bigg) x_{i_{j_1}} x_{i_{j_2}} x_{i_{j_3}} \Bigg).\notag
\end{align}

The \texttt{entropy-cubo} energy function in Eq.~\ref{eq:entropy-cubo} also takes in a few input arguments. The vector $\vec{x} \in \mathbb{Z}_2^N$ holds the binary variables $x_i$ that indicate whether the feature $f_i$ is included ($x_i=1$) in the feature set or not ($x_i=0$). The vector $\vec{\alpha} \in \mathbb{R}^3$ specifies a particular weighting of the first\nobreakdashes-, second\nobreakdashes-, and third-order terms in the \texttt{entropy-cubo} problem. We enforce the convention that $\sum_i \alpha_i = 1$.
After experimenting with different settings of $\vec{\alpha}$ and observing the impact on both the quality of the selected features and the difficulty of the problem from the perspective of classical solvers such as Gurobi, we ultimately found that putting all of the weight on the third-order terms (i.e., $\vec{\alpha} = (0,0,1)$) yielded the best results. 

All of the PCBO objective functions $Q$ described above may also be interpreted as Hamiltonians describing the energy landscape of a system of interacting spins, and there are a wide variety of quantum optimization algorithms that have been developed to find the ground states of such Hamiltonians. While Grover's algorithm provides a quadratic speedup to brute-force search methods, it is unclear whether such a speedup would translate into practical advantages once the overhead and latencies of error correction are taken into account~\cite{babbush2021focus, hoefler2023disentangling}. Here, we briefly provide an overview of two popular heuristic approaches to solving these problems before diving deeper into our approach and empirical results.

\paragraph{Adiabatic algorithms.}
Algorithms based on adiabatic evolution of quantum states have emerged as promising approaches to solving optimization problems~\cite{morstyn2022annealing, ferrari2022towards}.
\begin{equation}
    H_{tot}(s) = (1-s)H_{init} + s H_{Q}
\end{equation}
Such algorithms are based on the adiabatic theorem, which states that a quantum system, described by a Hamiltonian $H_{tot}$, prepared in the ground state of the Hamiltonian $H_{init}$ may be evolved into the ground state of a final Hamiltonian $H_Q$, so long as the evolution $s: 0 \rightarrow 1$ is performed adiabatically. Practically, the limiting factor on runtime is the minimum spectral gap along the path. In other words, the required runtime scales inversely with (at least) the square of this gap~\cite{albash2018adiabatic}.
Two examples of this approach are quantum annealing~\cite{hauke2020perspectives} and the quantum constrained Hamiltonian optimization algorithm (Q-CHOP)~\cite{perlin2024q}.
Purpose-built hardware devices, referred to as quantum annealers, have been constructed by companies such as D-Wave to implement this adiabatic evolution and have shown much promise in solving difficult optimization problems~\cite{morstyn2022annealing}.
The Q-CHOP algorithm, recently introduced in~\cite{perlin2024q} and based on techniques for secondary Hamiltonian evolution in~\cite{yu2021quantum}, is an adiabatic algorithm for solving constrained optimization problems.

\paragraph{Quantum Approximate Optimization Algorithm.}
The QAOA~\cite{farhi2014quantum} is a highly flexible method for solving optimization problems on a quantum computer with many proposed variants and adaptations to different problem classes. It can be thought of as a discretized version of the adiabatic algorithm, and in the limit of infinite circuit depth the two are equivalent. A general overview of the algorithm is as follows: given a cost function $Q(x)$ to optimize, a QAOA circuit consists of $p$ repeated applications of a mixing and cost Hamiltonian $M$ and $H_Q$, respectively, in alternation, parameterized by $2p$ parameters $(\vec{\gamma}, \vec{\beta})$:
\begin{equation}\label{eq:qaoa_ansatz}
    \ket{\psi(\vec{\gamma}, \vec{\beta})} = e^{-i\beta_pM}e^{-i\gamma_pH_Q}\cdots e^{-i\beta_1M}e^{-i\gamma_1H_Q} \ket{\psi_0},
\end{equation}
where $\ket{\psi_0}$ is an easily-prepared initial state. There is freedom in choosing the Hamiltonians $M$, and $H_Q$, although typically (as is followed in this work), $H_Q$ is chosen such that $H_Q\ket{x}=Q(x)\ket{x}$ for all $x$, and $M$ is the X mixer $M = \sum_{i=1}^n X_i$, where $n$ represents the total number of qubits in the circuit. $X_i = \begin{pmatrix}0 & 1 \\ 1 & 0\end{pmatrix}$ is the Pauli X operator, and the starting state $\ket{\psi_0}$ is chosen to be the uniform superposition over all $2^n$ bitstrings, as originally considered in~\cite{farhi2014quantum}. The goal of QAOA is then to produce a state $\psi(\vec{\gamma}, \vec{\beta})$ such that repeated sampling in the computational basis yields either an optimal or high-quality approximate solution to the problem. Finding such good parameters is the parameter-setting problem and may be approached in a number of ways with different tradeoffs, ranging from black-box optimization techniques to application specific approaches.

Interestingly, variants of the QAOA have also been proposed that utilize different mixing unitaries to enforce the problem constraints at the circuit level -- forcing the quantum state to evolve within a specific subspace that abides by the problem's constraints~\cite{saleem2023approaches}.
However, here we choose to forgo these constraint-preserving mixers due to the high gate cost overhead that they incur when mapped to quantum hardware~\cite{tomesh2024quantum}, and instead implement the unconstrained mixer unitary~\cite{farhi2014quantum}. This lowers resource requirements but expands the search to many constraint-violating bitstrings -- another example of a co-design opportunity in hybrid quantum-classical algorithm design.

Equation \ref{eq:entropy-cubo} defines a PCBO with a hard Hamming weight $\abs{\boldsymbol{x}}_1 = N$ constraint. Instead of incorporating this constraint at the circuit level with a constraint-preserving mixer, we can convert the PCBO to a  polynomial unconstrained binary optimization (PUBO) problem by penalizing constraint-violating bistrings at the level of the objective function.
We also choose to go one step further and convert the PUBO into a polynomial unconstrained \textit{spin} optimization (PUSO) problem for convenience.
This involves replacing all binary variables $b\in\{0,1\}$ with spins $s\in\{\pm1\}$ via $s=1-2b$, which induces only a constant shift and rescaling of energies, preserving the relative ordering of solutions.
The PCBO problem's hard constraint is then absorbed as a \textit{soft} quadratic penalty with a user‑set Lagrange multiplier~\cite{saleem2023approaches}. This preserves the energies of constraint‑abiding configurations while energetically suppressing infeasible ones. In practice, selecting the Lagrange multiplier $\lambda_c$ analytically is generally intractable; we set it empirically, finding that a value of $\lambda_c = 5$ is sufficient.

In the future, with access to large-scale fault-tolerant quantum computers (FTQCs), constraint-preserving unitaries are a very appealing and powerful approach to quantum optimization, but we leave this interesting topic for future research. 

\begin{figure}[t]
    \centering
    \includegraphics[width=0.7\linewidth]{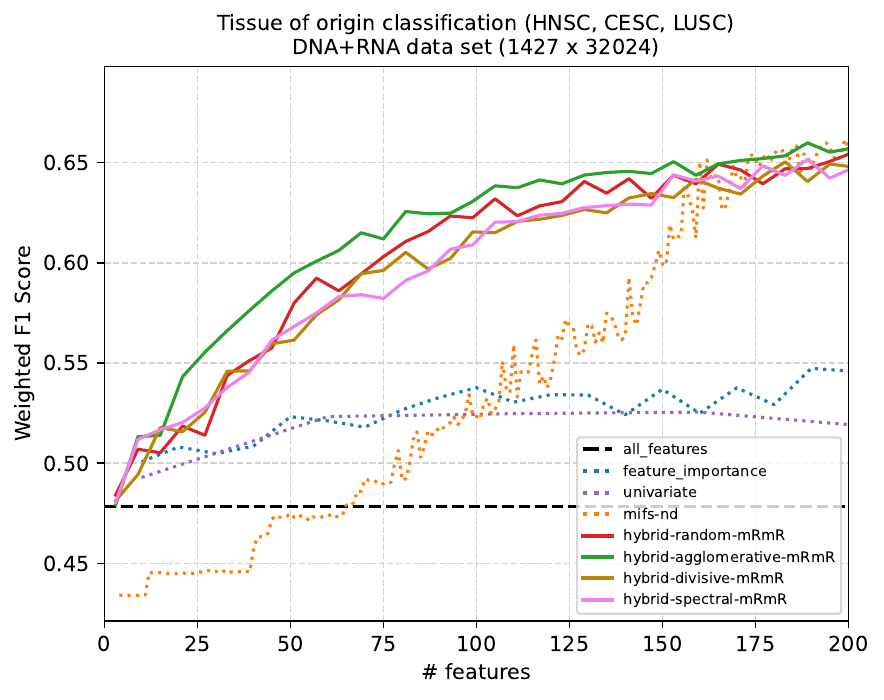}
    \caption{Evaluation of classical (dashed lines) and \texttt{mRmR}-based (solid lines) feature selection algorithms on a dataset containing 32,024 total features with three separate class labels. In this early dataset, the 13,525 mRNA variables were transformed and discretized so as to function as distractor variables. Each algorithm produces a feature set of a specific size (x-axis) which is used to train a logistic regression classifier. Five-fold cross-validation is performed to produce the reported F1 score (y-axis, higher is better).}
    \label{fig:f1-eval}
\end{figure}

While we may be biologically motivated to include higher-order terms in the feature selection PCBOs, this also causes an extremely rapid rise in computational complexity, making the feature selection optimization problem especially difficult for \textit{local} solvers, including both quantum and classical approaches~\cite{chen2019suboptimality, gamarnik2021overlap}. 
Specifically, $k$-spin Hamiltonians with random coefficients have been analyzed in physics literature. For $k=2$, the model is exactly solvable; however, for even $k\geq 4$, it is conjectured in~\cite{gamarnik2021overlap} that these problems are asymptotically difficult for local solvers due to a property known as the ``overlap gap property''.
Constant-depth QAOA is an example of a local solver which is known to perform poorly for such problems~\cite{basso2022performance}. 
Instead, it is suspected that in order for QAOA to find good solutions to these problems, the circuit depth must scale at least logarithmically with problem size~\cite{basso2022performance, farhi2020typical, farhi2020worst}.
Unfortunately, it is in this regime where trainability issues such as barren plateaus and local minima are known to appear~\cite{anschuetz2022quantum}.
Various strategies have been proposed to address the limitations of constant-depth QAOA. One of these methods is known as the recursive QAOA (RQAOA) which operates by recursively optimizing the QAOA ansatz and removing a problem variable in each iteration~\cite{bravyi2020obstacles}.
The RQAOA has been shown empirically to perform well, even beating classical algorithms on small instances of certain optimization problems~\cite{bae2022recursive, frederick2023benchmarking}, and over the course of our Q4Bio project we have focused on developing RQAOA with a particular emphasis on reducing resource requirements.

\subsubsection{Empirically Evaluating PCBO Feature Selection with Classical Solvers}

We evaluated the performance of hybrid and classical feature selection algorithms on a variety of datasets. One of the first experiments we conducted is shown in Fig.~\ref{fig:f1-eval} where a one-versus-rest logistic regression classifier was trained to predict the cancer tissue of origin for each patient, where there are three possible classes: cervical (CESC), head and neck (HNSC), and lung (LUSC) squamous cell carcinoma.
This dataset, containing 1,427 rows and 32,024 total features, is derived from TCGA, and the binary DNA mutation features are exactly as described in Sec.~\ref{sec:data-preprocessing}. However, this experiment was performed before we developed the mRNA preprocessing techniques described in Sec.~\ref{sec:data-preprocessing}, and therefore the mRNA features in this particular case function as distractor variables.

This is why the weighted F1 score (defined as the geometric mean of precision and recall weighted by the relative sizes of each class) of the model trained on all of the features performs substantially worse than the same model trained on a much smaller subset of selected features (Fig.~\ref{fig:f1-eval} shows feature set sizes ranging from 4 to 200 features). The solid lines correspond to different methods for clustering groups of features into smaller \texttt{mRmR} PCBO problems, and these techniques are able to select more accurate feature sets, primarily composed of DNA features, with sizes between approximately 25 and 150 features. The features were repeatedly clustered into groups of 100, corresponding to a 100-variable \texttt{mRmR} PCBO problem that was solved with the tabu search-based QBSolv algorithm~\cite{qbsolv2017}. The selected features from that problem were then passed on to the next round of clustering and solving until the desired feature set size was obtained (corresponding to the values along the x-axis). We compare against classical feature selection algorithms (dotted lines) including: feature importance (based on the \texttt{ExtraTreesClassifier} from \texttt{scikit-learn} with default parameters~\cite{scikit-learn}), a univariate approach based on feature-target mutual information (using the \texttt{SelectKBest(mutual\_info\_classif)} function from \texttt{scikit-learn} with default parameters), and \texttt{mifs-nd}, which is a greedy algorithm based on feature-feature and feature-target mutual information~\cite{hoque2014mifs}.

\begin{figure}[t]
    \centering
    \includegraphics[width=0.8\linewidth]{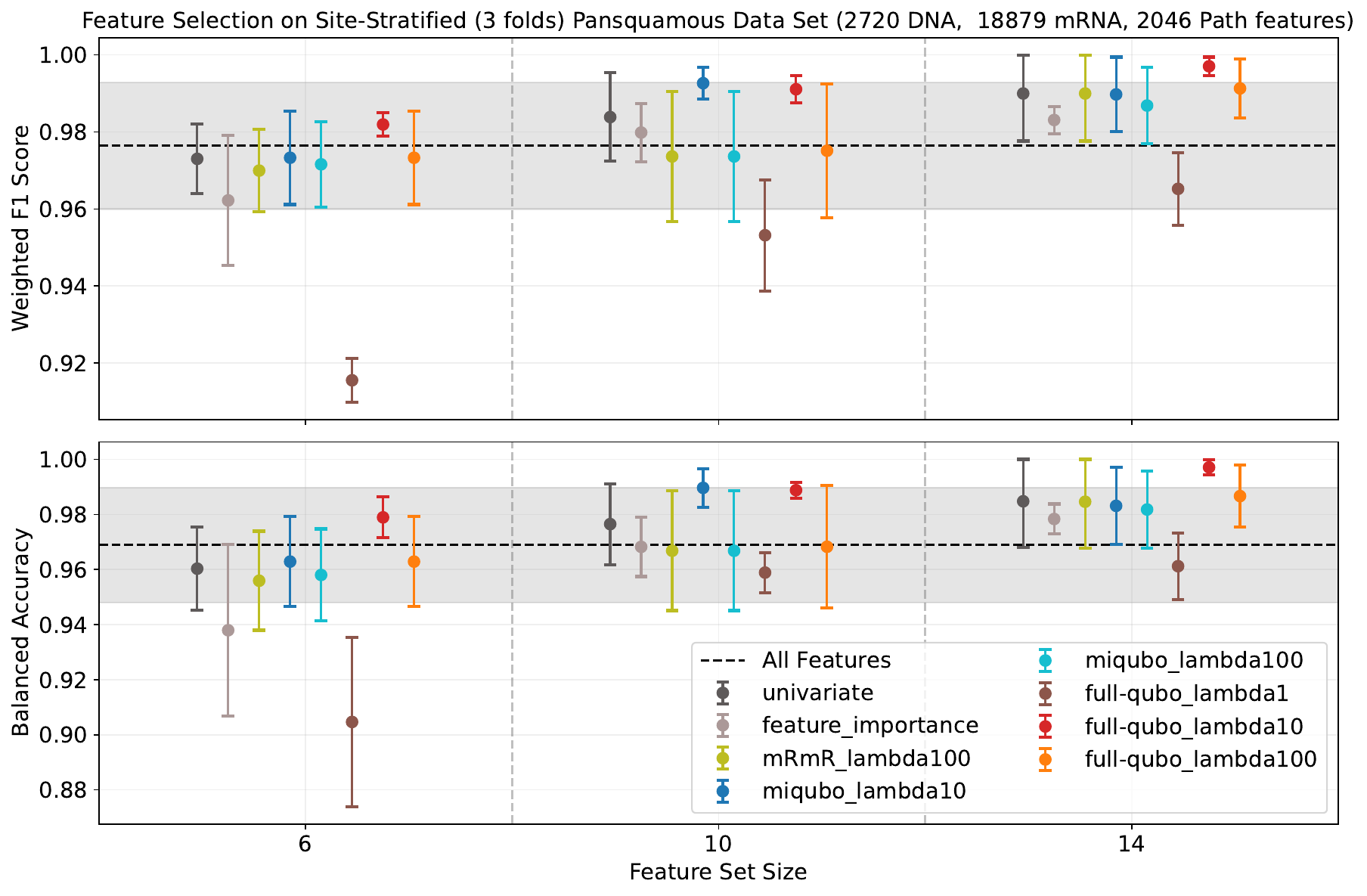}
    \caption{Average performance on the test split of a one-versus-rest logistic regression classifier after training on the train split of the site-stratified pansquamous dataset (composed of DNA, mRNA, and pathomics features and three tissue of origin class labels: HNSC, CESC, LUSC). Error bars denote the standard error of the mean. The \texttt{full-qubo} PCBO formulation with $\lambda=10$ selects high-performing feature sets of size 14 that significantly improve on the all-feature logistic regression classifier accuracy. }
    \label{fig:pansquamous-v2}
\end{figure}

Figure \ref{fig:pansquamous-v2} shows a later evaluation on a similar dataset, but, importantly, we incorporate the preprocessing techniques described in Sections \ref{sec:mrna-preprocessing} and \ref{sec:path-preprocessing} to produce more informative mRNA and pathomics features. We integrated all of the available DNA, mRNA, and pathomics features together into a dataset containing 660 rows, each corresponding to HNSC, CESC, or LUSC tissue of origin, and 23,645 total features.
The improved effect of the Sec.~\ref{sec:data-preprocessing} techniques can be seen in Fig.~\ref{fig:pansquamous-v2} by the improved performance of the logistic classifier trained on all features compared to what was achieved in Fig.~\ref{fig:f1-eval}.
To test and compare the different feature selection algorithms, first we addressed a known and problematic batch effect in the pathomic data that stems from site-specific signatures introduced during data collection~\cite{howard2021impact}. This batch effect was ameliorated by stratifying the full dataset into three separate folds, where the patient data from each site appear in exactly one of the three folds. These three disjoint folds were then combined to produce three different \textit{site-stratified} train-test splits of the data. Concretely, we start with disjoint subsets $\{\texttt{fold}_1, \texttt{fold}_2, \texttt{fold}_3\}$ of the full data such that for each site, all patients from the site are contained within the same fold. Then we combine these folds to produce train-test splits, such as $\texttt{split}_1 := \left( \{\texttt{fold}_2, \texttt{fold}_3\}_\textrm{train}, \{\texttt{fold}_1\}_\textrm{test} \right)$ with the other two splits generated by permuting the folds. Thus, if the learning model focuses on site-specific features in the training data, its performance suffers on the test data, where patient data are sourced from completely different sites. The stratification was also optimized to preserve, to the extent possible, the balance of cancer tissue labels (HNSC, CESC, and LUSC) seen in the full dataset.

For each train-test split, we evaluated each feature selection algorithm by asking it to select a small set of features from the training data. (in Fig.~\ref{fig:pansquamous-v2} feature set sizes range from 6 to 14.) Then, a one-versus-rest logistic regression classifier was trained to predict tissue label on the training data using the selected features. The trained model was then used to classify the test data. The feature selection, training, and testing was repeated for each algorithm five times, and the results corresponding to the best balanced accuracy on the test data were stored to represent that algorithm's performance for that specific train-test split. (Balanced accuracy is used to account for large differences in class sizes.) The performance across the three train-test splits was averaged to produce the data shown in Fig.~\ref{fig:pansquamous-v2}.
In addition to the \texttt{mRmR}, we also evaluated two other variants: \texttt{miqubo} and \texttt{full-qubo}. For all of the hybrid variants, groups of 1000 features were repeatedly generated and solved using the classical QBSolv algorithm (as described above) until the desired number of features was obtained.

The results of this evaluation, in Fig.~\ref{fig:pansquamous-v2}, show that the \texttt{full-qubo} PCBO with $\lambda=10$ is able to select very small, predictive feature sets, which outperform those selected by other algorithms. With 14 features, it significantly improves the model's performance beyond that of the classifier trained on all features. By successfully identifying a handful of highly informative features amidst the full feature space and leveraging them to enhance classifier performance, our approach not only simplifies the model but also yields tangible improvements in performance.

\begin{figure}[t]
    \centering
    \includegraphics[width=0.8\linewidth]{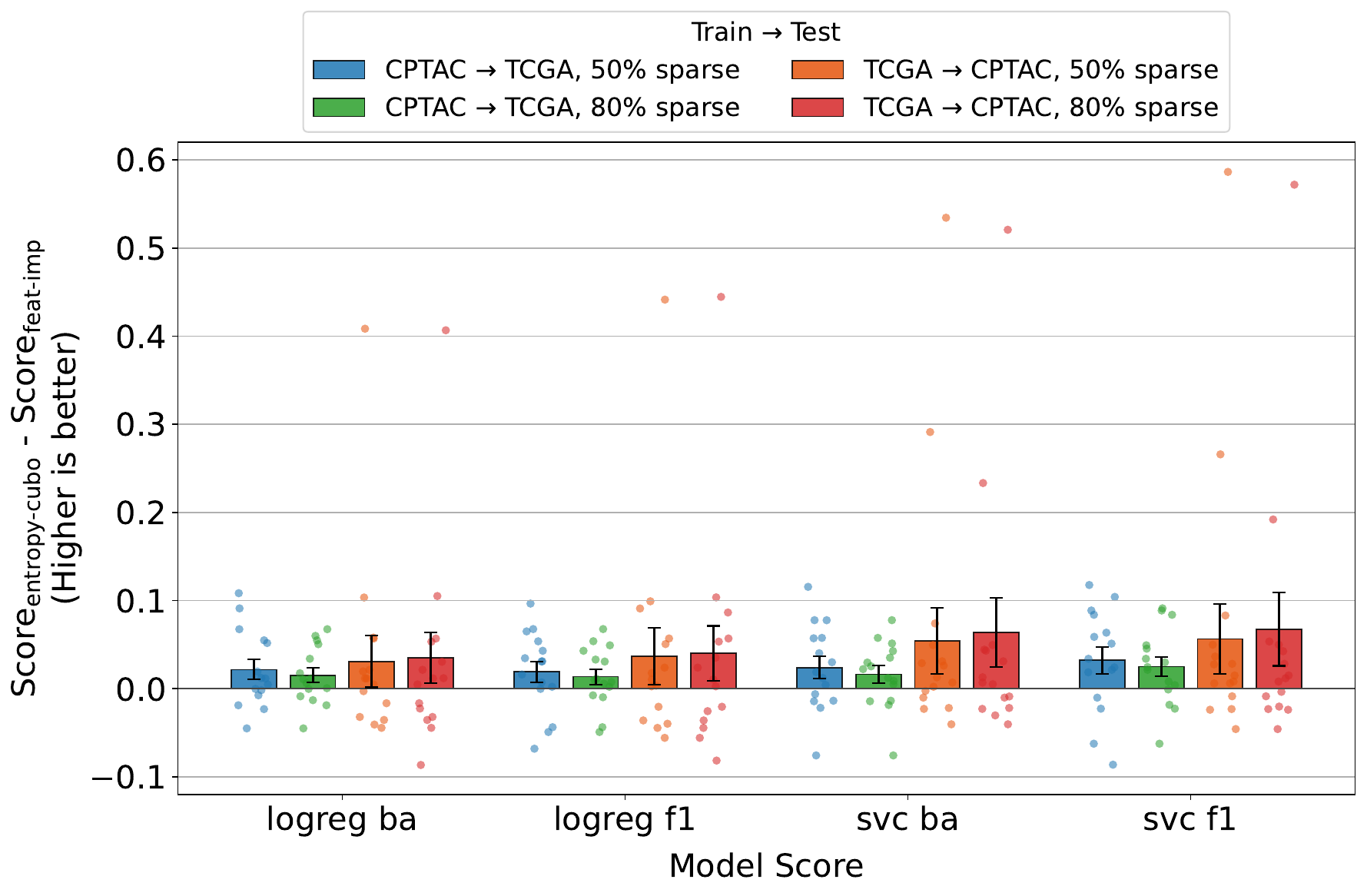}
    \caption{Average difference in model test scores after training on feature sets selected by \texttt{entropy-cubo} (after undergoing 50\% or 80\% edge sparsification) or feature-importance. In each problem instance, 10 mRNA features were selected from an original set of 40, then logistic regression (logreg) and support vector classifier (svc) models were trained on those features in the training dataset (either CPTAC or TCGA) and then tested in the other. Each dataset contains the same features and 8 different tissue of origin class labels. Individual problem instances are also shown scatter atop the bars with random jitter.}
    \label{fig:gurobi-cross-val}
\end{figure}

Finally, the results of one of our most recent evaluations are shown in Fig.~\ref{fig:gurobi-cross-val} where we show the difference in model score (both balanced accuracy and weighted F1) between the features selected by \texttt{entropy-cubo} and feature-importance.
In this evaluation, there are two separate datasets, TCGA (7,171 rows) and CPTAC (1,024 rows), both containing the same  mRNA features and a total of 8 different cancer tissues of origin as the class labels.
From these datasets, we generated 15 different sets of 40 randomly sampled features. Each set of 40 features defines a single feature selection problem where the objective is always to select 10 features. 
For each feature selection problem, feature-importance and \texttt{entropy-cubo} were used to select a set of 10 features. In this case, we evaluated two different variants of \texttt{entropy-cubo} (which always used the $\vec{\alpha} = (0,0,1)$ weighting) where either 50\% or 80\% of the terms in the problem were discarded before the problem was solved using the Gurobi Mixed-Integer-Programming (MIP) solver~\cite{gurobi2024}.

From the perspective of co-design, this evaluation was an important step towards understanding how well \texttt{entropy-cubo} can tolerate sparsification of terms -- which leads to reduced quantum gate cost when we consider how to map the problem to hardware. But it also served a practical purpose since we were unable to run Gurobi to termination for any 40-choose-10 problems containing all of the terms even after 48 hours of wall-clock runtime on 60 CPU cores with 500 GB of RAM.

\section{Bridging the Gap between Algorithm and Hardware} \label{sec:hardware}

To translate our formulated problem from the algorithmic level to the hardware level, we must address device-specific limitations. Specifically, implementations on near-term hardware are limited by a finite shot budget, which constrains the precision to which we are able to estimate expectation values of observables in a circuit. Imperfect gates and readout cause a deterioration in the output signal for larger circuits, effectively restricting the circuit depths we are able to scale to. For hardware systems such as the IBM Heron machines, we must also adapt to the available connectivity between qubits, as non-local connections require swapping qubits, which further contributes to the depth of the circuit.
In this section, we investigate parameter transfer, sparsification, and error mitigation methods to tackle these hardware constraints, iterating between experiment methodology design and running these experiments in simulation and hardware to study sensitivities and areas of improvement.

\subsection{Parameter Transfer}\label{sec:param-transfer}

\begin{figure}
    \centering
    \includegraphics[width=0.99\linewidth]{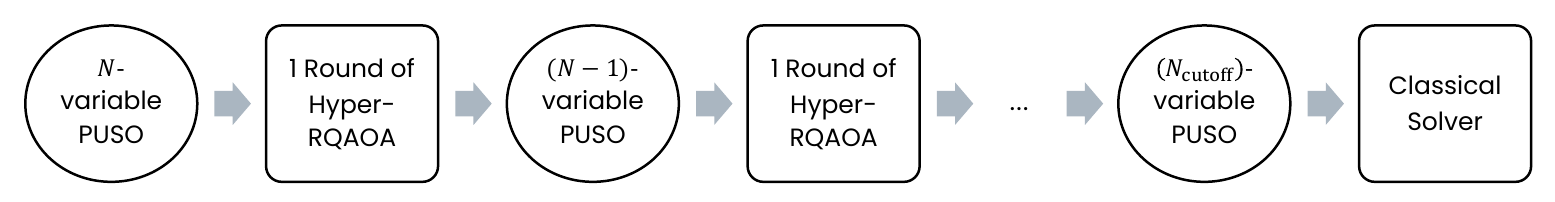}
    \caption{Overview of the hybrid feature selection algorithm applied to an $N$-variable input problem, with a user-specified value of $N_{\text{cutoff}} < N$. After the classical solver produces a solution for the $N_{\text{cutoff}}$-variable problem, a full solution is reconstructed for the original input problem by accounting for the constraints applied at the end of each round of f.}
    \label{fig:multi-round-overview}
\end{figure}

\begin{figure}[t]
    \centering
    \includegraphics[width=0.99\textwidth]{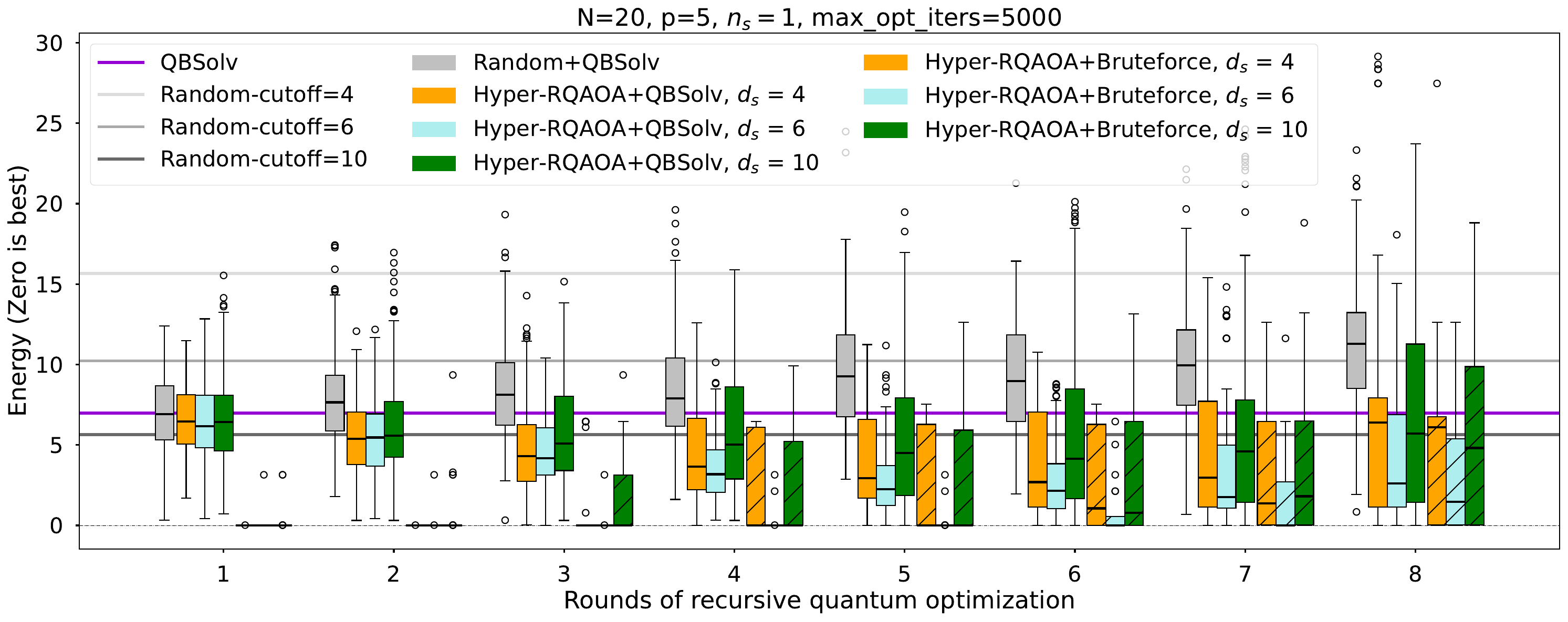}
    \caption{Sensitivity analysis of HRQAOA, combined with different classical solvers, analyzing performance as a function of the total rounds of recursive quantum optimization (i.e., edge fixing) and varying the size $d_s$ of the donor problem. The horizontal lines are the median values of the classical baselines. The HRQAOA+QBSolv (solid yellow, blue, and green) boxplots consistently achieve lower energies compared to the Random+QBSolv (gray) boxplots. The plots also reveal a trade-off between the number of recursion rounds and final energy, where fixing a few edges yields improved performance, but fixing too many edges can result in suboptimal performance if the RQAOA failed to fix the correct edges. The HRQAOA+Brute-force (hatched yellow, blue, and green) boxplots illustrate this point well. For the first few rounds they show that the ground state solution can be recovered with high probability, but as more edges are fixed, the median energy begins to increase as well.}
    \label{fig:hybrid-vs-classical}
\end{figure}

\begin{figure}[t]
    \centering
    \begin{subfigure}[b]{0.8\textwidth}
        \centering
        \includegraphics[width=\linewidth]{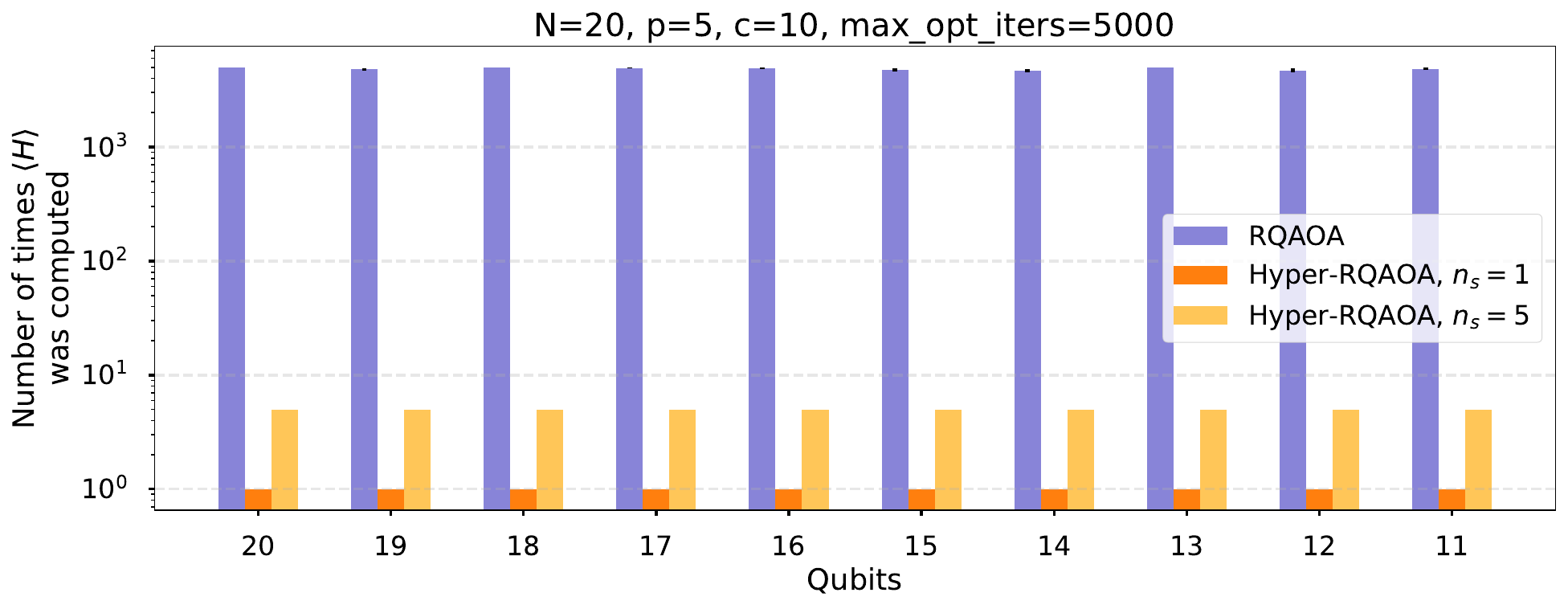}
        \caption{}
        \label{fig:resource-savings}
    \end{subfigure}
    \hfill
    \begin{subfigure}[b]{0.52\textwidth}
        \centering
        \includegraphics[width=\linewidth]{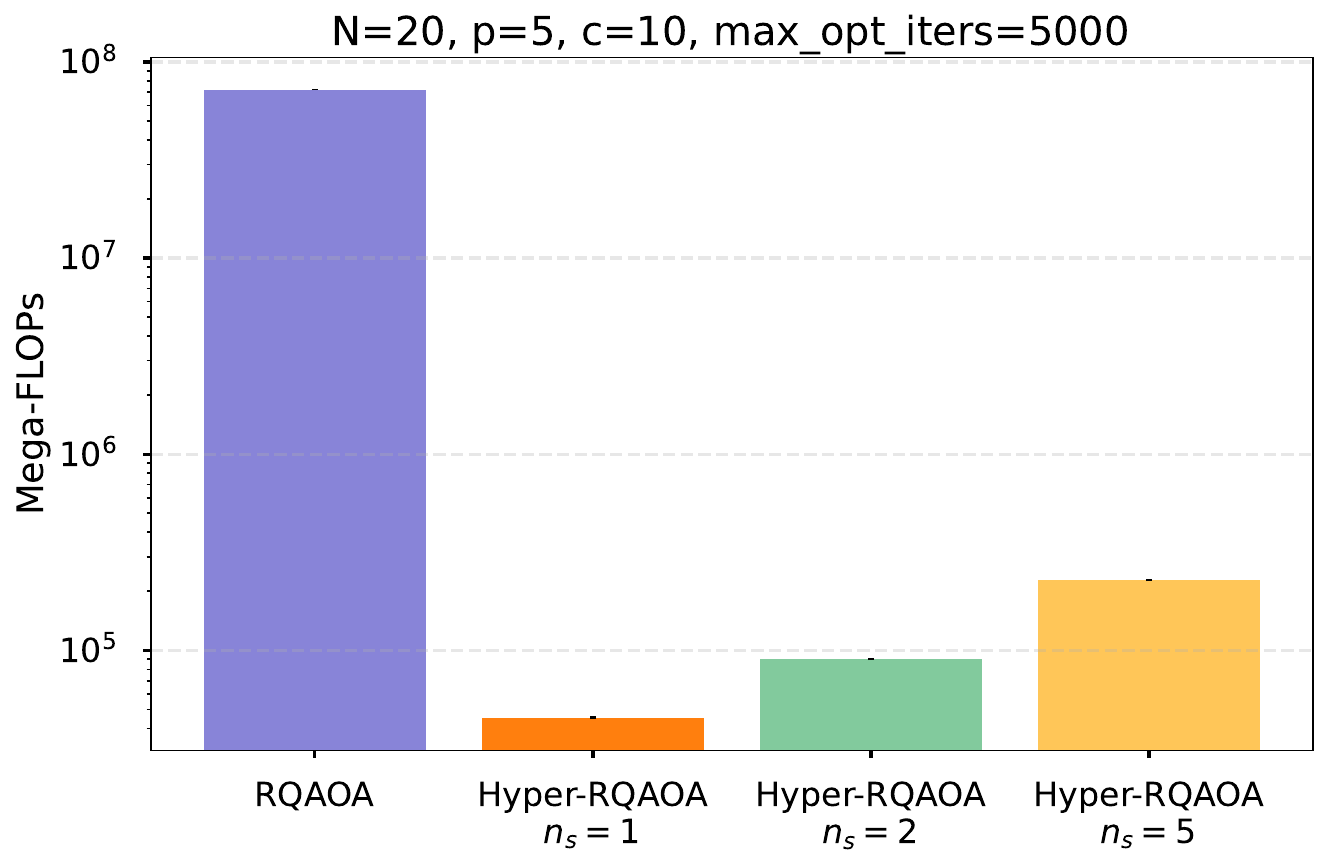}
        \caption{}
        \label{fig:mega-flops-comparison}
    \end{subfigure}
    \caption{(a) The average number (over 22 trials) of calls to the objective function $\langle H \rangle$ at each recursive step of the RQAOA. Note the log-scale on the y-axis; while RQAOA requires a full variational optimization at each circuit size, HRQAOA leverages parameter transfer from smaller (in this case $d_s=10$ qubits) circuits and only executes a fixed $n_s$ number of these larger circuits. These simulations used the COBYLA optimizer with the maximum number of allowed iterations set to 5,000. (b) Resource overhead of RQAOA and HRQAOA in terms of the total mega-FLOPs ($1\times10^6$ FLOPs) expended for each simulation. The bars denote average values obtained over 22 independent trials.}
    \label{fig:param-transfer-resource-reduction}
\end{figure}

One of the primary bottlenecks for variational quantum algorithms, including RQAOA, is the optimization loop. It alternates between running quantum circuits to collect enough samples to estimate a cost function and using a classical optimizer to choose the next parameter updates. To circumvent the prohibitive number of shots that would be required for this step, we employ parameter transfer, a technique for initializing large QAOA circuits using the variational parameters obtained from smaller QAOA circuits \cite{brandao_for_2018, galda_transferability_2021, shaydulin_parameter_2023, augustino2024strategies, hao2024end, pelofske2024scaling}.

In its original formulation \cite{bravyi2020obstacles}, the RQAOA is composed of iterative rounds of QAOA which recursively reduce the problem size by fixing edges within the problem until it reaches the user-designated ``cutoff'' size, as shown in Fig.~\ref{fig:multi-round-overview}. Once the cutoff point is reached, the reduced problem may be solved by another classical algorithm (e.g, a heuristic solver or exhaustive brute-force search), and a solution over the original input problem is obtained by working backwards through the edge constraints generated by each recursive application of QAOA. 

To avoid the expensive variational loop that appears during each recursive round of QAOA, we developed a solution that integrates the parameter transfer technique within the RQAOA, referring to this combination as the hyper-RQAOA (HRQAOA) algorithm. Instead of performing the variational optimization over the target problem, we first uniformly sample $n_s$ separate ``donor'' subproblems of size $d_s$ from the $N$-variable target problem. We strategically choose the size $d_s < N$ of the donor subproblems so that they can be easily simulated on a classical computer.
After each of the $n_s$ subproblems are trained (the optimizations are independent of one another so this can be performed in parallel) we obtain a list $[(\vec{\gamma_1}, \vec{\beta_1}), \dots, (\vec{\gamma_{n_s}}, \vec{\beta_{n_s}})]$ of potential donor parameters for the larger target problem.

To determine which set of $({\vec\gamma_i}, \vec{\beta_i})$ parameters will be used, we make $n_s$ calls to the quantum computer to compute $E_i = \langle \psi({\vec\gamma_i}, \vec{\beta_i}) | H_{\text{PCBO}} | \psi({\vec\gamma_i}, \vec{\beta_i})\rangle$ 
where 
\begin{equation} \label{eq:pcbo-hamiltonian}
    H_{\text{PCBO}} = \sum_l \alpha_l P_l = \sum_i C(i) Z_i + \sum_{i<j}C(i,j)Z_iZ_j + \sum_{i<j<k}C(i,j,k)Z_iZ_jZ_k,
\end{equation}
is obtained by converting the cost function in Equation \ref{eqn:pcbo} (or any of the other \texttt{mRmR}, \texttt{miqubo}, \texttt{full-qubo}, or \texttt{entropy-cubo} PCBOs) into a quantum operator under the transformation $x_i \rightarrow (1-Z_i) / 2$, with $Z_i$ being the Pauli-Z operator.
We select whichever set of parameters yields the lowest transferred energy to be the donor parameters $({\vec\gamma_{\text{don}}}, \vec{\beta_{\text{don}}}) = \argmin_{({\vec\gamma_i}, \vec{\beta_i})} \langle \psi({\vec\gamma_i}, \vec{\beta_i}) | H_{\text{PCBO}} | \psi({\vec\gamma_i}, \vec{\beta_i})\rangle$. 

Specifically, the chosen donor parameters  are used to prepare the $N$-qubit state $\ket{\psi({\vec\gamma_{\text{don}}}, \vec{\beta_{\text{don}}})}$, and the probability distribution defined by this quantum state is used to compute a ``correlation dictionary'' $\{(P_l): \langle \psi(\vec\gamma_{\text{don}}, \vec{\beta_{\text{don}}}) | P_l | \psi(\vec\gamma_{\text{don}}, \vec{\beta_{\text{don}}}) \rangle \}$, which describes the expectation value for each Pauli term, or edge, in $H_\text{PCBO}$. We refer to these Pauli terms as edges because this problem can be viewed as an instance of a weighted spin glass model on a complete hypergraph containing first\nobreakdashes-, second\nobreakdashes-, and third-order interactions. The application of QAOA to this family of problems has been previously studied in the context of finding optimal encodings \cite{campbell2022qaoa} as well as hardware evaluations \cite{pelofske2024short} and demonstrations of successful parameter transfer on IBM's superconducting processors \cite{pelofske2024scaling}.

At this point, the HRQAOA returns to the standard execution of RQAOA first presented in \cite{bravyi2020obstacles}.
The correlation dictionary contains information about the relative orientation of the problem's spin variables. We select whichever edge has the largest absolute value $\abs{\langle P_l \rangle}$ and fix the sign of that edge, which has the result of reducing the number of free variables within the problem by one. We refer to this step as the ``recursive step'' or ``edge fixing''. As an example, suppose the edge with the largest expectation value was $\langle Z_i Z_j Z_k \rangle = -1$. Without loss of generality, we randomly select one of the indices and assign it the value $Z_i = - Z_j Z_k$. Substituting this expression back into Equation \ref{eq:pcbo-hamiltonian} yields a new Hamiltonian with one fewer free variables, and the solution over the whole problem can be recovered once the subproblem is solved, which assigns definite $\{-1, +1\}$ values to $Z_j$ and $Z_k$, and allows us to infer the value of $Z_i$.

To our knowledge, this represents the first study of parameter transfer for RQAOA and the first demonstration of effective parameter transfer for problems derived from real-world datasets. The key enabling feature for successful parameter transfer within the RQAOA is the subsampling step that produces smaller, more tractable problem instances from the larger target problem itself, effectively creating smaller problems from the same problem class, that can be used to initialize the circuit parameters on the target problem.
We found that parameter transfer can effectively initialize larger RQAOA circuits using smaller classical simulations, resulting in lower energy solutions using over 1000x fewer resources. These resource savings suggest our hybrid feature selection algorithm may be viable on quantum hardware much sooner than previously anticipated, with significant implications for future hardware experiments and real-world applications in cancer research.

\paragraph{Boosting QBSolv's Performance in Simulation.}
Leveraging parameter transfer and classical simulation, we evaluated the performance of the HRQAOA by executing several rounds of edge fixing to reduce the effective problem size before handing the reduced problem off to a classical solver. The results are shown in Fig.~\ref{fig:hybrid-vs-classical}. We tested using both an optimal brute-force search and a classical heuristic for solving the edge-fixed subproblems. This hybrid approach demonstrated that the performance of the classical heuristic could be improved, compared to the classical heuristic operating alone, by first reducing the problem size with a few rounds of RQAOA. 
Specifically, in Fig.~\ref{fig:hybrid-vs-classical} we show the distribution of solution energies obtained by a variety of methods on a 20-variable PCBO problem, with $y=0$ denoting the ground state solution energy. The median energy obtained by QBSolv \cite{qbsolv2017} after 100 trials is denoted by the horizontal purple line. Similarly, the ``Random-cutoff=$C$'' baselines correspond to the median value after 100 trials of randomly fixing edges in the PCBO until the cutoff $C$ is reached, at which point the reduced problem is solved exactly via brute-force search to find the ground state. Therefore, the Random-cutoff=10 baseline achieves the lowest median energy since it involves the fewest number of random edge fixes. Next, the grey Random+QBSolv boxplots (representing a distribution of 100 trials) correspond to a number of rounds of random edge fixing (the number of rounds is denoted by the x-axis) after which the reduced problem is solved with QBSolv. The yellow, blue, and green boxplots correspond to different sized donor circuits ($d_s$) used to generate the transferred parameters in the HRQAOA, where each boxplot represents the distribution over 30 trials. The solid boxplots show the result after solving the reduced problem with QBSolv, whereas the hatched boxplots are solved with brute-force search.
What we see is that the HRQAOA+QBSolv is able to significantly decrease the median solution energy compared to QBSolv alone, up until approximately 6 rounds of edge fixing where the energy begins to trend upwards. The HRQAOA+Brute-force removes the variability of the classical heuristic so we are able to see the quality of the edge fixing procedure itself. We see that the $d_s=10$ trials have a higher probability of fixing an incorrect edge, followed by the $d_s=4$ trials.
Optimizing the specific values of $d_s$, $p$, and the number of rounds of edge fixing is an important and ongoing subject of research.

\paragraph{Quantifying Resource Savings.}
Figure \ref{fig:resource-savings} presents a comparative analysis of the quantum computational demands between the standard RQAOA algorithm and the proposed HRQAOA approach when applied to a problem with $N=20$ variables. 
The vertical axis, displayed on a logarithmic scale, quantifies the total number of times the energy expectation value $\langle H_{\text{PCBO}} \rangle$ was computed during the course of each algorithm.

The consistent number of $\langle H \rangle$ evaluations made by RQAOA (blue bars) across each recursive step (from 20- to 11-qubits) shows that it is repeatedly hitting the upper bound set for the COBYLA (constrained optimization by linear approximation) optimizer (here we set \texttt{max\_opt\_iters}=5,000). Despite the results being averaged over 22 independent trials, the minimal error bars indicate remarkably consistent behavior across different problem instances.
In stark contrast, the HRQAOA implementations demonstrate dramatically reduced quantum computational requirements. With $n_s=1$, the algorithm requires only a single evaluation of $\langle H \rangle$ per recursive round -- representing an approximately 5,000-fold reduction in quantum computational cost. Even when increasing to $n_s=5$, the quantum resource requirements remain approximately 1,000 times lower than the standard RQAOA approach. This dramatic reduction stems from the parameter transfer technique's ability to eliminate the need for direct variational optimization on the larger quantum circuits.

However, the results in Figure \ref{fig:resource-savings} do not capture HRQAOA's classical overhead associated with simulating $n_s$ different $d_s$-qubit donor circuits.
Figure \ref{fig:mega-flops-comparison} captures this information by quantifying the total classical computational cost in mega-FLOPs ($1\times10^6$ floating-point operations) for each algorithm variant. This metric encompasses all classical processing required during our simulations, including the classical overhead of simulating the $n_s$ donor circuits in the HRQAOA approach. In this data, each donor circuit contained 10 qubits.
The standard RQAOA algorithm exhibits a classical computational cost approaching $10^8$ mega-FLOPs, attributable to the expensive classical optimization loop over the large $q$-qubit quantum circuits that take on values $N \geq q > c$ during the course of RQAOA. In comparison, HRQAOA with $n_s=1$ requires approximately $4\times10^4$ mega-FLOPs -- a reduction of more than three orders of magnitude. While increasing the number of subsampled donor problems ($n_s$) to 2 and 5 incrementally raises the computational cost, even the most resource-intensive HRQAOA configuration ($n_s=5$) maintains significant advantage in classical efficiency over the standard approach.
These results collectively demonstrate that the parameter transfer technique delivers substantial computational advantages. This resource efficiency represents a significant advancement toward making quantum optimization algorithms practically viable on quantum hardware with limited gate speeds.

\subsection{Sparsification}
\label{sec:sparsify-mitigate}

\begin{figure}[t]
    \centering
    \includegraphics[width=0.6\linewidth]{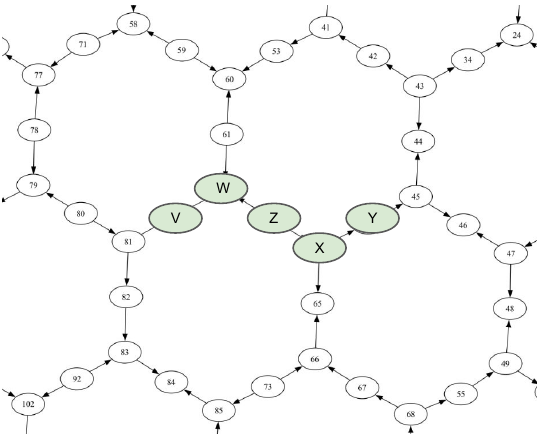}
    \caption{An example of the heavy-hex lattice connectivity between qubits on the IBM Heron-class quantum computers. Suppose an input \texttt{entropy-cubo} problem contains a set of third order edges $\{(V, W, Z), (W, X, Z), (X, Y, Z)\}$. The hardware-aware mapping technique we developed will place those program qubits on the physical qubits as shown in the diagram above. Then, those third-order edges can be implemented in the QAOA ansatz without incurring any SWAP operations.}
    \label{fig:heavy-hex}
\end{figure}

\begin{figure}[t]
    \centering
    \includegraphics[width=0.95\linewidth]{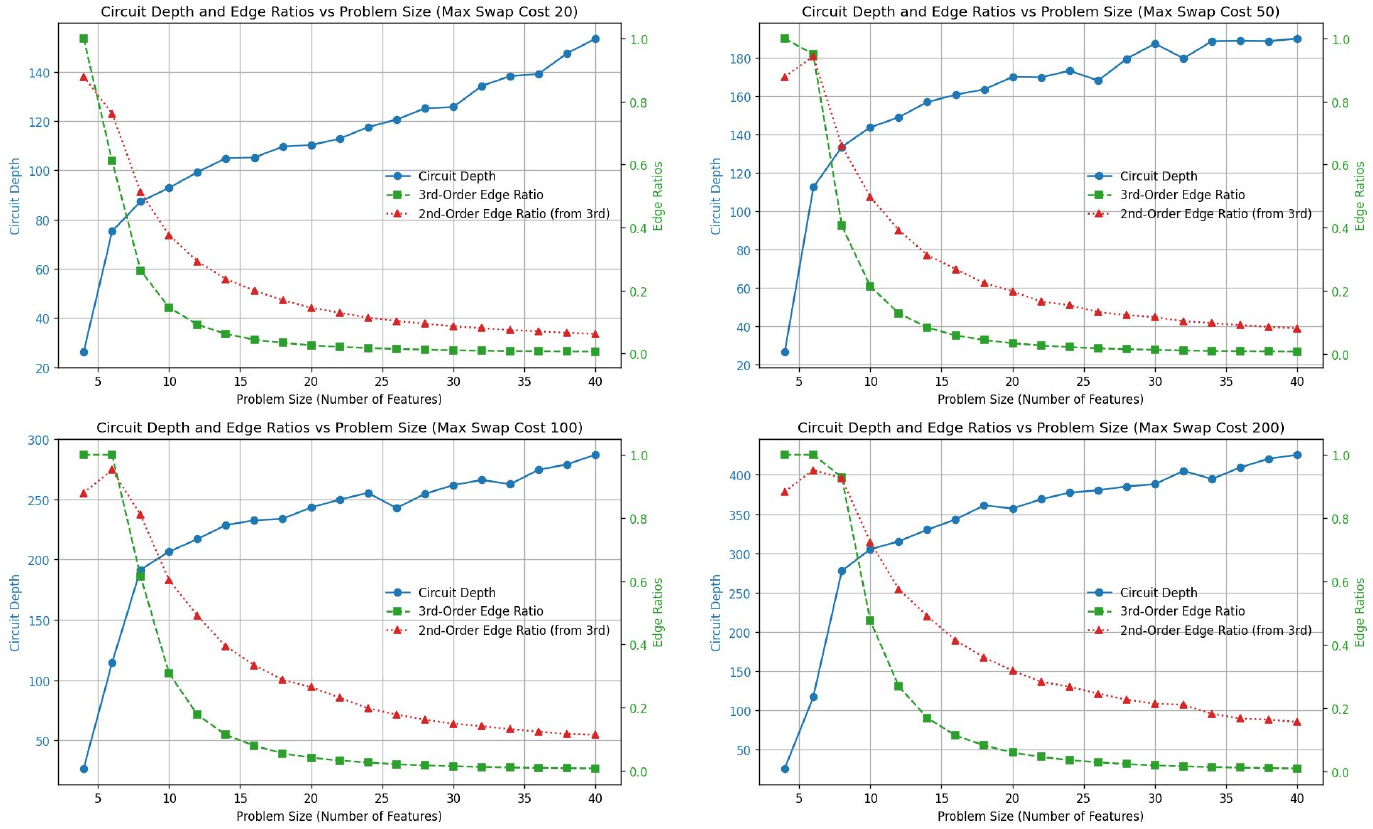}
    \caption{Resulting circuit depths (blue curve, left y-axis) after performing hardware-aware sparsification of an input \texttt{entropy-cubo} problem for different settings of the ``Max SWAP Cost'' parameter. We also report the ratio of second- (red) and third-order (green) edges that remain after sparsification.}
    \label{fig:hardware-aware-depths}
\end{figure}

\begin{figure}[t]
  \centering
  \begin{subfigure}[b]{0.49\textwidth}
      \centering
      \includegraphics[width=\linewidth]{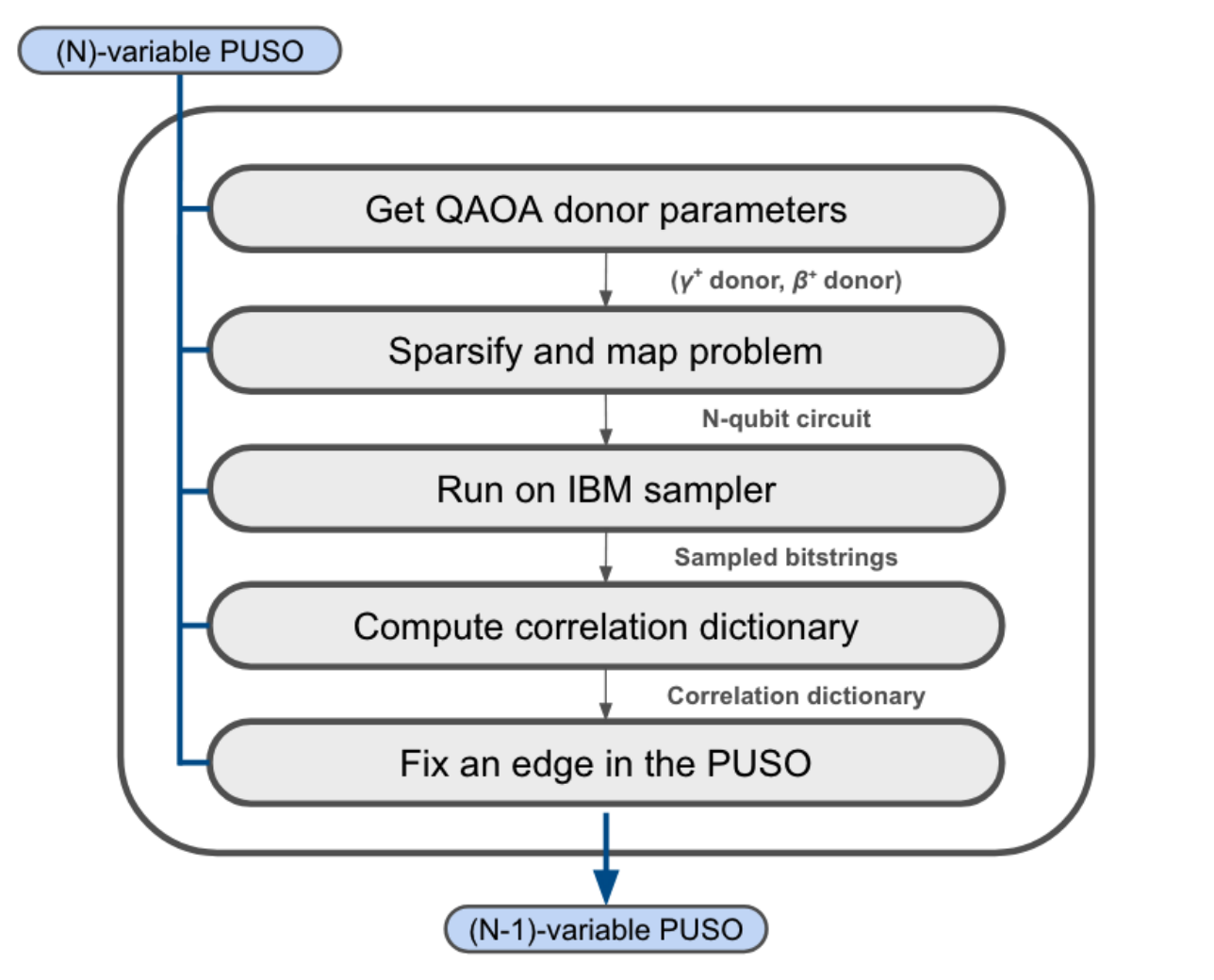}
      \caption{}
      \label{fig:sampler-flow}
  \end{subfigure}
  \hfill
  \begin{subfigure}[b]{0.49\textwidth}
      \centering
      \includegraphics[width=\linewidth]{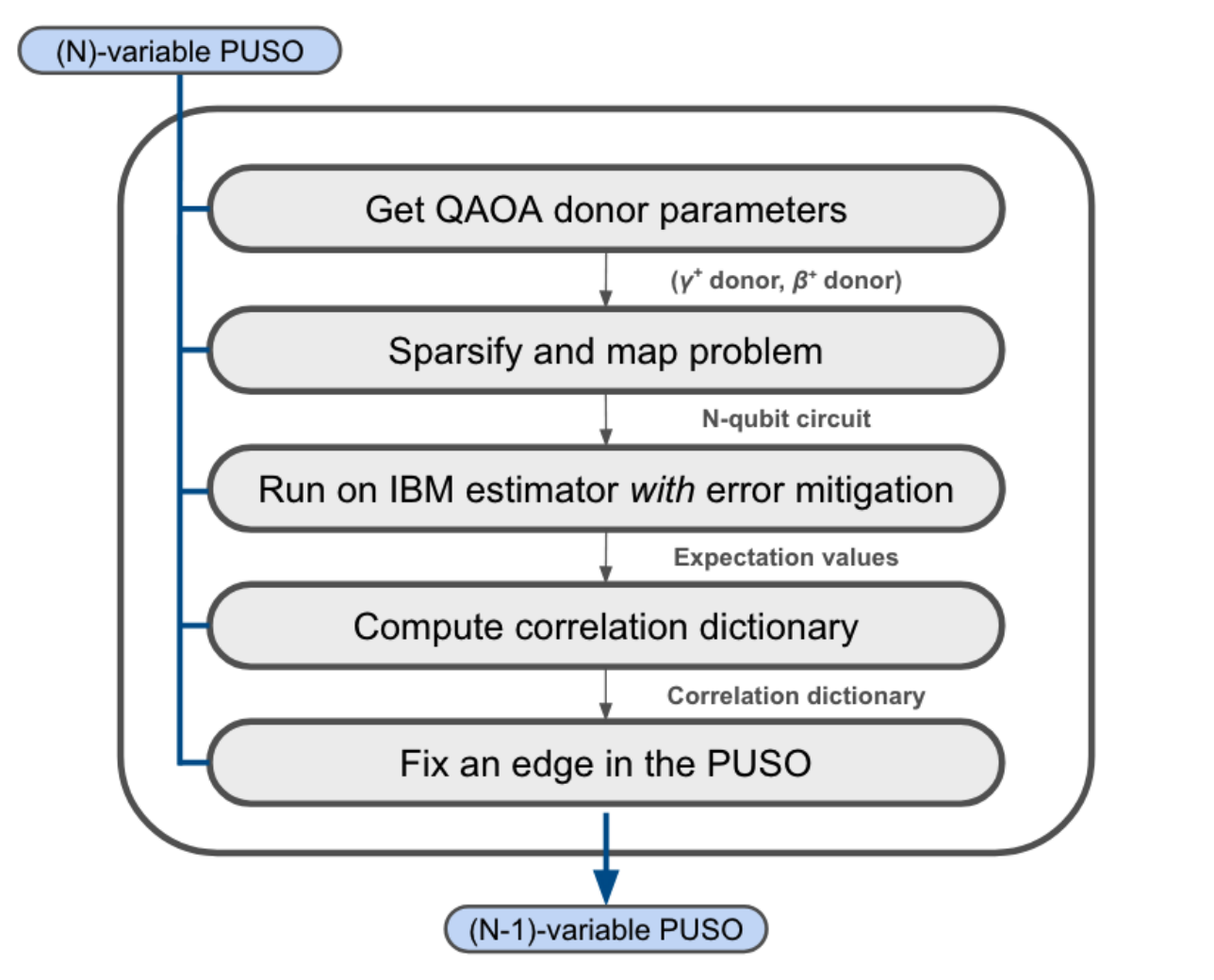}
      \caption{}
      \label{fig:estimator-flow}
  \end{subfigure}
  \caption{(Left) Sampler pipeline used in Experiment~1. Bitstrings from hardware are used to build a correlation dictionary, which drives a single edge-fix in the PCBO. (Right) Estimator pipeline used in Experiment~2. Expectation values of Pauli observables are estimated on hardware with noise mitigation applied and used to construct the correlation dictionary for edge-fixing.}
  \label{}
\end{figure}

\begin{figure}[t]
  \centering
  \includegraphics[width=0.5\linewidth]{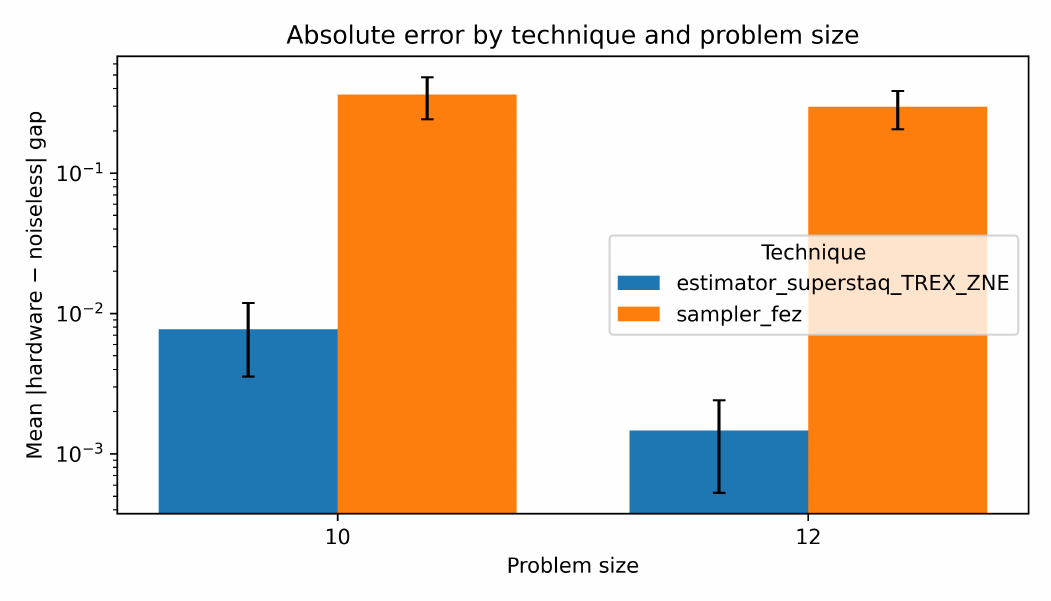}
  \caption{Mean absolute gap error: \texttt{Sampler} vs.\ \texttt{Estimator + TREX+ZNE}. Lower is better; the estimator with mitigation is orders of magnitude closer to the noiseless baseline.}
  \label{fig:mitigation-summary}
\end{figure}

Implementing the second- and third-order interactions of our unsparsified \texttt{entropy-cubo} PCBOs requires connecting every set of two and three qubits. However, IBM Heron-class devices use a heavy-hex connectivity graph -- executing all of the second- and third-order terms naively on this topology would result in circuits that are depth- and CNOT-prohibitive, both because of the $O(N^3)$ multi-qubit phase gates needed to implement the actual terms, and the SWAPs required to achieve the necessary connectivity. To reduce the depth and 2-qubit-gate overhead, we investigate methods for sparsifying the Hamiltonian before compilation. The goal is to produce a Hamiltonian that requires less connectivity while still preserving the information that drives edge-fixing and, critically, avoiding collapsing the instance into something that classical solvers find trivial. Here, we overview several potential sparsification methods.

\subsubsection{Simple and randomized sparsification}
\label{sec:simple-random-sparsify}

\paragraph{Simple (weight-ordered) sparsification.}
A practical baseline is to sort terms by absolute weight and retain only the largest ones, discarding a tail of small-magnitude terms. Empirically, moderate truncation often preserves the original ground state on our \texttt{entropy-cubo} problems, suggesting that the most useful signal concentrates in higher-weight edges. However, removing too much structure also reduces classical hardness, so we treat the retained fraction as a co-design knob and routinely sanity-check hardness after truncation.

\paragraph{Randomized sparsification of the small-angle tail.}
Rather than dropping all small terms, we replace the tail by a randomized surrogate that keeps large terms deterministically and applies a few larger surrogate rotations \emph{infrequently} to stand in for many tiny rotations. Conceptually, this is akin to qDRIFT-style \cite{Campbell_2019} randomized operator selection and probabilistic angle interpolation: it reduces two-qubit usage and depth while keeping expectation values close to the full model on the observables we care about. In practice, we choose a threshold that differentiates “large” versus “small” angles: all large-angle terms are kept deterministically, and the small-angle set is sampled with probabilities proportional to their weights, using a fixed surrogate angle. This yields a better accuracy–cost trade-off than deterministic truncation alone, especially when most weight resides in the head of the distribution but many small terms remain.

\paragraph{Practical recipe and guardrails.}
In runs to date, we first fix the budget of large-angle terms to retain (to cap depth and CNOTs), then layer in the randomized tail so that a limited number of additional surrogate rotations injects the small-term influence without exploding depth. We monitor two safeguards: (i) that the retained third-order content remains substantial (to preserve the interaction structure our edge-fixing exploits), and (ii) that classical difficulty does not collapse (measured in terms of Gurobi ``work'', a quantifier for computational effort required for the problem). The primary objective is to produce sparsified circuits that compile to heavy-hex with far fewer SWAPs and materially smaller depth, yet continue to support reliable edge-fix decisions in simulation.

\paragraph{Co-design perspective.}
Sparsification is a deliberate trade; we buy hardware feasibility with depth/CNOT savings, but we pay with some classical easiness. By measuring both sides -- circuit metrics on the quantum path and hardness on the classical path -- we steer to a regime that is executable on today’s machines while remaining nontrivial for exact classical solvers. This sets the stage for hardware-aware (heavy-hex) sparsification in the next subsection.

\subsubsection{Hardware-aware (heavy-hex) sparsification}
\label{sec:hw-aware-problems}
Hardware-aware sparsification aims to incorporate our knowledge of the device topology into our sparsification technique. Similar to simple and randomized sparsification, we prioritize high-weight terms, but also place importance on terms that can be mapped to connectivity-friendly layouts that minimize SWAP overhead and depth. 

\paragraph{Mapping strategy.}
We first sort all third-order terms by weight and place their qubits onto the heavy-hex graph so that each selected triplet occupies a short connected chain (Fig.~\ref{fig:heavy-hex}). The chain order is irrelevant for implementing $e^{i\theta Z_i Z_j Z_k}$, so we greedily choose placements that avoid detours and reduce two-qubit routing, until all qubits have been placed. We then sweep in second-order terms between physical neighbors (zero additional routing cost) and include all first-order terms as single-qubit $Z$ rotations. Conflicts between overlapping triplets are resolved by preferring higher-weight terms and reusing already-placed qubits when it does not increase route length.
The small-angle tail can either be omitted or represented via randomized truncation (Sec.~\ref{sec:simple-random-sparsify})
, which preserves expectation values relevant to our correlation dictionary with far fewer CNOTs.

\paragraph{Practical impact.}
Figure~\ref{fig:hardware-aware-depths} summarizes the outcome across several routing budgets: circuit depth decreases substantially compared to na\"ive all-to-all compilation, while a meaningful fraction of third-order content is retained. This co-design dial (how many high-weight triplets survive versus total depth) is tuned jointly with classical-hardness checks (e.g., Gurobi work/mip-gap) to avoid over-simplifying the instance. In practice, these mapped circuits transpile reliably on heavy-hex and support stable edge-fixing in noiseless simulation, setting up the hardware results in Sec.~\ref{sec:hw-experiments}.

\subsection{Hardware results with heavy-hex sparsification}
\label{sec:hw-experiments}
All hardware runs used our topology-aware heavy-hex mapper on IBM Heron R2 (“Fez”) devices. We evaluated performance with an \emph{energy-gap} metric: after the hybrid loop fixes one edge in the PCBO, we compute the difference between the ground-truth energy and the energy of the reconstructed PCBO. Smaller gaps indicate more reliable edge-fixing. To control circuit depth and two-qubit usage, we apply a heavy-hex–aware mapper for these experiments and sparsify higher-order terms before circuit construction.

\paragraph{Sampler pipeline.}
In the first study we executed the pipeline with \texttt{Qiskit Sampler} (see Fig.~\ref{fig:sampler-flow}). Starting from an \(N\)-variable PCBO, we extract the donor $(\vec\gamma_{\text{don}}, \vec{\beta_{\text{don}}})$ parameters, build the receiver circuit to run it on hardware to collect bitstrings, form a correlation dictionary from those samples, and use the highest correlation in the dictionary to fix one edge. While simple, this route estimates correlations indirectly from finite samples and is sensitive to readout bias and shot noise, which can inflate the reconstructed energy gap and increase variability.

\paragraph{Estimator pipeline.}
In the second study we replaced sampling with \texttt{Qiskit Estimator} (Fig.~\ref{fig:estimator-flow}), directly estimating the expectation values of the Pauli observables that define the correlation dictionary. We pair this with lightweight mitigation -- twirled readout error extinction (TREX) and zero-noise extrapolation (ZNE) -- applied at the observable level. In our studies, moving from sampling to direct expectation estimation materially reduces the energy gap after an edge fix, and adding TREX+ZNE further tightens agreement with noiseless baselines while reducing run-to-run variance. This aligns with the design goal: edge-fixing depends on accurate multi-qubit expectations, which the estimator pipeline targets natively.

\paragraph{Error-mitigation comparison.}
Figure~\ref{fig:mitigation-summary} summarizes the effect of mitigation by comparing the mean absolute gap error for the \texttt{Sampler} pipeline versus the \texttt{Estimator} pipeline with TREX+ZNE at the same problem sizes. The Estimator-with-mitigation bars are lower by orders of magnitude, indicating that combining \texttt{Estimator} with hardware-native mitigation yields substantially more faithful correlation dictionaries and more reliable edge-fix decisions on Heron.

\section{Resource Analysis for Empirical Quantum Advantage}\label{sec:resources}

\begin{figure}[t]
    \centering
    \begin{subfigure}[b]{0.49\textwidth}
        \centering
        \includegraphics[width=\textwidth]{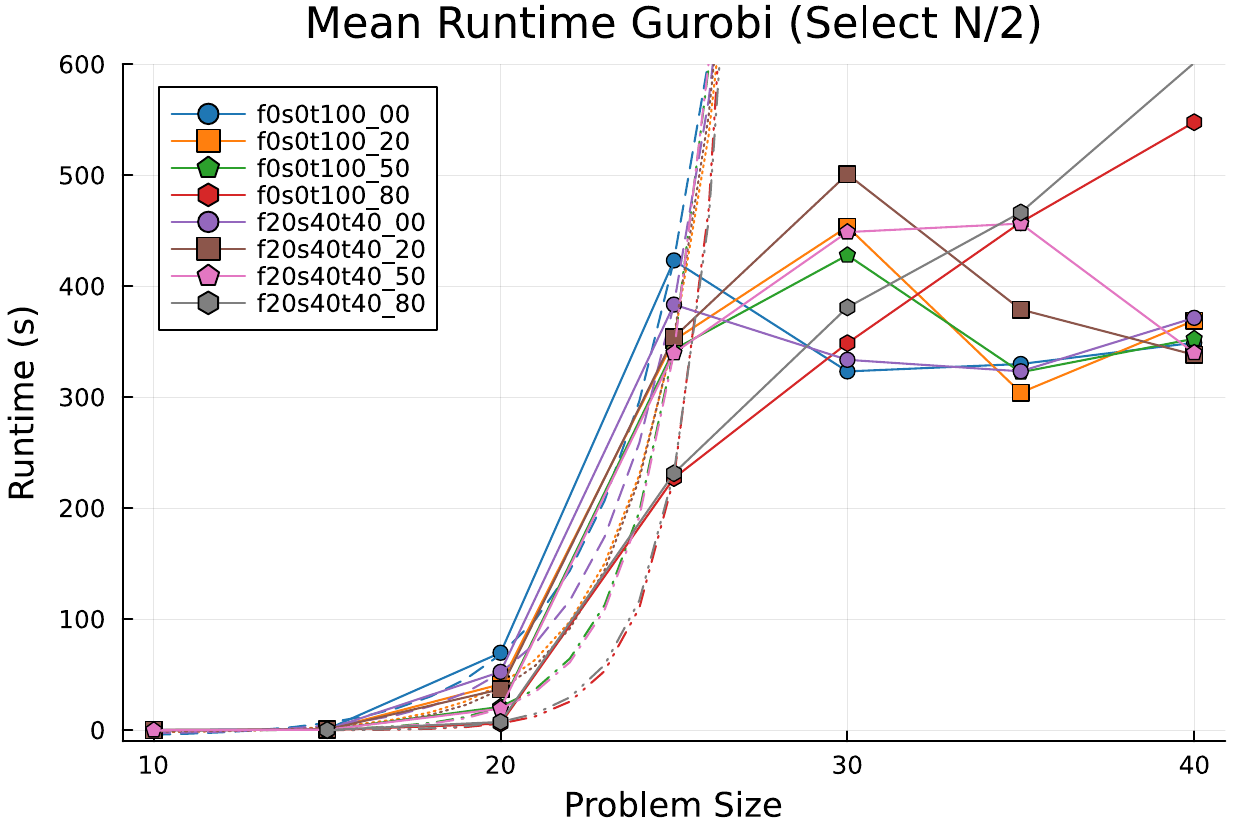}
        \caption{}
        \label{fig:gurobi-runtime}
    \end{subfigure}
    \hfill
    \begin{subfigure}[b]{0.49\textwidth}
        \centering
        \includegraphics[width=\textwidth]{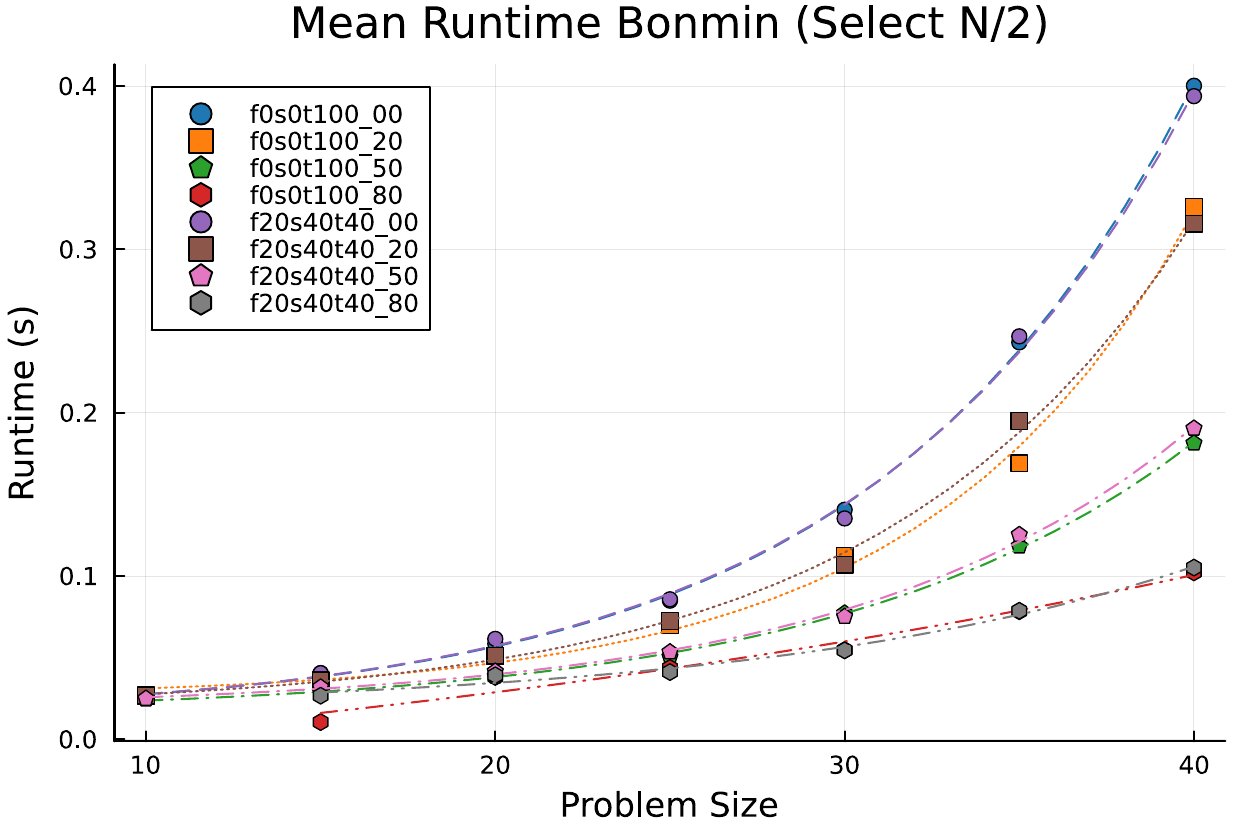}
        \caption{}
        \label{fig:bonmin-runtime}
    \end{subfigure}
    \caption{Mean runtime scaling study of classical solvers on \texttt{entropy-cubo} PCBOs as a function of problem size, where $N/2$ features are to be selected since these combinatorially have the largest search space. Each data point in this plot is averaged over ten different problem instances.  For each problem size, Gurobi was given a termination criterion that would trigger if the incumbent solution had not improved within 5 minutes. At $N\geq25$ it hit this termination criterion, and therefore, in (a), we only fit the curve up to $N=25$ with the exponential function $y=ae^{bx}+c$. The data is shown for multiple term weightings and sparsification.  For example "f0s0t100\_00" indicates a problem with 0\% weight on the first order terms, 0\% weight on the second order terms, and 100\% weight on the third order terms, with 0\% sparsification.  Due to the low problem size, we did not consider an 80 \% sparsification for problems of size $N = 10$. (a) We report the runtime (in units of seconds) Gurobi used before it terminated.  Dashed lines indicate an exponential fit for data to $N = 25$, where Gurobi often, though not always, proved optimality before the time based termination criterion was reached. (b) We report the runtime (in units of seconds) of the Bonmin heuristic solver.  Dashed lines indicate an exponential fit, with the red dashed line appearing near linear, due to the outlying low mean runtime at size 15.}
    \label{fig:classical-runtime-scaling}
\end{figure}

\begin{figure}[t]
    \centering
    \includegraphics[width=0.6\linewidth]{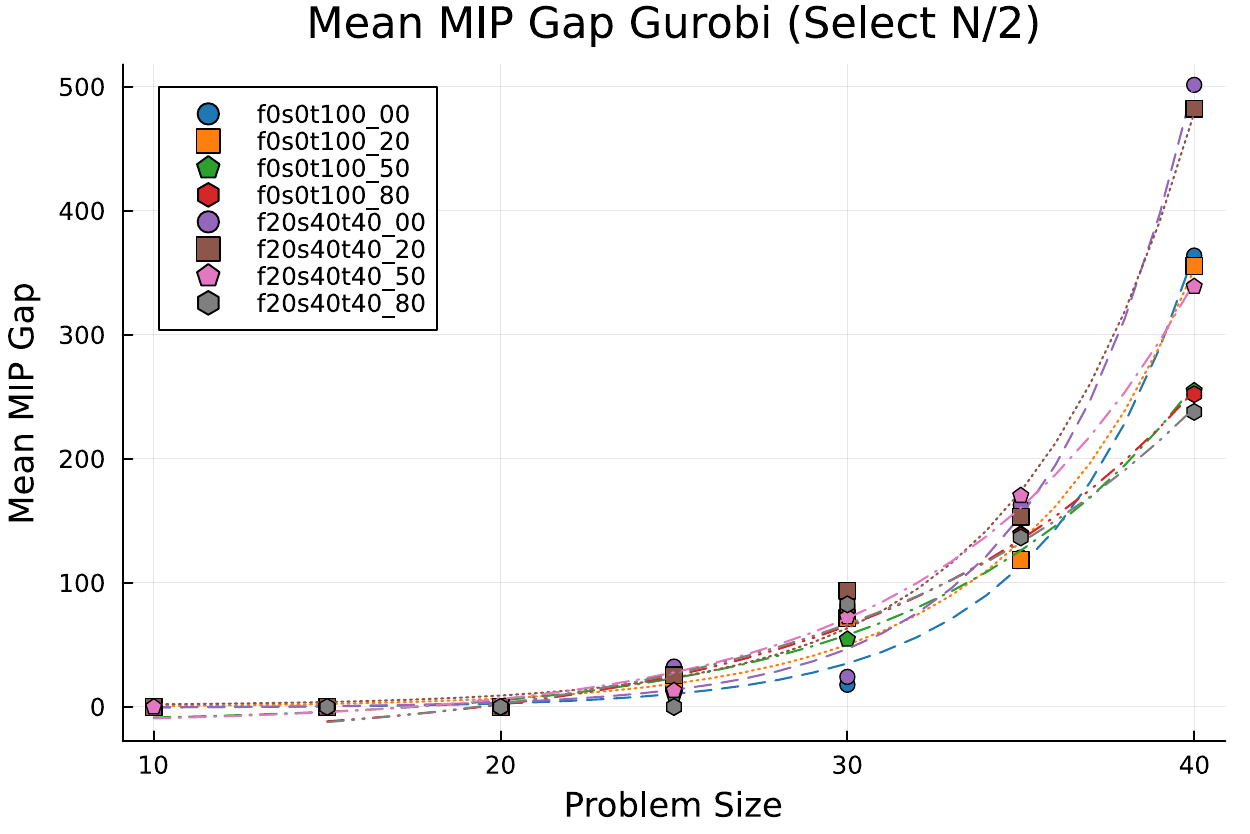}
    \caption{Mean MIP gap (frequently referred to as the optimality gap) reported by Gurobi at the time of termination for the various problem instances.  The exponential scaling of the MIP gap further shows the necessity of good heuristic solvers for large instances of these \texttt{entropy-cubo} PCBOs.}
    \label{fig:gurobi-mip-gap}
\end{figure}

\begin{figure}[t]
    \centering
    \begin{subfigure}[b]{0.49\textwidth}
        \centering
        \includegraphics[width=\textwidth]{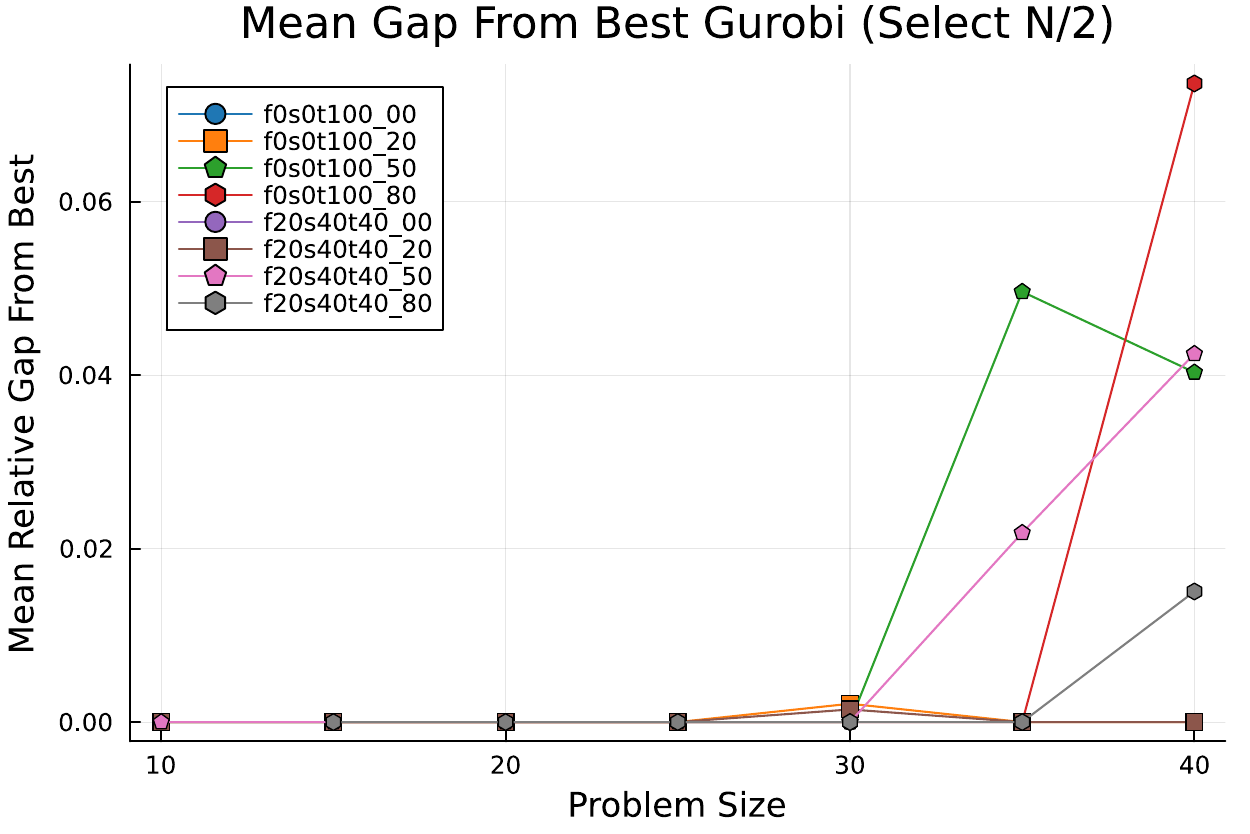}
        \caption{}
        \label{fig:gurobi-best-gap}
    \end{subfigure}
    \hfill
    \begin{subfigure}[b]{0.49\textwidth}
        \centering
        \includegraphics[width=\textwidth]{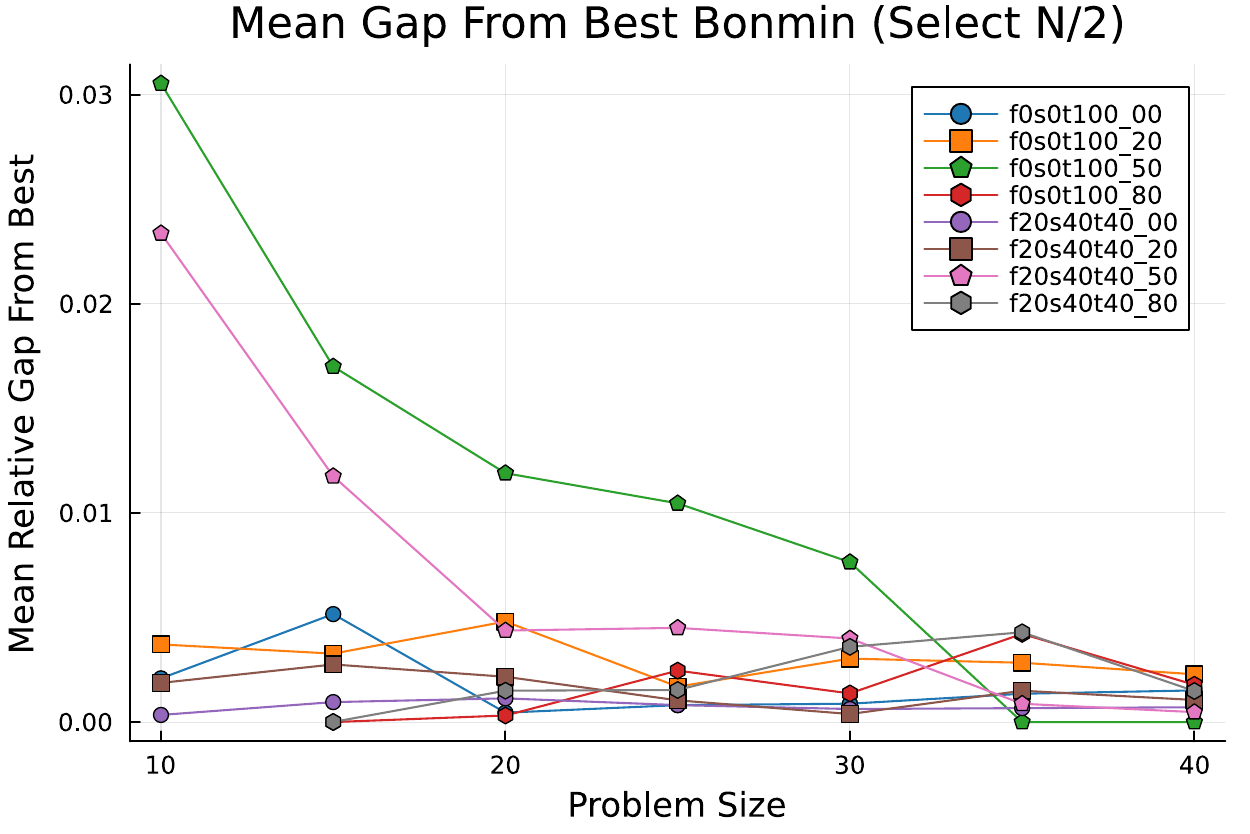}
        \caption{}
        \label{fig:bonmin-best-gap}
    \end{subfigure}
    \caption{We report the mean relative gap between the solutions found by (a) Gurobi and (b) Bonmin to reach the best found solution. This best solution is obtained by taking the minimum objective value between those provided by Gurobi and Bonmin.  For problems of size $N \leq 20$ with sparsification $ \leq 50$\%, and problems of size $N \leq 25$ with sparsification $= 80$ \%, this is equivalent to the optimality gap, since Gurobi was able to solve all of these problem instances to prove optimality for all of these instances. After this point, it no longer functions as guarantee of solution quality, but rather an indication of how well Gurobi and Bonmin perform relative to each other.  Gurobi tends to perform very well, but on certain problem instances with larger problem sizes, it struggles to find solutions of a similar quality to the Bonmin solutions.}
    \label{fig:gurobi-bonmin-best-gap}
\end{figure}

\begin{figure}[t]
    \centering
    \includegraphics[width=0.6\linewidth]{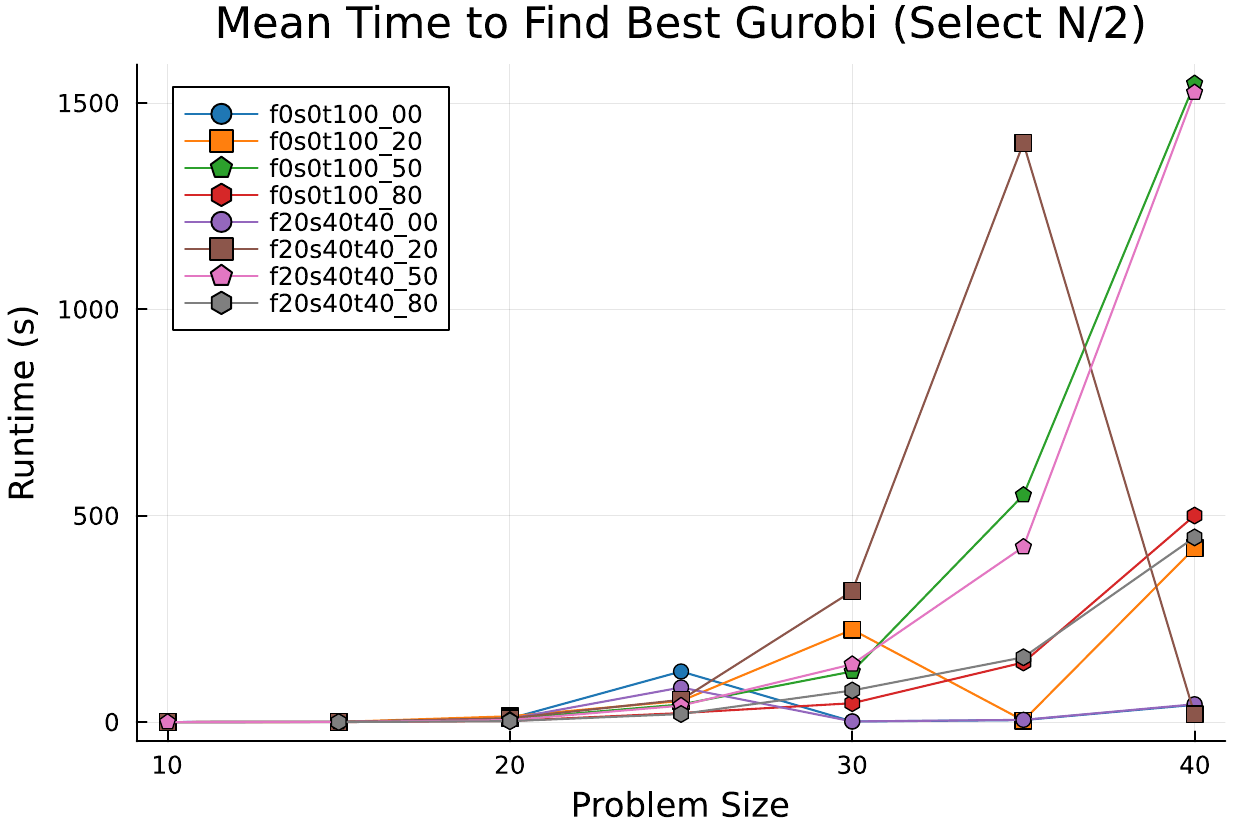}
    \caption{The mean time required for Gurobi to find the best solution found between Gurobi with a time-based solution quality improvement termination condition and Bonmin for a given problem instance.  In this case, Gurobi terminates immediately when a solution with the appropriate objective value is found.  This demonstrates the increasing difficulty to even find high quality solutions, independent of proofs of optimality, for Gurobi.  For large problem instances where the best solution is found quickly, it is necessary to recall that these solutions are not necessarily globally optimal.  It is likely that if Gurobi were given an intractable amount of time, it would find better solutions, but it is getting stuck quickly in local minima that are challenging to escape.}
    \label{fig:gurobi-time-to-best}
\end{figure}

Higher-order PCBO instances appear promising for feature selection, but today’s quantum devices are still shy of the scale and fidelity needed for a decisive advantage. In this section, we analyze the asymptotic scaling of state-of-the-art classical solvers, showing that two such solvers, Gurobi and Bonmin, face exponential difficulty beyond $100$ variables. Meanwhile, HRQAOA is capable of reducing the problem size at an asymptotically linear rate with polynomial computational overhead. If parameter concentration holds, this opens up a regime where a quantum and classical system working together may achieve an empirical advantage over a purely classical system. As quantum computing hardware advances, cross-layer optimizations at the compilation and error correction levels provide additional co-design levers to reach this regime. At the same time, classical computing capabilities continue to progress, potentially curtailing the advantage of a hybrid algorithm. The path to empirical advantage thus depends on the dynamic balance between advances in both classical and quantum computing.

\subsection{Benchmarking the Scalability of Classical Optimization Solvers} \label{subsec:classicalsolvers}
To evaluate the scalability of classical solvers, specifically mixed integer nonlinear programming (MINLP) solvers, on these problems, we consider the scaling of two solvers: Gurobi \cite{gurobi2024} and Bonmin \cite{bonami2008}. We chose these solvers after sweeping through all of the MINLP solvers provided on an academic license from AMPL (A Mathematical Programming Language)\cite{ampl1989}, a commercial mathematical programming language with support for many open source and commercial optimizers. The initial performance evaluation determined that, of the heuristic solvers, Bonmin consistently yielded surprisingly high-quality solutions very quickly. Of the exact solvers, Gurobi was the only solver that could consistently prove optimality for small problem instances (fewer than 30 variables), though it did struggle to prove optimality for larger problem instances (30 variables or more) in a reasonable amount of time.  This computational bottleneck is to be expected, as the problems in this study are NP-hard, so the resources required to prove optimality scale super-polynomially.  We first provide a brief overview of the classical solvers that we use:
\begin{itemize}
    \item{\bf{Gurobi}}: Gurobi is a commercial solver that is very robust and can provide provable globally optimal results.  As it is proprietary, it is impossible to state exactly how Gurobi works; however, it is known to be a branch and bound solver with cutting planes at its core.  It minimizes a primal problem, while maximizing a dual problem, providing an upper and lower bound for the globally optimal solution.  When the primal and dual solutions match, optimality is proven.  Gurobi can work with linear and quadratic models, but it is not inherently able to work with cubic models, so a dimensionality reduction is required.  For our dimensionality reduction, we apply methods similar to those shown in \cite{anthony2017quadratic}, introducing ancillary variables and quadratic constraints to reformulate the cubic model as a quadratic model.  To aid Gurobi's performance, we reformulate the quadratic constraints as the convex hull of the binary points of the constraint functions.  This both linearizes the constraint functions and allows Gurobi to use cutting planes more efficiently, yielding higher performance than the more naive implementation with quadratically defined constraints.  Gurobi provides its own modeling formulation, Gurobipy, which we use to invoke Gurobi with more fine-grained control relative to a separate mathematical modelling language.
    \item {\bf{Bonmin}}: Bonmin is a heuristic MINLP solver, so it does not prove optimality. Bonmin has six different algorithms built into it, but for the purpose of this work, we only consider the default algorithm (B-BB) as it yields high-quality solutions quickly, and has been shown to be the most robust solver \cite{kronqvist2019review} included in the package.  Bonmin works using an underlying branch and cut solver acting on continuous relaxations of the integer program \cite{bonami2008}. When applied to convex problems, the results from Bonmin are exact, but for non-convex problems, the results are not necessarily globally optimal.  Bonmin is able to directly work with cubic models so no dimensionality reduction is required, unlike Gurobi.  We simply formulate our model in Pyomo (a mathematical programming package in Python) and solve the problem using Bonmin.
\end{itemize}

Due to the high computational resources required to prove optimality for larger problems, we use a termination condition to keep the computational resources in check.  We define a termination condition such that Gurobi will return the incumbent (current best) solution if it has not improved in 5 minutes.  A higher allowed time would allow for larger problems to be proved, and a lower time limit would yield approximate solutions more quickly.  The time cutoff used here is somewhat arbitrary, but in general it provides enough time to prove optimality and terminate early for the smaller problem instances, while terminating in a reasonable amount of time for large problems with a relatively high-quality solution with a known bound on the optimality gap.  The results of this runtime scaling study for both Gurobi and Bonmin are shown in Figure \ref{fig:classical-runtime-scaling}.  Our analysis of Gurobi confirms asymptotically exponential scaling for these problems for when exact solutions are proved (i.e. for problems of size $N \leq 20$, or $N \leq 25$ with a sparsification of 80 \%).  Due to Bonmin always reaching its internal termination criteria quickly, its exponential scaling is much better behaved than Gurobi's and the trend can be fit for all problem sizes.

Similarly, the exponential increase in problem difficulty can be seen by the scaling of Gurobi's MIP Gap, a parameter that shows how close Gurobi has gotten to proving optimality.  We report this metric in Figure \ref{fig:gurobi-mip-gap}, and observe that the MIP Gap scales exponentially as problem complexity increases.  For problems with 30, 35, and 40 variables, Gurobi consistently reached the solution improvement time limit before proving optimality, so the MIP gap scales exponentially. 

Despite Bonmin's high speed, it doesn't guarantee optimal solutions.  To evaluate the comparative solution quality between Gurobi and Bonmin, we report the relative gap from the best found solution given by either algorithm in Figure \ref{fig:gurobi-bonmin-best-gap}.  This is given according to the formula
\begin{equation}
\Delta_{\text{best}} ({{\bf{x}}, {\bf{x^*}}}) = \frac{|{{\bf{x}}} - {\bf{x^*}}|}{|{\bf{x^*}}|}
\end{equation}
where ${\bf{x}}$ is the solution given by an algorithm, and ${\bf{x^*}}$ is the best solution found by either algorithm.  For instances where Gurobi could prove optimality, Gurobi inherently has a relative gap of $0$, while for problems where it cannot prove optimality it sometimes has a non-zero gap.  Therefore, for large problem sizes it is best to consider this metric as a useful tool for comparing Gurobi and Bonmin's relative performance, since it does not yield useful information about the overall solution quality beyond the MIP gap.

To provide a more detailed assessment of the time it takes Gurobi to find a high-quality solution, after running both Bonmin and Gurobi with the time-since-improvement termination condition described above, we also run Gurobi for an unbounded amount of time until a solution that matches the best found solution by the previous methods is achieved, with results observed in Figure \ref{fig:gurobi-time-to-best}. We see that for small problems, the time to find the best solution grows before falling off for larger problem instances.  This is likely caused by Gurobi getting trapped in a local minima with a better objective value than Bonmin could find quickly.  We observe that for some individual problem instances where Bonmin finds high-quality solutions quickly, it takes Gurobi several hours to find a solution of the same quality.  Therefore, higher quality solutions likely exist than those found by either algorithm, but Gurobi was just not able to find them in a reasonable amount of time, making the exponential scaling less apparent in this figure despite its presence.

The behavior we observe suggests that for sufficiently large problems the quantum portion of the hybrid feature selection approach may have a long enough window to reduce the problem size by a meaningful amount. The exact amount of time necessary will depend on the specifics of the quantum computing architecture under consideration.  It will also be necessary for the algorithm to yield near-optimal solutions consistently, otherwise it may be outperformed by other heuristic methods like Bonmin.
The primary runtime bottleneck will be the speed with which a fault-tolerant quantum computer, of sufficient scale, can execute the RQAOA edge fixing routine.

\subsection{Heuristic Scaling Analysis for Hybrid Approach}\label{subsection:qaoa_scaling}
Empirically, we have observed that optimally solving an $N$-variable PCBO problem scales exponentially in the problem size, $N$, using a state-of-the-art classical solver such as Gurobi. As an NP-hard problem, this scaling is expected (without resorting to further approximate classical heuristics). We have also seen that the RQAOA is a potential alternative in achieving (near) optimal solutions at sufficient depth, $p$. However, it does not necessarily offer a guaranteed optimal solution for some arbitrary problem size at finite $p$. Instead, we consider a hybrid approach: provide the classical solver a smaller problem instance to optimally solve by harnessing the problem size reduction from the RQAOA (where there is reason to believe that classical solver optimality remains with the reduced problem, evidenced by the ability of the RQAOA to find optimal solutions when used on its own). This is in the spirit of what the RQAOA is generally expected to do, breaking down the problem small enough to be, for example, easy to solve via brute-force. In particular, the significant appeal of this approach is the potential runtime savings to be attained in light of the observed exponential scaling of the classical solver and the much less stringent requirement on the QAOA subroutine to find an optimal solution itself. Notably, we find that if the cost incurred from using the RQAOA as a subroutine remains as polynomial/polylogarithmic for edge fixing, then there is potential for an exponential reduction in the optimal time-to-solution for the classical solver. Heuristically, we observe this by considering the runtime scaling of the hybrid approach vs. the purely classical approach. Suppose that the classical runtime is given by $T_{\mathrm{classical}}(N) \sim Ae^{bN}$ (for some $b>0$ that depends on the type of problem and the solver used). After using the RQAOA, the classical solver now runs on reduced size $N_c$ ($< N$), with (worst case) time $T_{\mathrm{classical}}(N_c)\sim A\,e^{b N_c}$. The total runtime of the hybrid approach is therefore
\begin{equation}
\label{eq:total}
T_{\mathrm{hybrid}}(N)
\;\sim\;
T_{\mathrm{RQAOA}}(N\!\to\!N_c)
\;+\;
A\,e^{b N_c}.
\end{equation}
If we then consider the ratio of the runtimes, the exponential term dominates, and the effective improvement over a direct classical run in the asymptotic limit of large $N$ is naturally:
\begin{equation}
\label{eq:ratio}
\frac{T_{\mathrm{hybrid}}(N)}{T_{\mathrm{classical}}(N)}
\;=\;
\frac{\mathrm{poly}(N) + A\,e^{b N_c}}{A\,e^{b N}}
\;\sim\;
e^{-b(N-N_c)}
\end{equation}
To substantiate the polynomial crux of this argument, we consider the runtime scaling of a single RQAOA iteration, $t_q(N)$, where
\begin{equation}\label{eq:rqaoa_time_sum}
T_{\mathrm{RQAOA}}(N\!\to\!N_c) = \sum_{i=0}^{N-N_c-1} t_{q}(N-i).    
\end{equation}
Similarly to \cite{weidenfeller2022scaling}, we model the runtime of RQAOA as $t_q = n_\mathrm{iters}\times(n_{\mathrm{shots}}\times t_{\mathrm{shot}}) + t_\mathrm{opt}$. With HRQAOA, we do not have to worry about iterative evaluations of the QAOA circuit for a given $N$, since we leverage parameter transfer and absorb any associated classical optimization cost into a roughly constant baseline $t_\mathrm{opt}$ for simplicity.
To answer the scaling of $n_{\mathrm{shots}}$, recall that with HRQAOA, we are not strictly relying on an accurate estimate of the true energy, $\langle H\rangle$, but rather on estimating and ranking correlation expectation values, $z$, from the set $\mathcal{P}$:
\begin{equation}
\mathcal{P} \subseteq \{\langle Z_i\rangle\}\cup\{\langle Z_iZ_j\rangle\}\cup\{\langle Z_iZ_jZ_k\rangle\},    
\end{equation}
and trying to determine the strongest absolute correlation magnitude. To estimate a sampled expectation value, $\hat{z} \in \mathcal{P}$, we calculate the average parity over the $n_{\mathrm{shots}}$ independently sampled bitstrings of the prepared QAOA state. As all observables commute, each bitstring gives a simultaneous sample for all the observables of interest in $\mathcal{P}$. The question is: how many shots, $n_{\mathrm{shots}}$, are needed to ensure that all observables are estimated within a tolerance $\epsilon$ simultaneously? In practice, we often only need to rank the largest expectation values, so this requirement is conservative. To do so, we can get an estimate for an observable through the Hoeffding inequality, 
\begin{equation}
\mathrm{Pr}(\lvert \hat{z} - z \rvert \geq \epsilon) \leq 2e^{-2 \epsilon^2 n_{\mathrm{shots}}}.
\end{equation}
Given that the Hamiltonian $H$ we consider can contain up to three-body terms, we may have $\mathcal{O}(N^3)$ observables (Pauli strings) to rank. Taking the union bound across all observables, the probability that any observable's error exceeds $\epsilon$ is then at most $2N^3 e^{-2 \epsilon^2 n_{\mathrm{shots}}}$. If we require that this should be less than some desired confidence failure rate, $\delta$, then we can get an estimate for $n_{\mathrm{shots}}$ as:
\begin{equation}
n_{\mathrm{shots}} \geq \frac{1}{2\epsilon^2}\ln(2N^{3}\delta^{-1}).
\end{equation}
The remaining quantity to estimate is the time it takes to get a shot of the sampled bitstring. This mainly involves the time to prepare the optimized QAOA state, which we take to be dominated by the preparation of the $e^{-i\gamma H}$ cost layer. With a dense 3-body Hamiltonian and pessimistic serial gates, the number of operations is at most proportional to $\mathcal{O}(N^3)$. Taking for simplicity an average dominating gate time, $t_g$, and scaling for up-to $p$-layers, we estimate that in leading order,
\begin{equation}
t_{\mathrm{shot}} \sim pN^3t_g + t_{p}
\end{equation}
where any additional time factors are absorbed into $t_{p}$ (such as state initialization time for example). Thus, we can get a reasonable heuristic estimate for $t_q$ as:
\begin{equation}
\label{eq:one-fix}
t_{q}(N) \;\sim\; 
\underbrace{\frac{\ln(2N^3/\delta)}{2\epsilon^2}}_{\text{shots}}
\;\times\;
\underbrace{pN^3t_g + t_{p}}_{\text{time per shot}}
\;+\; t_{\mathrm{opt}},
\end{equation}
Finally, taking the sum from Eq.~\ref{eq:rqaoa_time_sum} for the cumulative time needed for all $(N-N_c)$ rounds used for the RQAOA, and dropping some constant and logarithmic factors, we find our estimate to be on the order of
\begin{equation}
T_{\mathrm{RQAOA}}(N\!\to\!N_c)
\;\sim\;
\mathcal{\tilde{O}}\!\left(
\frac{p\,N^4\,t_{g} }{\epsilon^2}
\right),
\end{equation}
where, if we assume that the QAOA depth $p \sim N$ is sufficient to attain good performance for the RQAOA, our estimate for $T_{\mathrm{RQAOA}}$ remains in $\mathrm{poly}(N)$. This suggests the potential for an empirical quantum advantage via a hybrid quantum-classical approach.
The exact crossover point where the hybrid approach may outperform a purely classical approach is primarily bottlenecked by the speed with which a fault-tolerant quantum computer can perform the edge-fixing subroutine within the RQAOA.
We turn to this question in more detail in the following section, highlighting the main components that make the largest contributions to the runtime overhead.

\subsection{Future Hardware Resources and Opportunities for Cross-Layer Optimizations}
Determining the problem sizes at which empirical advantage may be possible depends on the capabilities of both classical and quantum computing. Looking to our classical resource analysis in Sec.~\ref{subsec:classicalsolvers}, it is likely that the window for empirical advantage will open up for dense, third-order problems of at least 100 features. At the circuit depths that HRQAOA would require for such a problem, each feature would likely have to be encoded into an error correcting code. While the capability for operating $>100$ logical qubits is beyond current devices, there is an industry-wide effort to achieve this scale, with roadmaps outlining estimates as early as 2028 (Infleqtion) \cite{infleqtion2025_qc_roadmap} and 2029 (IBM) \cite{ibm2025_dev_innov_roadmap}.

As hardware progresses, more methods become feasible for meaningful cross-layer optimization, enabling better quantum runtimes.
As seen in the previous section, the primary rate limiting step in the hybrid approach to feature selection is the runtime of the RQAOA. 
Much of this bottleneck is already handled by parameter transfer: as discussed in Sec.~\ref{sec:param-transfer}, the need to execute the variational loop directly on quantum hardware can be effectively removed, reducing resource overheads by orders of magnitude. 
While parameter concentration has been observed empirically for the PCBO problems considered in this work -- we demonstrated parameter transfers spanning tens of qubits between the donor and receiver problems, reducing resource requirements while often discovering higher-quality solutions. Future work remains to analytically show concentration as the problem size scales \cite{akshay2021parameter}, and this is a topic of ongoing investigation.
It is possible that the direct subsampling leveraged by the current approach constructs good donor problems that retain enough structural problem information to enable effective parameter transfer.

Our objective is then to optimize the runtime of a single, fixed QAOA circuit instance, assuming the parameters have been set via efficient parameter transfer.
 Even under fault-tolerance -- where clock rates are still slow \cite{babbush2021focus} -- effective parameter transfer can be a significant first step towards optimizing this runtime bottleneck.

\paragraph{Trading space for time}One way to minimize the runtime, or equivalently the depth of the quantum circuit, is to trade more space for less time.
When circuit depth, not qubits (i.e., circuit width), is the binding constraint, mid-circuit measurement and teleportation enable a depth-for-width tradeoff that the compiler may choose to exploit. This enables us to recompile the cubic-depth, linear-width QAOA circuits to an equivalent linear-depth, cubic-width representation.
This tradeoff is well-suited to qubit modalities with the ability to rapidly scale the number of qubits, such as neutral-atom quantum computers. 
Recently, Infleqtion has demonstrated the use of this compiler technique to execute constant-depth CNOT ladders on logical qubits \cite{rines2025demonstration}.

\begin{figure}[h]
    \centering
    \includegraphics[width=.95\linewidth]{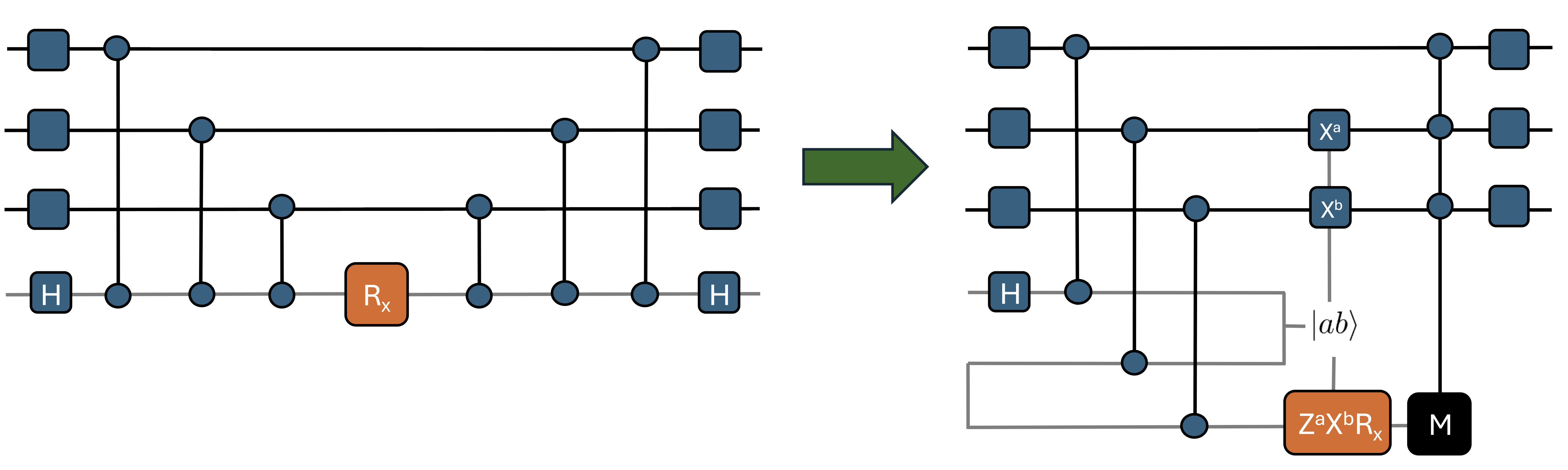}
    \caption{Weight 3 constant-depth Pauli-phasor example. The naive implementation of the Pauli-phasor (left) can instead be implemented in constant depth (right) through the introduction of ancilla qubits (gray) and mid-circuit measurement.}
    \label{fig:cdpp}
\end{figure}

This same compilation technique is relevant to the circuit implementation of the Pauli-phasors (i.e., unitary evolution by a real angle under the $Z$, $ZZ$, and $ZZZ$ Pauli strings in the PCBO Hamiltonian, see Eq.~\ref{eq:pcbo-hamiltonian}) that dominate the QAOA ansatz corresponding to the PCBO feature selection problems.
The relative time savings afforded by this method depend on the weight of the Pauli strings. In this work, we consider Pauli strings up to weight three, but our method may be generally extended to higher-order terms as well.
For small-weight Pauli-phasors ($\leq3$), compiling the CNOT ladders in constant depth does not reduce the overall depth of each Pauli-phasor, due to a constant overhead cost incurred by the recompilation. However, ansatzes with small-weight Pauli-phasors still benefit from this method as it compiles Pauli-phasors with arbitrary grid connectivity to constant depth without requiring costly SWAP networks or mid-circuit atom motion (Figure \ref{fig:cdpp}). Large-weight Pauli-phasors ($\geq3$) will enjoy the additional benefit of having their depth reduced in addition to being compiled to grid connectivity. 
Additional depth (and entangling gate) savings are accomplished by compiling CNOT ladders to constant depth when a large percentage of the Pauli-phasors are Clifford.
In fault-tolerant quantum circuits, the gate overhead incurred by fault-tolerant decompositions of small-angle single-qubit rotations can be quite substantial. However, different methods, such as partial fault-tolerance \cite{akahoshi2024partially} or direct synthesis of small-angle rotations \cite{choi2023fault}, are under development to help address this challenge that aims to enact single-qubit rotation gates without requiring prohibitively expensive Solovay-Kitaev decompositions \cite{kitaev1997quantum, nielsen2010quantum} at the expense of increasing the logical error rate and thereby reducing the overall shot rate. 

\paragraph{Choosing an error-correcting code}While the surface code remains a plausible encoder for a fault-tolerant implementation of the underlying QAOA \cite{omanakuttan2025threshold} required for our hybrid algorithm, it does not exhaust the potential considerations associated with the additional quantum error correction (QEC) overhead. qLDPC codes \cite{leverrier2022quantum, bravyi2024high} represent an alternative path for asymptotic scaling advantage; however, various practical questions \cite{baspin2022quantifying} remain for these codes, such as their significant amount of nonlocality and comparatively time-inefficient gates \cite{yoder2025tour}. Recently, concatenated codes have also emerged as another promising realization for fault-tolerant quantum computation \cite{yoshida2025concatenate, litinski2025blocklet, goto2024high, yamasaki2024time}. In particular, implementations such as the concatenated quantum Hamming codes of Ref.~\cite{yamasaki2024time} yield a quasi-polylogarithmic time overhead with an $\mathcal{O}(1)$ space cost to fault-tolerantly simulate a $\mathcal{O}(\mathrm{poly}(N))$-sized quantum circuit. The use of transversal gates with techniques such as Knill-style error correction and gate teleportation for non-Cliffords, whilst possessing an efficient decoder, can allow the QEC overhead for such concatenated codes to stay minimal for retention of the advantage conjectured in Sec.~\ref{subsection:qaoa_scaling}. While proposed at an architectural level, experimental realizations are needed for practical validation (since an advantage that concatenated codes lack, when compared to, say, the surface code, is that of relatively high thresholds). However, their higher encoding rates make them potentially quite appealing for more near-term hardware realizations of RQAOA on advantageous problem sizes.

\subsection{Dequantization}

There exist multiple examples of new quantum algorithms being \emph{dequantized} whereby new classical algorithms are developed which match the performance and scalability of the original quantum algorithm, erasing the potential for any quantum advantage \cite{tang2021quantum}. Often this is achieved by exploiting special structure within the problem or the quantum algorithm used to solve it. For example, the quantum algorithm may be efficiently simulatable if it exhibits a low degree of entanglement and magic.

While the PCBO problems we consider for feature selection do resemble the Sherrington-Kirkpatrick model, where known polynomial time classical algorithms exist \cite{montanari2025optimization}, the extension to higher-order hyperedges increases their complexity to be NP-Hard. 
Therefore, it is likely that neither classical nor quantum algorithms can exactly solve these problems in polynomial time. It is also known that properties such as the overlap gap property preclude constant-depth QAOA from finding even approximately good solutions \cite{chen2019suboptimality, gamarnik2021overlap}. However, this is exactly the issue that the RQAOA addresses, breaking the constant-depth barrier by effectively increasing the circuit depth at each iteration \cite{bravyi2020obstacles}. 
At each step of the RQAOA, the decision of which edge to fix is made after measuring the expectation value of a low-weight Pauli operator. However, for sufficiently high depth circuits that generate highly entangled states, efficiently estimating the expectation value may require access to an exponential number of basis states. Furthermore, the PCBO problems we consider are extremely dense, therefore their quantum circuit representations are highly connected and so even low-depth QAOA ansatzes would ``see the whole graph'' \cite{farhi2020typical, farhi2020worst}.
These observations suggest that directly simulating the quantum portion of the hybrid feature selection algorithm on a quantum computer is unlikely to scale. It is still possible that an alternative, completely classical approach to edge fixing may outperform the RQAOA approach, and this is an intriguing subject of ongoing investigation. 

\section{Biological Insights and Further Applications} \label{sec:applications}
The biological and clinical relevance of our higher-order combinatorial approach to feature selection lies in its ability to surface synergistic relationships within and across modalities that conventional methods may miss. In this section, we consider how applying this approach to multimodal cancer datasets (Table~\ref{tab:final-p2-datasets}) translates into clinically useful artifacts, advancing precision medicine~\cite{boehm2022harnessing}. Fully integrated computational stacks spanning genomics, transcriptomics, proteomics, and metabolomics, together with clinical and imaging data, remain costly and logistically challenging to assemble. Our team addressed this challenge head-on, leveraging the University of Chicago's institutional resources and the support of Wellcome Leap's Q4Bio program to collate harmonized, patient-level multimodal treatment-response datasets. Although sample sizes remain limited by cost, these resources provide a meaningful testbed to evaluate the clinical utility of our methods and serve as a stepping stone for further precision oncology research in the future.

The hybrid feature selection algorithm is designed to simultaneously address the challenges of data scarcity, capture higher-order relationships among features, and remain agnostic to both modality and disease. In rare diseases, where inherently small cohorts preclude the large-sample regimes assumed by many modern machine learning approaches, our method avoids over-parameterized models by targeting compact, information-rich subsets of features that train simpler models with better generalization on held-out test datasets. At the same time, explicitly modeling higher-order relationships and detecting cross-modality synergy is particularly valuable in oncology, where integrating DNA mutations, RNA expression, and pathology images remains an active but difficult goal. To handle the combinatorial complexity of this feature selection problem, we employ a hybrid computational strategy in which a quantum computer first reduces the size of the input problem before handing it off to a classical solver. Finally, by being modality- and disease-agnostic, the algorithm surfaces context-dependent interactions, such as gene expression conditioned on mutations, immune microenvironment, and histopathological features, providing a more principled route to identifying interactions, pathways, and features implicated in tumorigenesis while generating testable hypotheses for personalized interventions.

In a harmonized UChicago cohort, we are assembling data from 257 patients with matched treatment outcomes, digital pathology, DNA whole-exome sequencing (WES), and RNA-seq. The dataset captures combination therapies (immunotherapy, chemotherapy, and targeted therapy) with rigorous response annotations. Within this setting, our hybrid approach can be applied to identify multimodal features most predictive of differential expression patterns and response versus non-response across treatment modalities. We focus on this future application in more detail in Sec.~\ref{sec:treatment-response}.
Narrowing the training set of features to a small set of interpretable biomarkers improves clinical interpretability and reduces screening costs, which is essential for broadening access to precision oncology beyond major academic centers into community practices. For example, in drug discovery and clinical-trial biomarker design, teams could ingest efficient multimodal patient profiles (genomic, transcriptomic, pathomic) to produce deployable, generalizable biomarkers that rank candidate treatment combinations with estimated response probabilities, supporting trial stratification and enrichment.

As the capabilities of quantum hardware continue to advance, we anticipate running larger experiments to test for the presence of EQA by evaluating whether a quantum computer is able to significantly improve the performance of state-of-the-art classical solvers. Additionally, the hybrid feature selection algorithm may be scaled up to tackle larger multimodal datasets and account for even higher-order interactions, better matching the complexity of contemporary oncology datasets and, more generally, modern medical data.

\subsection{Applications to treatment response in oncology}\label{sec:treatment-response}
\paragraph{Why treatment response.}
The most impactful near-term application of our hybrid feature-selection pipeline in clinical oncology is to generate a biomarker to predict \emph{treatment response}. Our team has published numerous studies developing prognostic signatures from multiple data modes~\cite{huang2021hpvneg_hnscc_proteogenomic,izumchenko2015notch1,broner2021dclk1,peri2017nsd1_nsd2,ferrarotto2022al101,mkrtchyan2022dna_repair_multiomics,pun2023ai_dual_purpose_targets,steinobrien2018timecourse_omics,hayashi2020gulp1_nrf2}, with a focus on treatment response outcomes and clinical expertise in head/neck squamous cell carcinoma (HNSCC). For example, we found that loss of \textit{SMAD4} expression in HNSCC is a determinant of cetuximab -- an FDA-approved endothelial growth factor receptor antibody -- resistant phenotype. Furthermore, analysis of multi-modal data including RNA-seq and DNA methylation revealed that genes associated with cetuximab resistance include \textit{FGFR1}, a known driver of resistance to epidermal growth factor receptor (EGFR) inhibitors~\cite{steinobrien2018timecourse_omics}. We also demonstrated that KEAP1-NRF2 signaling conferred cisplatin resistance in urothelial cancer~\cite{hayashi2020gulp1_nrf2}, providing strong rationale for an equivalent role of the NRF2 axis in HNSCC~\cite{osman2023nrf2_cisplatin_hnscc}. Furthermore, even with the complexity of signaling networks regulating continuous cross talk between tumor, stromal and immune cells within the tumor microenvironment (TME), genomic patterns associated with clinical response/resistance to immune checkpoint inhibitors (ICI) have been reported in HNSCC~\cite{zhu2022immune_infiltration_hnscc,chen2022immune_ssgsea_oscc}.

In Phase~3 our application goal is to develop biomarkers of systemic treatment response and resistance in HNSCC. To this end we are curating a unique harmonized dataset designed explicitly for this task, integrating high-quality transcriptomics with detailed treatment regimens and outcome annotations. Our primary target cohort is an IRB-approved UChicago head-and-neck cancer cohort (\textit{n} = 257) treated with platinum doublet chemotherapy, EGFR-targeted therapy, PD-1 immunotherapy, or combinations thereof, with RECIST~1.1 response recorded 8--12 weeks post-treatment. This resource provides a clinically grounded testbed that goes beyond earlier tissue of origin experiments and enables evaluation of the hybrid algorithm on a more challenging outcome of direct clinical relevance.

\paragraph{From technical results (tissue of origin) to clinical outlook (response).}
The Phase~2 work established the end-to-end pipeline and demonstrated feasibility on cancer tissue of origin, setting a foundation for Phase~3’s treatment-response focus and hardware evaluations. In this outlook paper, we emphasize forward-looking clinical applications, especially response prediction, while reserving detailed tissue of origin analyses and their validations for the technical manuscript.

\paragraph{Decision support and cost.}
In the near term, a realistic impact is in R\&D: multi-omic biomarkers to stratify likely responders vs.\ non-responders, helping enrich cohorts and manage trial size and cost. Because the selector yields compact, interpretable panels, candidate signatures can be implemented as targeted assays (e.g., focused DNA panels, small RNA sets, or IHC surrogates) rather than routine WES + RNA-seq.

\paragraph{Linking selected features back to pathways and mechanisms.}
Having narrowed down the feature set to the most relevant ones, the selected features can be mapped back to specific biological pathways and mechanisms, enabling deeper exploration of relationships between these features and treatment response. Further, whereas previous cancer research largely focuses on data of a single mode, our work collecting and studying data across all the omics layers could inspire novel connections between these multimodal features and their joint interactions with treatment response.

\subsection{Applications Beyond Oncology}

While our hybrid quantum-classical algorithm was initially developed and validated using cancer datasets, its application extends beyond oncology. In general, the algorithm is well-suited for feature selection problems characterized by a limited number of samples but an abundance of potential features -- a scenario frequently encountered in biomedical research.
A key strength of our approach is its inherent flexibility and modularity that allow it to easily incorporate additional data modalities as they emerge. The PCBO framework is a filter method that does not make assumptions about the specific nature of the features being analyzed. This data-agnostic quality makes our approach particularly valuable in an era of rapidly evolving biomedical technologies that continuously generate new types of high-dimensional data.

For instance, as single-cell technologies, spatial transcriptomics, and advanced imaging modalities become more prevalent, our method can readily integrate these new data streams without fundamental restructuring. When researchers develop novel biomarker assays or data collection methodologies, they can be incorporated into existing analytical frameworks by simply defining appropriate preprocessing steps for the new modality while preserving the core PCBO architecture. This allows research teams to incrementally enhance their multimodal datasets without discarding previous investments or requiring complete analytical redesigns.

To illustrate the modularity of our approach, we first provide a brief overview of its possible applications across a diverse array of research settings:

\begin{enumerate}
    \item \textbf{Clinical Trial Networks:} Multicenter trials often struggle with heterogeneous data; our approach surfaces cross-modal biomarkers that are stable across sites, improving stratification and outcome prediction. Given the high cost and failure rates of trials, biomarker-stratified screening and companion-diagnostic development are active priorities; compact, deployable signatures from our selector can enrich for likely responders, streamline screening, and help reduce cost while improving study power and probability of success.
    \item \textbf{Pharmaceutical R\&D:} Drug discovery spans high-throughput screens, structural and phenotypic assays, and clinical samples. Our selector integrates these modalities to identify assay combinations that best predict efficacy, safety, and surface dysregulated pathways that delineate molecular subtypes. This can focus experiments, reduce costly tests, and guide targeted therapies rather than treating heterogeneous cohorts with a single agent.
    \item \textbf{Global Health Monitoring:} Public health surveillance systems collect heterogeneous data from clinical reports, environmental monitoring, population mobility tracking, and genomic surveillance. Our approach could identify which cross-modal feature combinations best predict disease outbreaks, enabling more efficient resource allocation with limited data in emerging epidemic situations.
    \item \textbf{Agricultural Research:} Crop improvement programs collect phenotypic, genetic, microbiomic, and environmental data. Our algorithm could help identify which combinations of features across these domains best predict crop yield or disease resistance, accelerating breeding programs with limited experimental plots.
    \item \textbf{Environmental Conservation:} Biodiversity monitoring generates data through field observations, remote sensing, genetic sampling, and acoustic recordings. Our approach could identify the most informative combinations of these data types to efficiently track endangered species or ecosystem health with minimal sampling.
\end{enumerate}

Considering more disease-related applications, we find that the fundamental challenge of identifying optimal feature subsets from high-dimensional, low-sample datasets appears consistently across numerous clinical research areas. Based on our assessment of publicly available multimodal datasets in the European Genome-Phenome Archive (EGA), we have identified several promising domains where our hybrid quantum-classical approach could offer significant advantages:

\begin{enumerate}
    \item \textbf{Neurodegenerative Disorders:} Conditions such as Parkinson's disease~\cite{ega-parkinsons}, progressive multifocal leukoencephalopathy~\cite{ega-pml}, multiple sclerosis~\cite{ega-ms}, and amyotrophic lateral sclerosis (ALS)~\cite{ega-als} often involve complex, multifactorial pathogenesis with heterogeneous presentation. These conditions typically have relatively limited cohort sizes compared to the vast array of molecular, imaging, and clinical variables collected. For example, the Parkinson's disease dataset in the EGA contains rich multimodal data where feature selection could potentially reveal novel biomarker combinations and improve diagnostic accuracy~\cite{ega-parkinsons}. Understanding the interplay between genetic risk factors, protein aggregation patterns, neuroimaging findings, and clinical manifestations requires precisely the kind of cross-modal feature selection our algorithm excels at.
    \item \textbf{Rare Diseases:} Conditions like inclusion body myositis, an inflammatory muscle disease, present particular challenges for conventional machine learning approaches due to their rarity and consequent small sample sizes~\cite{ega-myositis}. Our hybrid algorithm's capacity to efficiently navigate expansive feature spaces with limited samples makes it especially valuable for identifying meaningful patterns in such datasets. For rare diseases, where large cohorts are simply not possible to assemble, the ability to extract maximum information from limited multimodal data is particularly valuable for developing diagnostic tools and therapeutic approaches.
    \item \textbf{Emerging Clinical Challenges:} Novel clinical entities such as Long COVID represent scenarios where researchers are collecting extensive multimodal data (genomic, transcriptomic, proteomic, clinical, and imaging) from relatively small patient cohorts~\cite{ega-covid}. These emerging conditions create an urgent need for advanced computational methods that can effectively extract insights from high-dimensional, limited-sample datasets. Our approach can help identify which combinations of features across modalities most effectively distinguish different disease subtypes and predict response to interventions, potentially accelerating clinical progress.
\end{enumerate}

Our hybrid quantum-classical approach offers several specific advantages in these clinical contexts. The RQAOA-based edge-fixing can simultaneously consider complex interactions between features while systematically reducing problem dimensionality. This is particularly valuable in clinical datasets where critical biomarkers may only be detectable through the interaction of multiple features.
As demonstrated in our scaling analysis, classical approaches quickly become computationally prohibitive when exploring feature combinations in high-dimensional spaces. For diseases with complex molecular mechanisms, the ability to efficiently evaluate combinatorial feature sets could reveal previously undetectable patterns.

The binary optimization formulation of our approach naturally accommodates the integration of heterogeneous data types, allowing researchers to identify optimal feature combinations across genomic, transcriptomic, imaging, and clinical data simultaneously -- a capability particularly relevant for complex, multifactorial conditions. This integration is essential for precision medicine approaches that aim to characterize a patient's unique profile holistically rather than through isolated biomarkers.

The potential applications extend beyond the clinical examples listed above to any domain facing data scarcity challenges -- where samples are limited but features are abundant. This includes environmental monitoring, materials science, and various other fields where data collection is expensive or limited, but the potential feature space is vast.
As quantum computing hardware continues to advance, we anticipate that our hybrid approach will enable more comprehensive and insightful analysis of these challenging datasets, potentially accelerating discovery in fields currently constrained by computational limitations of classical feature selection methods.

\section{Conclusion} \label{sec:conclusion}
In this outlook paper, we present a case study on a hybrid quantum-classical pipeline for cancer biomarker discovery, illustrating the importance of co-design in the development of useful hybrid algorithms. Even before a definitive advantage is shown, framing biomarker discovery as high-order PCBOs plus hybrid reduction has already yielded biological and methodological insights (e.g., emphasis on third-order information, pipeline integration, and rigor around train/test evaluation). This is emblematic of our \emph{co-design} philosophy -- starting from the application and pushing algorithms, compilers, and hardware together toward clinically meaningful endpoints.

To construct and implement our pipeline on current devices, co-design is necessary at every step. In devising the algorithm, we choose a component likely to benefit from a quantum subroutine, deciding to implement the feature selection component using a quantum combinatorial optimization algorithm. To construct a problem that was implementable on a quantum device, we phrase the feature selection problem as a PCBO, enabling us to apply the RQAOA algorithm. 

We further tailor the problem construction to make the algorithm feasible to run on current hardware. Addressing shot budget constraints, we leverage parameter transfer to reduce the computational cost of the variational loop within QAOA, leading to order-of-magnitude reductions in computational cost between RQAOA and Hyper-RQAOA. To reduce circuit depths, we investigate problem sparsification methods, with the aim of truncating the number of terms in the problem Hamiltonian while retaining enough third-order structure to keep edge-fixing informative and the instance nontrivial for classical solvers. In preliminary test runs of these sparsified circuits, we also employ error mitigation to substantially reduce the effect of imperfect gates and readout on our results. 

Our resource analysis investigates pathways and highlights obstacles to empirical quantum advantage for our hybrid biomarker discovery pipeline. Benchmarking state-of-the-art classical solvers on our PCBO problems shows exponential growth in runtime with problem size, whereas HRQAOA reduces problem size with only polynomial overhead, creating a potential crossover window. Looking forward, the main objective is to minimize the wall-clock time of building the correlation dictionary -- i.e., estimating the multi-qubit expectation values that drive edge-fixing -- now that parameter transfer largely removes the variational-loop bottleneck. This will require cross-layer optimization of fault-tolerant protocols and compilation: compiling Pauli-phasors to (effectively) constant depth via ancilla, mid-circuit measurement, and teleportation to trade depth for width; improving logical shot rate by optimizing rotation synthesis and T-state throughput (or selectively using partial fault tolerance/direct small-angle synthesis where acceptable); and scheduling circuits to grid connectivity to avoid SWAP overheads. By tuning rotation-approximation error to the sampling error budget and co-designing these layers around the Estimator-style expectation-value workflow, we aim to cut per-round dictionary time and shift the EQA window closer. At the same time, computational advances are continuously being made: improvements in classical algorithms impact the location of this window for empirical advantage and present risks for dequantization of this hybrid algorithm.

 Beyond computational advances, our study clarifies how hybrid, higher-order feature selection translates into biologically meaningful artifacts. First, discretization strategies -- notably quintile binning for mRNA -- preserve predictive signal while stabilizing information-theoretic estimates, providing a simple, domain-agnostic pre-processing step that can stand alone or feed hybrid selectors. Second, by explicitly scoring higher-order and cross-modal interactions, the pipeline identifies compact, non-redundant panels that support simpler, more interpretable models and improved generalization in data-scarce regimes. This reduces over-parameterization, surfaces candidate interactions spanning mutations, expression, and histopathology. Finally, the resulting panels and hypotheses create a practical path toward response-focused biomarkers in head-and-neck cancer, aligning with our curated, harmonized UChicago cohorts: lean signatures lower assay cost and facilitate deployment beyond major centers, while mechanistic links support prospective validation and clinical translation.

This work was catalyzed by Wellcome Leap’s Quantum for Bio (Q4Bio) program, which brought together quantum computing, genetics, and oncology teams around a single, clinically grounded application, focused our attention on biomarker discovery, and provided the impetus for the co-design practices developed throughout the paper. Ultimately, empirical quantum advantage is a goal within a changing computational landscape, shifting as the fields of quantum and classical computation continue to advance. Regardless, there is scientific and practical value in pushing the boundaries of what is possible with current computational tools and working closely with domain experts to solve immediate computational problems. Although the pipeline outlined in this case study may evolve as computational fields progress, our work highlights the importance of co-design in translating hardware capabilities into useful applications and offers a template for continued efforts in hybrid algorithm design.

\addcontentsline{toc}{section}{Acknowledgements}
\section*{Acknowledgments}
We thank Ali Javadi-Abhari, Nate Earnest-Noble, and Kevin Sung for their insightful discussions and assistance with executing circuits on IBM quantum hardware. We also thank Josephine Lee, Kyla Gonzalez, Asha Thomas, and Stephanie Gavilanes for developing a suite of magic state distillation circuits that advanced our study of running Q4Bio circuits on large-scale quantum computers. Finally, we thank Kerry Zhou for his support with Q4Bio simulations.
This work is supported in part by Wellcome Leap as part of the ‘Quantum Biomarker Algorithms for Multimodal Cancer Data’ research project within the Quantum for Bio (Q4Bio) Program, and in part by the IBM-UChicago Quantum Collaboration, under agreement number MAS000364, with access to the fleet of IBM Quantum computers.
This research used resources of the National Energy Research Scientific Computing Center, a DOE Office of Science User Facility supported by the Office of Science of the U.S. Department of Energy under Contract No. DE-AC02-05CH11231 using NERSC award NERSC DDR-ERCAP0032212 and DDR-ERCAP0030280.

\newpage

\addcontentsline{toc}{section}{References}
\bibliographystyle{unsrt}
\bibliography{refs}

\begin{thebibliography}{100}

\bibitem{dakal2024emerging}
Tikam~Chand Dakal, Ramgopal Dhakar, Abhijit Beura, Kareena Moar, Pawan~Kumar Maurya, Narendra~Kumar Sharma, Vipin Ranga, and Abhishek Kumar.
\newblock Emerging methods and techniques for cancer biomarker discovery.
\newblock {\em Pathology - Research and Practice}, 262:155567, October 2024.
\newblock Epub 2024-08-29; Review; PMID: 39232287.

\bibitem{dacosta_byfield_2025_biomarker}
Stacey DaCosta~Byfield, Bela Bapat, Laura Becker, Carolina Reyes, Ismini Chatzitheofilou, Brock~E. Schroeder, Damon Hostin, and John Fox.
\newblock Biomarker testing approaches, treatment selection, and cost of care among adults with advanced cancer.
\newblock {\em JAMA Network Open}, 8(7):e2519963, jul 2025.

\bibitem{roche2025_trop2_bdd}
{Roche Diagnostics}.
\newblock Roche granted {FDA} breakthrough device designation for first {AI}-driven companion diagnostic for non-small cell lung cancer.
\newblock News release (Roche Diagnostics US), April 2025.
\newblock Published 2025-04-29. Accessed 2025-09-25.

\bibitem{tomesh2021quantum}
Teague Tomesh and Margaret Martonosi.
\newblock Quantum codesign.
\newblock {\em IEEE Micro}, 41(5):33--40, 2021.

\bibitem{bravyi2020obstacles}
Sergey Bravyi, Alexander Kliesch, Robert Koenig, and Eugene Tang.
\newblock Obstacles to variational quantum optimization from symmetry protection.
\newblock {\em Physical Review Letters}, 125(26):260505, 2020.

\bibitem{harrow2020small}
Aram~W Harrow.
\newblock Small quantum computers and large classical data sets.
\newblock {\em arXiv preprint arXiv:2004.00026}, 2020.

\bibitem{tomesh2021coreset}
Teague Tomesh, Pranav Gokhale, Eric~R Anschuetz, and Frederic~T Chong.
\newblock Coreset clustering on small quantum computers.
\newblock {\em Electronics}, 10(14):1690, 2021.

\bibitem{gurobi2024}
{Gurobi Optimization, LLC (2024)}.
\newblock Gurobi optimizer reference manual, 2024.
\newblock Software. Documentation available at: \url{https://www.gurobi.com}.

\bibitem{Shor_1997}
Peter~W. Shor.
\newblock Polynomial-time algorithms for prime factorization and discrete logarithms on a quantum computer.
\newblock {\em SIAM Journal on Computing}, 26(5):1484–1509, October 1997.

\bibitem{holevo1973bounds}
A.~S. Holevo.
\newblock Bounds for the quantity of information transmitted by a quantum communication channel.
\newblock {\em Problems of Information Transmission}, 9(3):177--183, 1973.
\newblock English translation of the Russian original.

\bibitem{tomczak2015review}
Katarzyna Tomczak, Patrycja Czerwi{\'n}ska, and Maciej Wiznerowicz.
\newblock {Review The Cancer Genome Atlas (TCGA): an immeasurable source of knowledge}.
\newblock {\em Contemporary Oncology/Wsp{\'o}{\l}czesna Onkologia}, 2015(1):68--77, 2015.

\bibitem{li2023proteogenomic}
Yize Li, Yongchao Dou, Felipe Da~Veiga Leprevost, Yifat Geffen, Anna~P Calinawan, Fran{\c{c}}ois Aguet, Yo~Akiyama, Shankara Anand, Chet Birger, Song Cao, et~al.
\newblock Proteogenomic data and resources for pan-cancer analysis.
\newblock {\em Cancer cell}, 41(8):1397--1406, 2023.

\bibitem{hieromnimon_building_2025}
Hanna~M. Hieromnimon, James Dolezal, Kristina Doytcheva, Frederick~M. Howard, Sara Kochanny, Zhenyu Zhang, Robert~L. Grossman, Kevin Tanager, Cindy Wang, Jakob~Nikolas Kather, Evgeny Izumchenko, Nicole~A. Cipriani, Elana~J. Fertig, Alexander~T. Pearson, and Samantha~J. Riesenfeld.
\newblock Building digital histology models of transcriptional tumor programs with generative deep learning for pathology-based precision medicine.
\newblock 17(1):87.

\bibitem{muzellec2023pydeseq2}
Boris Muzellec, Maria Telenczuk, Vincent Cabeli, and Mathieu Andreux.
\newblock Pydeseq2: a python package for bulk rna-seq differential expression analysis.
\newblock {\em Bioinformatics}, 2023.

\bibitem{teo2025k}
Mariesa~H Teo, Willers Yang, James Sud, Teague Tomesh, Frederic~T Chong, and Eric~R Anschuetz.
\newblock k-contextuality as a heuristic for memory separations in learning.
\newblock {\em arXiv preprint arXiv:2507.11604}, 2025.

\bibitem{shimony1984contextual}
Abner Shimony.
\newblock Contextual hidden variables theories and bell's inequalities.
\newblock {\em The British Journal for the Philosophy of Science}, 35(1):25--45, 1984.

\bibitem{howard2014contextuality}
Mark Howard, Joel Wallman, Victor Veitch, and Joseph Emerson.
\newblock Contextuality supplies the ‘magic’for quantum computation.
\newblock {\em Nature}, 510(7505):351--355, 2014.

\bibitem{gao2022enhancing}
Xun Gao, Eric~R Anschuetz, Sheng-Tao Wang, J~Ignacio Cirac, and Mikhail~D Lukin.
\newblock Enhancing generative models via quantum correlations.
\newblock {\em Physical Review X}, 12(2):021037, 2022.

\bibitem{anschuetz2022interpretable}
Eric~R Anschuetz, Hong-Ye Hu, Jin-Long Huang, and Xun Gao.
\newblock Interpretable quantum advantage in neural sequence learning.
\newblock {\em PRX Quantum}, 4(2):020338, 2023.

\bibitem{sakamoto_1987}
Y.~Sakamoto, M.~Ishiguro, and G.~Kitagawa.
\newblock {\em Akaike Information Criterion Statistics}.
\newblock Kluwer Academic Publishers, 1987.

\bibitem{wainwright2019high}
Martin~J Wainwright.
\newblock {\em High-dimensional statistics: A non-asymptotic viewpoint}, volume~48.
\newblock Cambridge university press, 2019.

\bibitem{wang2009rna}
Zhong Wang, Mark Gerstein, and Michael Snyder.
\newblock Rna-seq: a revolutionary tool for transcriptomics.
\newblock {\em Nature Reviews Genetics}, 10(1):57--63, 2009.

\bibitem{sun2014feature}
Lin Sun and Jiucheng Xu.
\newblock Feature selection using mutual information based uncertainty measures for tumor classification.
\newblock {\em Bio-Medical Materials and Engineering}, 24(1):763--770, 2014.

\bibitem{dolezal2024slideflow}
James~M Dolezal, Sara Kochanny, Emma Dyer, Siddhi Ramesh, Andrew Srisuwananukorn, Matteo Sacco, Frederick~M Howard, Anran Li, Prajval Mohan, and Alexander~T Pearson.
\newblock Slideflow: deep learning for digital histopathology with real-time whole-slide visualization.
\newblock {\em BMC Bioinformatics}, 25(1):134, 2024.

\bibitem{xu2024whole}
Hanwen Xu, Naoto Usuyama, Jaspreet Bagga, Sheng Zhang, Rajesh Rao, Tristan Naumann, Cliff Wong, Zelalem Gero, Javier González, Yu~Gu, et~al.
\newblock A whole-slide foundation model for digital pathology from real-world data.
\newblock {\em Nature}, 630(7993):181--188, 2024.

\bibitem{guyon_2006}
Isabelle Guyon, Masoud Nikravesh, Steve Gunn, and Lotfi~A. Zadeh.
\newblock {\em Feature Extraction: Foundations and Applications}.
\newblock Studies in Fuzziness and Soft Computing. Springer Berlin, 2006.

\bibitem{ros_2024}
Frederic Ros and Rabia Riad.
\newblock {\em Dimensionality reduction}, pages 11--25.
\newblock Springer Nature Switzerland, Cham, 2024.

\bibitem{watkinson2008identification}
John Watkinson, Xiaodong Wang, Tian Zheng, and Dimitris Anastassiou.
\newblock Identification of gene interactions associated with disease from gene expression data using synergy networks.
\newblock {\em BMC systems biology}, 2:1--16, 2008.

\bibitem{watkinson2009inference}
John Watkinson, Kuo-ching Liang, Xiadong Wang, Tian Zheng, and Dimitris Anastassiou.
\newblock Inference of regulatory gene interactions from expression data using three-way mutual information.
\newblock {\em Annals of the New York Academy of Sciences}, 1158(1):302--313, 2009.

\bibitem{abdelwahab2022feature}
Omar Abdelwahab, Nourelislam Awad, Menattallah Elserafy, and Eman Badr.
\newblock A feature selection-based framework to identify biomarkers for cancer diagnosis: A focus on lung adenocarcinoma.
\newblock {\em PLoS One}, 17(9):e0269126, 2022.

\bibitem{ash2012information}
Robert~B Ash.
\newblock {\em Information theory}.
\newblock Courier Corporation, 2012.

\bibitem{kuo_1962}
Hu~Kuo Ting.
\newblock On the amount of information.
\newblock {\em Theory of Probability \& Its Applications}, 7(4):439--447, 1962.

\bibitem{tapia2018neurotransmitter}
M{\'o}nica Tapia, Pierre Baudot, Christine Formisano-Tr{\'e}ziny, Martial~A Dufour, Simone Temporal, Manon Lasserre, B{\'e}atrice Marqu{\`e}ze-Pouey, Jean Gabert, Kazuto Kobayashi, and Jean-Marc Goaillard.
\newblock Neurotransmitter identity and electrophysiological phenotype are genetically coupled in midbrain dopaminergic neurons.
\newblock {\em Scientific reports}, 8(1):13637, 2018.

\bibitem{baudot2019topological}
Pierre Baudot, Monica Tapia, Daniel Bennequin, and Jean-Marc Goaillard.
\newblock Topological information data analysis.
\newblock {\em Entropy}, 21(9):869, 2019.

\bibitem{chen2019suboptimality}
Wei-Kuo Chen, David Gamarnik, Dmitry Panchenko, and Mustazee Rahman.
\newblock {Suboptimality of local algorithms for a class of max-cut problems}.
\newblock {\em The Annals of Probability}, 47(3):1587 -- 1618, 2019.

\bibitem{gamarnik2021overlap}
David Gamarnik and Aukosh Jagannath.
\newblock {The overlap gap property and approximate message passing algorithms for $p$-spin models}.
\newblock {\em The Annals of Probability}, 49(1):180 -- 205, 2021.

\bibitem{grover1996fast}
Lov~K Grover.
\newblock A fast quantum mechanical algorithm for database search.
\newblock In {\em Proceedings of the twenty-eighth annual ACM symposium on Theory of computing}, pages 212--219, 1996.

\bibitem{farhi2014quantum}
Edward Farhi, Jeffrey Goldstone, and Sam Gutmann.
\newblock A quantum approximate optimization algorithm.
\newblock {\em arXiv preprint arXiv:1411.4028}, 2014.

\bibitem{morstyn2022annealing}
Thomas Morstyn.
\newblock Annealing-based quantum computing for combinatorial optimal power flow.
\newblock {\em IEEE Transactions on Smart Grid}, 14(2):1093--1102, 2022.

\bibitem{ferrari2022towards}
Maurizio Ferrari~Dacrema, Fabio Moroni, Riccardo Nembrini, Nicola Ferro, Guglielmo Faggioli, and Paolo Cremonesi.
\newblock Towards feature selection for ranking and classification exploiting quantum annealers.
\newblock In {\em Proceedings of the 45th International ACM SIGIR Conference on Research and Development in Information Retrieval}, pages 2814--2824, 2022.

\bibitem{babbush2021focus}
Ryan Babbush, Jarrod~R McClean, Michael Newman, Craig Gidney, Sergio Boixo, and Hartmut Neven.
\newblock Focus beyond quadratic speedups for error-corrected quantum advantage.
\newblock {\em PRX quantum}, 2(1):010103, 2021.

\bibitem{grossi2022mixed}
Michele Grossi, Noelle Ibrahim, Voica Radescu, Robert Loredo, Kirsten Voigt, Constantin Von~Altrock, and Andreas Rudnik.
\newblock Mixed quantum--classical method for fraud detection with quantum feature selection.
\newblock {\em IEEE Transactions on Quantum Engineering}, 3:1--12, 2022.

\bibitem{mucke2023feature}
Sascha M{\"u}cke, Raoul Heese, Sabine M{\"u}ller, Moritz Wolter, and Nico Piatkowski.
\newblock Feature selection on quantum computers.
\newblock {\em Quantum Machine Intelligence}, 5(1):11, 2023.

\bibitem{sklearn2024featimp}
{Feature importances with a forest of trees}.
\newblock \url{https://scikit-learn.org/stable/auto_examples/ensemble/plot_forest_importances.html}.
\newblock Accessed: 2024-02-29.

\bibitem{hoefler2023disentangling}
Torsten Hoefler, Thomas H{\"a}ner, and Matthias Troyer.
\newblock Disentangling hype from practicality: On realistically achieving quantum advantage.
\newblock {\em Communications of the ACM}, 66(5):82--87, 2023.

\bibitem{albash2018adiabatic}
Tameem Albash and Daniel~A Lidar.
\newblock Adiabatic quantum computation.
\newblock {\em Reviews of Modern Physics}, 90(1):015002, 2018.

\bibitem{hauke2020perspectives}
Philipp Hauke, Helmut~G Katzgraber, Wolfgang Lechner, Hidetoshi Nishimori, and William~D Oliver.
\newblock {Perspectives of quantum annealing: Methods and implementations}.
\newblock {\em Reports on Progress in Physics}, 83(5):054401, 2020.

\bibitem{perlin2024q}
Michael~A Perlin, Ruslan Shaydulin, Benjamin~P Hall, Pierre Minssen, Changhao Li, Kabir Dubey, Rich Rines, Eric~R Anschuetz, Marco Pistoia, and Pranav Gokhale.
\newblock Q-chop: Quantum constrained hamiltonian optimization.
\newblock {\em arXiv preprint arXiv:2403.05653}, 2024.

\bibitem{yu2021quantum}
Hongye Yu, Frank Wilczek, and Biao Wu.
\newblock Quantum algorithm for approximating maximum independent sets.
\newblock {\em Chinese Physics Letters}, 38(3):030304, 2021.

\bibitem{saleem2023approaches}
Zain~H Saleem, Teague Tomesh, Bilal Tariq, and Martin Suchara.
\newblock Approaches to constrained quantum approximate optimization.
\newblock {\em SN Computer Science}, 4(2):183, 2023.

\bibitem{tomesh2024quantum}
Teague Tomesh, Nicholas Allen, Daniel Dilley, and Zain Saleem.
\newblock Quantum-classical tradeoffs and multi-controlled quantum gate decompositions in variational algorithms.
\newblock {\em Quantum}, 8:1493, 2024.

\bibitem{basso2022performance}
Joao Basso, David Gamarnik, Song Mei, and Leo Zhou.
\newblock Performance and limitations of the qaoa at constant levels on large sparse hypergraphs and spin glass models.
\newblock In {\em 2022 IEEE 63rd Annual Symposium on Foundations of Computer Science (FOCS)}, pages 335--343. IEEE, 2022.

\bibitem{farhi2020typical}
Edward Farhi, David Gamarnik, and Sam Gutmann.
\newblock The quantum approximate optimization algorithm needs to see the whole graph: A typical case.
\newblock {\em arXiv preprint arXiv:2004.09002}, 2020.

\bibitem{farhi2020worst}
Edward Farhi, David Gamarnik, and Sam Gutmann.
\newblock The quantum approximate optimization algorithm needs to see the whole graph: Worst case examples.
\newblock {\em arXiv preprint arXiv:2005.08747}, 2020.

\bibitem{anschuetz2022quantum}
Eric~R Anschuetz and Bobak~T Kiani.
\newblock Quantum variational algorithms are swamped with traps.
\newblock {\em Nature Communications}, 13(1):7760, 2022.

\bibitem{bae2022recursive}
Eunok Bae and Soojoon Lee.
\newblock {Recursive QAOA outperforms the original QAOA for the MAX-CUT problem on complete graphs}.
\newblock {\em arXiv}, 11 2022.

\bibitem{frederick2023benchmarking}
Paige Frederick, Rich Rines, FT~Chong, and Pranav Gokhale.
\newblock Benchmarking the recursive quantum approximate optimization algorithm.
\newblock In {\em Quantum Computing, Communication, and Simulation III}, volume 12446, pages 46--52. SPIE, 2023.

\bibitem{qbsolv2017}
Michael Booth, Steven Reinhardt, and Aidan Roy.
\newblock {Partitioning Optimization Problems for Hybrid Classical/Quantum Execution}.
\newblock Technical Report 14-1006A-A, D-Wave Systems Inc., Burnaby, Canada, Oct 2017.

\bibitem{scikit-learn}
F.~Pedregosa, G.~Varoquaux, A.~Gramfort, V.~Michel, B.~Thirion, O.~Grisel, M.~Blondel, P.~Prettenhofer, R.~Weiss, V.~Dubourg, J.~Vanderplas, A.~Passos, D.~Cournapeau, M.~Brucher, M.~Perrot, and E.~Duchesnay.
\newblock Scikit-learn: Machine learning in {P}ython.
\newblock {\em Journal of Machine Learning Research}, 12:2825--2830, 2011.

\bibitem{hoque2014mifs}
Nazrul Hoque, Dhruba~K Bhattacharyya, and Jugal~K Kalita.
\newblock Mifs-nd: A mutual information-based feature selection method.
\newblock {\em Expert Systems with Applications}, 41(14):6371--6385, 2014.

\bibitem{howard2021impact}
Frederick~M Howard, James Dolezal, Sara Kochanny, Jefree Schulte, Heather Chen, Lara Heij, Dezheng Huo, Rita Nanda, Olufunmilayo~I Olopade, Jakob~N Kather, et~al.
\newblock The impact of site-specific digital histology signatures on deep learning model accuracy and bias.
\newblock {\em Nature communications}, 12(1):4423, 2021.

\bibitem{brandao_for_2018}
Fernando G. S.~L. Brandao, Michael Broughton, Edward Farhi, Sam Gutmann, and Hartmut Neven.
\newblock For {Fixed} {Control} {Parameters} the {Quantum} {Approximate} {Optimization} {Algorithm}'s {Objective} {Function} {Value} {Concentrates} for {Typical} {Instances}.
\newblock {\em arXiv:1812.04170 [quant-ph]}, December 2018.
\newblock arXiv: 1812.04170.

\bibitem{galda_transferability_2021}
Alexey Galda, Xiaoyuan Liu, Danylo Lykov, Yuri Alexeev, and Ilya Safro.
\newblock Transferability of optimal qaoa parameters between random graphs.
\newblock In {\em 2021 IEEE International Conference on Quantum Computing and Engineering (QCE)}, pages 171--180. IEEE, 2021.

\bibitem{shaydulin_parameter_2023}
Ruslan Shaydulin, Phillip~C. Lotshaw, Jeffrey Larson, James Ostrowski, and Travis~S. Humble.
\newblock Parameter {Transfer} for {Quantum} {Approximate} {Optimization} of {Weighted} {MaxCut}.
\newblock {\em ACM Transactions on Quantum Computing}, 4(3):1--15, September 2023.
\newblock arXiv:2201.11785 [quant-ph].

\bibitem{augustino2024strategies}
Brandon Augustino, Madelyn Cain, Edward Farhi, Swati Gupta, Sam Gutmann, Daniel Ranard, Eugene Tang, and Katherine Van~Kirk.
\newblock Strategies for running the qaoa at hundreds of qubits.
\newblock {\em arXiv preprint arXiv:2410.03015}, 2024.

\bibitem{hao2024end}
Tianyi Hao, Zichang He, Ruslan Shaydulin, Jeffrey Larson, and Marco Pistoia.
\newblock End-to-end protocol for high-quality qaoa parameters with few shots.
\newblock {\em arXiv preprint arXiv:2408.00557}, 2024.

\bibitem{pelofske2024scaling}
Elijah Pelofske, Andreas B{\"a}rtschi, Lukasz Cincio, John Golden, and Stephan Eidenbenz.
\newblock Scaling whole-chip qaoa for higher-order ising spin glass models on heavy-hex graphs.
\newblock {\em npj Quantum Information}, 10(1):109, 2024.

\bibitem{campbell2022qaoa}
Colin Campbell and Edward Dahl.
\newblock Qaoa of the highest order.
\newblock In {\em 2022 IEEE 19th International Conference on Software Architecture Companion (ICSA-C)}, pages 141--146. IEEE, 2022.

\bibitem{pelofske2024short}
Elijah Pelofske, Andreas B{\"a}rtschi, and Stephan Eidenbenz.
\newblock Short-depth qaoa circuits and quantum annealing on higher-order ising models.
\newblock {\em npj Quantum Information}, 10(1):30, 2024.

\bibitem{Campbell_2019}
Earl Campbell.
\newblock Random compiler for fast hamiltonian simulation.
\newblock {\em Phys. Rev. Lett.}, 123:070503, Aug 2019.

\bibitem{bonami2008}
Pierre Bonami, Lorenz~T. Biegler, Andrew~R. Conn, Gérard Cornuéjols, Ignacio~E. Grossmann, Carl~D. Laird, Jon Lee, Andrea Lodi, François Margot, Nicolas Sawaya, and Andreas Wächter.
\newblock An algorithmic framework for convex mixed integer nonlinear programs.
\newblock {\em Discrete Optimization}, 5(2):186--204, 2008.
\newblock In Memory of George B. Dantzig.

\bibitem{ampl1989}
Robert Fourer, David~M. Gay, and Brian~W. Kernighan.
\newblock Ampl: A mathematical programing language.
\newblock In Stein~W. Wallace, editor, {\em Algorithms and Model Formulations in Mathematical Programming}, pages 150--151, Berlin, Heidelberg, 1989. Springer Berlin Heidelberg.

\bibitem{anthony2017quadratic}
Martin Anthony, Endre Boros, Yves Crama, and Aritanan Gruber.
\newblock Quadratic reformulations of nonlinear binary optimization problems.
\newblock {\em Mathematical Programming}, 162(1):115--144, 2017.

\bibitem{kronqvist2019review}
Jan Kronqvist, David~E Bernal, Andreas Lundell, and Ignacio~E Grossmann.
\newblock A review and comparison of solvers for convex minlp.
\newblock {\em Optimization and Engineering}, 20(2):397--455, 2019.

\bibitem{weidenfeller2022scaling}
Johannes Weidenfeller, Lucia~C Valor, Julien Gacon, Caroline Tornow, Luciano Bello, Stefan Woerner, and Daniel~J Egger.
\newblock Scaling of the quantum approximate optimization algorithm on superconducting qubit based hardware.
\newblock {\em Quantum}, 6:870, 2022.

\bibitem{infleqtion2025_qc_roadmap}
{Infleqtion}.
\newblock Quantum computer roadmap.
\newblock PDF, September 2025.
\newblock Last update: September 2025.

\bibitem{ibm2025_dev_innov_roadmap}
{IBM Quantum}.
\newblock Ibm quantum development \& innovation roadmap.
\newblock PDF, 2025.
\newblock Accessed 2025-09-29.

\bibitem{akshay2021parameter}
Vishwanathan Akshay, Daniil Rabinovich, Ernesto Campos, and Jacob Biamonte.
\newblock Parameter concentrations in quantum approximate optimization.
\newblock {\em Physical Review A}, 104(1):L010401, 2021.

\bibitem{rines2025demonstration}
Rich Rines, Benjamin Hall, Mariesa~H Teo, Joshua Viszlai, Daniel~C Cole, David Mason, Cameron Barker, Matt~J Bedalov, Matt Blakely, Tobias Bothwell, et~al.
\newblock Demonstration of a logical architecture uniting motion and in-place entanglement: Shor's algorithm, constant-depth cnot ladder, and many-hypercube code.
\newblock {\em arXiv preprint arXiv:2509.13247}, 2025.

\bibitem{akahoshi2024partially}
Yutaro Akahoshi, Kazunori Maruyama, Hirotaka Oshima, Shintaro Sato, and Keisuke Fujii.
\newblock Partially fault-tolerant quantum computing architecture with error-corrected clifford gates and space-time efficient analog rotations.
\newblock {\em PRX quantum}, 5(1):010337, 2024.

\bibitem{choi2023fault}
Hyeongrak Choi, Frederic~T Chong, Dirk Englund, and Yongshan Ding.
\newblock Fault tolerant non-clifford state preparation for arbitrary rotations.
\newblock {\em arXiv preprint arXiv:2303.17380}, 2023.

\bibitem{kitaev1997quantum}
A~Yu Kitaev.
\newblock Quantum computations: algorithms and error correction.
\newblock {\em Russian Mathematical Surveys}, 52(6):1191, 1997.

\bibitem{nielsen2010quantum}
Michael~A Nielsen and Isaac~L Chuang.
\newblock {\em Quantum computation and quantum information}.
\newblock Cambridge university press, 2010.

\bibitem{omanakuttan2025threshold}
Sivaprasad Omanakuttan, Zichang He, Zhiwei Zhang, Tianyi Hao, Arman Babakhani, Sami Boulebnane, Shouvanik Chakrabarti, Dylan Herman, Joseph Sullivan, Michael~A Perlin, et~al.
\newblock Threshold for fault-tolerant quantum advantage with the quantum approximate optimization algorithm.
\newblock {\em arXiv preprint arXiv:2504.01897}, 2025.

\bibitem{leverrier2022quantum}
Anthony Leverrier and Gilles Z{\'e}mor.
\newblock Quantum tanner codes.
\newblock In {\em 2022 IEEE 63rd Annual Symposium on Foundations of Computer Science (FOCS)}, pages 872--883. IEEE, 2022.

\bibitem{bravyi2024high}
Sergey Bravyi, Andrew~W Cross, Jay~M Gambetta, Dmitri Maslov, Patrick Rall, and Theodore~J Yoder.
\newblock High-threshold and low-overhead fault-tolerant quantum memory.
\newblock {\em Nature}, 627(8005):778--782, 2024.

\bibitem{baspin2022quantifying}
Nou{\'e}dyn Baspin and Anirudh Krishna.
\newblock Quantifying nonlocality: How outperforming local quantum codes is expensive.
\newblock {\em Physical review letters}, 129(5):050505, 2022.

\bibitem{yoder2025tour}
Theodore~J Yoder, Eddie Schoute, Patrick Rall, Emily Pritchett, Jay~M Gambetta, Andrew~W Cross, Malcolm Carroll, and Michael~E Beverland.
\newblock Tour de gross: A modular quantum computer based on bivariate bicycle codes.
\newblock {\em arXiv preprint arXiv:2506.03094}, 2025.

\bibitem{yoshida2025concatenate}
Satoshi Yoshida, Shiro Tamiya, and Hayata Yamasaki.
\newblock Concatenate codes, save qubits.
\newblock {\em npj Quantum Information}, 11(1):88, 2025.

\bibitem{litinski2025blocklet}
Daniel Litinski.
\newblock Blocklet concatenation: Low-overhead fault-tolerant protocols for fusion-based quantum computation.
\newblock {\em arXiv preprint arXiv:2506.13619}, 2025.

\bibitem{goto2024high}
Hayato Goto.
\newblock High-performance fault-tolerant quantum computing with many-hypercube codes.
\newblock {\em Science Advances}, 10(36):eadp6388, 2024.

\bibitem{yamasaki2024time}
Hayata Yamasaki and Masato Koashi.
\newblock Time-efficient constant-space-overhead fault-tolerant quantum computation.
\newblock {\em Nature Physics}, 20(2):247--253, 2024.

\bibitem{tang2021quantum}
Ewin Tang.
\newblock Quantum principal component analysis only achieves an exponential speedup because of its state preparation assumptions.
\newblock {\em Physical Review Letters}, 127(6):060503, 2021.

\bibitem{montanari2025optimization}
Andrea Montanari.
\newblock Optimization of the sherrington--kirkpatrick hamiltonian.
\newblock {\em SIAM Journal on Computing}, 54(4):FOCS19--1, 2025.

\bibitem{boehm2022harnessing}
Kevin~M Boehm, Pegah Khosravi, Rami Vanguri, Jianjiong Gao, and Sohrab~P Shah.
\newblock Harnessing multimodal data integration to advance precision oncology.
\newblock {\em Nature Reviews Cancer}, 22(2):114--126, 2022.

\bibitem{huang2021hpvneg_hnscc_proteogenomic}
Chen Huang, Lijun Chen, Sara~R Savage, Rodrigo Vargas~Eguez, Yongchao Dou, Yize Li, Felipe da~Veiga~Leprevost, Eric~J Jaehnig, Jonathan~T Lei, Bo~Wen, et~al.
\newblock Proteogenomic insights into the biology and treatment of hpv-negative head and neck squamous cell carcinoma.
\newblock {\em Cancer Cell}, 39(3):361--379.e16, March 2021.
\newblock Epub 2021-01-07; PMID: 33417831.

\bibitem{izumchenko2015notch1}
E~Izumchenko, K~Sun, S~Jones, M~Brait, N~Agrawal, W~Koch, C~L McCord, D~R Riley, S~V Angiuoli, V~E Velculescu, W~W Jiang, and D~Sidransky.
\newblock {NOTCH1} mutations are drivers of oral tumorigenesis.
\newblock {\em Cancer Prevention Research (Phila)}, 8(4):277--286, April 2015.
\newblock Epub 2014-11-18; PMID: 25406187; PMCID: PMC4383685.

\bibitem{broner2021dclk1}
E.~C. Broner, J.~A. Trujillo, M.~Korzinkin, T.~Subbannayya, N.~Agrawal, I.~V. Ozerov, A.~Zhavoronkov, L.~Rooper, N.~Kotlov, L.~Shen, A.~T. Pearson, A.~J. Rosenberg, P.~A. Savage, V.~Mishra, A.~Chatterjee, D.~Sidransky, and E.~Izumchenko.
\newblock Doublecortin-like kinase 1 ({DCLK1}) is a novel {NOTCH} pathway signaling regulator in head and neck squamous cell carcinoma.
\newblock {\em Frontiers in Oncology}, 11:677051, July 2021.

\bibitem{peri2017nsd1_nsd2}
S~Peri, E~Izumchenko, A.~D. Schubert, M.~J. Slifker, K~Ruth, I.~G. Serebriiskii, T~Guo, B.~A. Burtness, R~Mehra, E.~A. Ross, D~Sidransky, and E.~A. Golemis.
\newblock {NSD1}- and {NSD2}-damaging mutations define a subset of laryngeal tumors with favorable prognosis.
\newblock {\em Nature Communications}, 8(1):1772, November 2017.

\bibitem{ferrarotto2022al101}
R.~Ferrarotto, V.~Mishra, E.~Herz, A.~Yaacov, O.~Solomon, R.~Rauch, A.~Mondshine, M.~Motin, T.~Leibovich-Rivkin, M.~Davis, J.~Kaye, C.~R. Weber, L.~Shen, A.~T. Pearson, A.~J. Rosenberg, X.~Chen, A.~Singh, J.~C. Aster, N.~Agrawal, and E.~Izumchenko.
\newblock {AL101}, a gamma-secretase inhibitor, has potent antitumor activity against adenoid cystic carcinoma with activated {NOTCH} signaling.
\newblock {\em Cell Death \& Disease}, 13(8):678, August 2022.

\bibitem{mkrtchyan2022dna_repair_multiomics}
G.~V. Mkrtchyan, A.~Veviorskiy, E.~Izumchenko, A.~Shneyderman, F.~W. Pun, I.~V. Ozerov, A.~Aliper, A.~Zhavoronkov, and M.~Scheibye-Knudsen.
\newblock High-confidence cancer patient stratification through multiomics investigation of {DNA} repair disorders.
\newblock {\em Cell Death \& Disease}, 13(11):999, November 2022.

\bibitem{pun2023ai_dual_purpose_targets}
F.~W. Pun, G.~H.~D. Leung, H.~W. Leung, J.~Rice, T.~Schmauck-Medina, S.~Lautrup, X.~Long, B.~H.~M. Liu, C.~W. Wong, I.~V. Ozerov, A.~Aliper, F.~Ren, A.~J. Rosenberg, N.~Agrawal, E.~Izumchenko, E.~F. Fang, and A.~Zhavoronkov.
\newblock A comprehensive {AI}-driven analysis of large-scale omic datasets reveals novel dual-purpose targets for the treatment of cancer and aging.
\newblock {\em Aging Cell}, 22(12):e14017, December 2023.
\newblock Epub 2023-10-27; PMID: 37888486; PMCID: PMC10726874.

\bibitem{steinobrien2018timecourse_omics}
G.~Stein-O'Brien, L.~T. Kagohara, S.~Li, M.~Thakar, R.~Ranaweera, H.~Ozawa, H.~Cheng, M.~Considine, S.~Schmitz, A.~V. Favorov, L.~V. Danilova, J.~A. Califano, E.~Izumchenko, D.~A. Gaykalova, C.~H. Chung, and E.~J. Fertig.
\newblock Integrated time course omics analysis distinguishes immediate therapeutic response from acquired resistance.
\newblock {\em Genome Medicine}, 10(1):37, May 2018.

\bibitem{hayashi2020gulp1_nrf2}
M.~Hayashi, E.~Guida, Y.~Inokawa, R.~Goldberg, L.~O. Reis, A.~Ooki, M.~Pilli, P.~Sadhukhan, J.~Woo, W.~Choi, E.~Izumchenko, L.~M. Gonzalez, L.~Marchionni, A.~Zhavoronkov, M.~Brait, T.~Bivalacqua, A.~Baras, G.~J. Netto, W.~Koch, A.~Singh, and M.~O. Hoque.
\newblock {GULP1} regulates the {NRF2}-{KEAP1} signaling axis in urothelial carcinoma.
\newblock {\em Science Signaling}, 13(645):eaba0443, August 2020.

\bibitem{osman2023nrf2_cisplatin_hnscc}
A.~A. Osman, E.~Arslan, M.~Bartels, C.~Michikawa, A.~Lindemann, K.~Tomczak, W.~Yu, V.~Sandulache, W.~Ma, L.~Shen, J.~Wang, A.~K. Singh, M.~J. Frederick, N.~D. Spencer, J.~Kovacs, T.~Heffernan, W.~F. Symmans, K.~Rai, and J.~N. Myers.
\newblock Dysregulation and epigenetic reprogramming of {NRF2} signaling axis promote acquisition of cisplatin resistance and metastasis in head and neck squamous cell carcinoma.
\newblock {\em Clinical Cancer Research}, 29(7):1344--1359, April 2023.

\bibitem{zhu2022immune_infiltration_hnscc}
C.~Zhu, Q.~Wu, N.~Yang, Z.~Zheng, F.~Zhou, and Y.~Zhou.
\newblock Immune infiltration characteristics and a gene prognostic signature associated with the immune infiltration in head and neck squamous cell carcinoma.
\newblock {\em Frontiers in Genetics}, 13:848841, May 2022.

\bibitem{chen2022immune_ssgsea_oscc}
Y.~Chen, Y.~Feng, F.~Yan, Y.~Zhao, H.~Zhao, and Y.~Guo.
\newblock A novel immune-related gene signature to identify the tumor microenvironment and prognose disease among patients with oral squamous cell carcinoma patients using {ssGSEA}: A bioinformatics and biological validation study.
\newblock {\em Frontiers in Immunology}, 13:922195, July 2022.

\bibitem{ega-parkinsons}
European Genome-Phenome Archive.
\newblock {Multiomics analyses of Parkinson's disease midbrains}.
\newblock Online EGA Database, 2023.
\newblock Accessed 2025-03-08 at: \url{https://ega-archive.org/studies/EGAS00001004966}.

\bibitem{ega-pml}
European Genome-Phenome Archive.
\newblock {Progressive multifocal leukoencephalopathy (PML)}.
\newblock Online EGA Database, 2023.
\newblock Accessed 2025-03-08 at: \url{https://ega-archive.org/studies/EGAS50000000139}.

\bibitem{ega-ms}
European Genome-Phenome Archive.
\newblock {Epigenomic priming of immune genes implicates oligodendroglia in multiple sclerosissusceptibility}.
\newblock Online EGA Database, 2023.
\newblock Accessed 2025-03-08 at: \url{https://ega-archive.org/studies/EGAS00001005911}.

\bibitem{ega-als}
European Genome-Phenome Archive.
\newblock {Multiomics analyses of ALS prefrontal cortex tissue}.
\newblock Online EGA Database, 2023.
\newblock Accessed 2025-03-08 at: \url{https://ega-archive.org/studies/EGAS00001007318}.

\bibitem{ega-myositis}
European Genome-Phenome Archive.
\newblock {Cell type mapping of inflammatory muscle diseases highlights selective myofiber vulnerability in inclusion body myositis}.
\newblock Online EGA Database, 2023.
\newblock Accessed 2025-03-08 at: \url{https://ega-archive.org/studies/EGAS50000000310}.

\bibitem{ega-covid}
European Genome-Phenome Archive.
\newblock {Multi-omics characterisation of immune cells in Long Covid}.
\newblock Online EGA Database, 2023.
\newblock Accessed 2025-03-08 at: \url{https://ega-archive.org/studies/EGAS50000000142}.

\end{thebibliography}

\end{document}